\pdfoutput=1
\documentclass[a4paper,11pt]{article}
\usepackage{amsmath}
\usepackage{amssymb}
\usepackage{graphicx}
\usepackage{xfrac}

\usepackage[utf8]{inputenc}

\usepackage[square,numbers,comma,sort&compress]{natbib}
\usepackage[colorlinks=true,citecolor=blue,linkcolor=purple,urlcolor=purple]{hyperref}
\usepackage{geometry}
\usepackage[usenames,dvipsnames,svgnames,table]{xcolor}
\usepackage{booktabs}
\usepackage{slashed}

\def\varabstract{ }
\def\varkeywords{ }
\def\vararxivnumber{ }
\def\vartitle{ }
\def\varsubtitle{ }
\renewcommand{\title}[1]{\gdef\vartitle{#1}}

\renewcommand{\abstract}[1]{\gdef\varabstract{#1}}
\newcommand{\keywords}[1]{\gdef\varkeywords{#1}}

\newtoks\authtoks
\renewcommand{\author}[2][]{%
	\authtoks=\expandafter{\the\authtoks#2$^{#1}$\ }%
}
\newtoks\affiltoks
\newcommand{\affiliation}[2][]{%
    \affiltoks=\expandafter{\the\affiltoks{\item[$^{#1}$]#2}}%
}
\newtoks\emailtoks\newcounter{emailcounter}%
\newcommand{\emailAdd}[1]{%
\ifnum\theemailcounter>0\emailtoks=\expandafter{\the\emailtoks, \typeemail{#1}}%
\else\emailtoks=\expandafter{\typeemail{#1}}%
\fi
\stepcounter{emailcounter}}
\newcommand{\typeemail}[1]{\href{mailto:#1}{\tt #1}}

\renewcommand\maketitle{
	\newgeometry{margin=2cm}
	\pagestyle{empty}\setcounter{page}{0}
	{\huge\flushleft\sffamily\bfseries\vartitle\\\Large\varsubtitle\par}
\vskip6ex
{\large\bfseries\raggedright\sffamily\the\authtoks\par}
\vskip2ex
\begin{list}{}{%
\setlength{\leftmargin}{0.28cm}%
\setlength{\labelsep}{0pt}%
\setlength{\itemsep}{-3pt}%
\setlength{\topsep}{-\parskip}}
\itshape\small%
\the\affiltoks
\end{list}
\vskip2ex
\noindent\hspace{0.28cm}\begin{minipage}[l]{.9\textwidth}
\begin{flushleft}
\textit{E-mail:} \the\emailtoks
\end{flushleft}
\end{minipage}
\vskip5ex
\noindent{\renewcommand\baselinestretch{.9}\textsc{Abstract:}}\ \varabstract
\vskip5ex
\if!\varkeywords!\else\noindent{\textsc{Keywords:}}\ \varkeywords \vskip2ex\fi
\if!\vararxivnumber!\else\noindent{\textsc{ArXiv ePrint:}} \href{http://arxiv.org/abs/\vararxivnumber}{\vararxivnumber}\vskip2ex\fi

\newpage
\restoregeometry
\pagestyle{plain}

\setcounter{footnote}{0}
}

\usepackage[square,numbers,comma,sort&compress]{natbib}

\definecolor{MS}{rgb}{0,0,1}

\usepackage{ifthen}

	\newcommand{\barlimc}[7]{
  \pgfmathparse{\mypos+0.3}
  \edef\mypos{\pgfmathresult}
		\node[left,scale=0.6] at (0,\mypos) {#1};
		\pgfmathparse{#3 > 5 ? 1 : 0}
		\ifthenelse{\pgfmathresult=1}{
			\fill[#2] ($(0,\mypos)+(0,-0.1)$) rectangle +(5,0.2);
			\fill[white] ($(0,\mypos)+(3.5,-0.1)$) rectangle +(0.3,0.2);
			\draw[decoration={zigzag},decorate,#2,very thick] (3.4,\mypos) to +(0.5,0);
			\node[left,scale=0.6] at (5,\mypos) {#3};
			}{
			\fill[#2] ($(0,\mypos)+(0,-0.1)$) rectangle +(#3,0.2);
			\node[left,scale=0.6] at (#3,\mypos) {#3};
		}
		\fill[#4] ($(0,\mypos)+(0,-0.1)$) rectangle +(#5,0.2);
		\node[left,scale=0.6] at (#5,\mypos) {#5};
		\fill[#6] ($(0,\mypos)+(0,-0.1)$) rectangle +(#7,0.2);
		\pgfmathparse{#7 <0.3 ? 1 : 0}
		\ifthenelse{\pgfmathresult=1}{
			\node[right,scale=0.6] at (0,\mypos) {#7};
		}{
		\node[left,scale=0.6] at (#7,\mypos) {#7};
	}
}

\title{Anatomy of a six-parameter fit to the $b\to s \ell^+\ell^-$ anomalies}

\author[1]{Bernat Capdevila,}\emailAdd{bcapdevila@ifae.es}
\author[2,3]{Ursula Laa,}\emailAdd{ursula.laa@monash.edu}
\author[2]{and German Valencia}\emailAdd{german.valencia@monash.edu}

\affiliation[1]{Universitat Aut\`onoma de Barcelona, Institut de F\'isica d'Altes\\ Energies (IFAE), The Barcelona Institute of Science and Technology, Campus UAB, 08193 Bellaterra (Barcelona).}

\affiliation[2]{School of Physics and Astronomy, Monash University,
Melbourne VIC-3800}

\affiliation[3]{School of Econometrics and Business Statistics, Monash University,
Melbourne VIC-3800}

\abstract{
Discrepancies between measurements of decay modes with an underlying quark level transition
$b\to s \ell^+\ell^-$ and standard model (SM) predictions have persisted for several years, particularly for the muon channels.
The inadequacy of the SM becomes more compelling in a global fit.
For example, Ref.~\cite{Capdevila:2017bsm} described 175 observables by six parameters encoding new physics and quantified the disagreement
with the SM at about the $5\sigma$ level.
While certain one and two parameter fits have previously been considered in detail, we establish a framework for the detailed discussion of the full 6d fit.
We visualize and quantify the 6d $1\sigma$ region around the best fit point and define fit uncertainties for both current and future observables.
We then define metrics quantifying the deviations between measurements and both SM and best fit predictions.
These metrics relate observables to directions in parameter space, revealing their precise role in the fit,
thus providing guidance for future theoretical and experimental work.
Some metrics further quantify the role of correlated uncertainties, which turns out to be significant.
For example the relevance of angular observables such as $P_5^\prime$ is reduced in this context.
Finally, studying the space of observables allows us to discuss the internal tensions in the fit.
}


\begin{document}

\maketitle

{\hypersetup{linkcolor=black}
  \tableofcontents}
  
\newpage

\section{Introduction}

Many measurements have been performed in recent years on decay modes with an underlying quark transition $b\to s \ell^+\ell^-$, where $\ell$ includes muons and electrons. Not surprisingly, some of these measurements show deviations from the standard model (SM) predictions by a few standard deviations. More interesting is the fact that several of these deviations from the SM appear to be `in the same direction', in such a way that when quantified by a global fit the discrepancy with the SM is at just over the 5~$\sigma$ level.

Several global fits that allow for the possibility of new physics (NP) in a model independent fashion described in the framework of effective field theory have been performed in the literature \cite{Capdevila:2017bsm,Descotes-Genon:2015uva,Alonso:2015sja,DAmico:2017mtc,Altmannshofer:2017fio,Altmannshofer:2017yso,Geng:2017svp,Arbey:2018ics,Ciuchini:2017mik}. The basis for our study will be the fit of Ref.~\cite{Capdevila:2017bsm} which allows for NP by floating six Wilson coefficients of the effective low energy Hamiltonian responsible for the quark-level transition:
\begin{eqnarray}
{\cal H}_{\rm eff} = -\frac{4G_F}{\sqrt{2}}V_{tb}V^\star_{ts}\sum_iC_i{\cal O}_i.
\label{effH}
\end{eqnarray}
The six operators in question are
\begin{eqnarray}
{\cal O}_7 = \frac{e}{16\pi^2}m_b(\bar{s}\sigma_{\mu\nu}P_Rb)F^{\mu\nu},&&{\cal O}_{7^\prime} = \frac{e}{16\pi^2}m_b(\bar{s}\sigma_{\mu\nu}P_Lb)F^{\mu\nu}, \nonumber \\
{\cal O}_{9\ell} = \frac{e^2}{16\pi^2}(\bar{s}\gamma_{\mu}P_Lb)(\bar\ell \gamma^\mu\ell),&&{\cal O}_{9^\prime\ell} = \frac{e^2}{16\pi^2}(\bar{s}\gamma_{\mu}P_Rb)(\bar\ell \gamma^\mu\ell), \nonumber \\
{\cal O}_{10\ell} = \frac{e^2}{16\pi^2}(\bar{s}\gamma_{\mu}P_Lb)(\bar\ell \gamma^\mu\gamma_5\ell),&&{\cal O}_{10^\prime\ell} = \frac{e^2}{16\pi^2}(\bar{s}\gamma_{\mu}P_Rb)(\bar\ell \gamma^\mu\gamma_5\ell).
\end{eqnarray}
The factorization scale is taken at 4.8~GeV, so that $m_b(\mu_b)$ is the $\bar{\text{MS}}$ running b-quark mass at that scale, and the SM values of the Wilson coefficients are $C^\text{SM}_{7,9,10}=-0.29,4.07,-4.31$ and $C^\text{SM}_{7^\prime,9^\prime,10^\prime}=0$. New physics would be parametrized in a model independent way by deviations in these coefficients from their SM values, $C_{i\ell} \equiv C_i^\text{SM}+C^\text{NP}_{i\ell}$ ($i=7^{(')},9^{(')},10^{(')}$, $\ell=\mu$), that is, we only treat the muon coefficients as free parameters. We assume these deviations to be real and drop any  CP violating observables. We note that this fit does not include (pseudo) scalar or tensor operators.

The full fit includes all available results for the following decay channels (see appendix for a full list of the 175 observables with references): 
\begin{itemize}
\item $B^{(0,+)}\to K^{*(0,+)}\mu^+\mu^-$, $B^{(0,+)}\to K^{*(0,+)}e^+e^-$, $B^{(0,+)}\to K^{*(0,+)}\gamma$,
\item $B^{(0,+)}\to K^{(0,+)}\mu^+\mu^-$, $B^+\to K^+e^+e^-$ (through the $R_K$ observable),
\item $B_s\to \phi\mu^+\mu^-$, $B_s\to \phi\gamma$,
\item $B\to X_s\mu^+\mu^-$, $B\to X_s\gamma$ and $B_s\to \mu^+\mu^-$.
\end{itemize} 
Although experimental data on the baryonic decay $\Lambda_b \to \Lambda\mu^+\mu^-$ is available, it is not included in the fit  
because for the low-$q^2$ region the QCD factorization is poorly understood~\cite{Wang:2015ndk},
while at high-$q^2$, where a recent determination of the $\Lambda_b\to\Lambda$ form factors from lattice QCD~\cite{Detmold:2016pkz}
reduces theory uncertainties, experimental errors are large~\cite{Aaij:2015xza} (see discussion in~\cite{Descotes-Genon:2015uva,Meinel:2016grj}).

Ref.~\cite{Capdevila:2017bsm} discussed several scenarios with one or two non-zero NP contributions at a time in detail and analysed the general scenario where NP contributions are allowed to all six muonic operators. Updating their numbers\footnote{The experimental results used here were up to date as of February 2019.} we find the best fit point (BF) to be,\footnote{The minimisation of the $\chi^2$ function leading to the BF above is performed by means of the Markov-Chain Monte Carlo Metropolis-Hastings algorithm. The small differences between the numbers quoted here and the results of Ref.~\cite{Capdevila:2017bsm} are a manifestation of the intrinsic error that a numerical minimisation routine always carries.}
\begin{eqnarray}
C^\text{NP}_7 = 4.3 \times 10^{-3},&& C^\text{NP}_{7^\prime}=0.019, \nonumber \\
C^\text{NP}_{9\mu} = -1.06, && C^\text{NP}_{9^\prime\mu}=0.37,  \nonumber \\
C^\text{NP}_{10\mu} = 0.34, && C^\text{NP}_{10^\prime\mu}=-0.04\ .
\label{bfp}
\end{eqnarray}

These coefficients lead to a $\chi^2$ lower than that of the SM by 39.9, indicating that the best fit differs from the SM at the level of 5~$\sigma$. In this particular fit, the coefficients $C^\text{NP}_{9\mu}$, $C^\text{NP}_{9^\prime\mu}$, $C^\text{NP}_{10\mu} $ and $C^\text{NP}_{10^\prime\mu}$ are so labelled because they are only allowed to differ from the SM for the muons, breaking lepton flavor universality. Bearing in mind that the six coefficients identified explicitly in Eq.~\ref{bfp} are the {\it only} ones allowed to differ from their SM value in this study, we will proceed to drop the superscript NP and the muon label for notational simplicity.

From this starting point, our paper provides a comprehensive statistical analysis in the full six-dimensional space,
by visualizing aspects of it with the aid of the grand and guided tour \cite{doi:10.1080/10618600.1995.10474674},
a systematic comparison between predictions of the SM and BF point against measurements,
and a quantitative analysis relating parameter directions and individual observables, based on the Hessian approximation to the $\chi^2$ function.
Throughout this analysis we show for the first time how knowledge of correlated uncertainties is important in judging the relative importance of individual observables to the fit.

Our paper is organized as follows. In Section~\ref{s:bestfit} we visualize the neighborhood of the best fit point,  we present the Hessian matrix of second derivatives at the minimum of the $\chi^2$ function and use it to construct twelve points characterizing the boundary of the one sigma region. These points will serve as benchmarks for our quantitative studies. We also use the Hessian to quantify the difference between the six-dimensional BF point and simple one and two parameter fits studied previously in the literature. In Section~\ref{s:quanti} we define metrics to compare the predictions of the fit at different parameter sets and apply them to all the observables. In Section~\ref{s:newob} we discuss future observables designed to further test lepton flavor universality and assess their potential impact on the global fit. Finally, in Section~\ref{s:con} we summarize our results and conclude.

\section{The best fit point and its neighborhood}\label{s:bestfit}

The parameter space in our problem corresponds to the set of six Wilson coefficients $C^{}_i$ encoding physics beyond the SM as per Eq.~\ref{effH}. Predictions for the experimental observables, as well as their theoretical uncertainties, are functions of these six parameters. 
The minimization of the $\chi^2$ function has been performed numerically, and in fact the analytic form of the function is not known. It is thus desirable to have an approximate analytic form for this function which we construct in terms of the Hessian matrix of second derivatives of  $\chi^2$ at the minimum. This information will form the basis of our analysis in this section. 

We obtain this matrix numerically in two different ways.  
We first use the minimization procedure  to provide us with a set of six-dimensional points (4959) near the minimum, $S_{\chi^2}(C^{}_i)$,  along with  their $\chi^2$.  With this set of points we construct an approximation to the $\chi^2$ function near its minimum and we  derive from it the Hessian.Alternatively, once the BF is determined, the Hessian matrix can also be computed by explicit evaluation of the corresponding second partial derivatives of the $\chi^2$ function with numerical routines. We verify agreement between these two numerical Hessian matrices within the uncertainty of the approximations. 

\subsection{Visualization of the BF region}

We begin by visualising the six dimensional parameter region using the sample of points $S_{1\sigma}$ (points of $S_{\chi^2}(C^{}_i)$ which are at most $1\sigma$ away from the BF)  and using the tour algorithm to obtain a sequence of 2-d projections that result in an animation of scatter plots  \href{https://uschilaa.github.io/animations/points/animation.html}{(as shown here)}.\footnote{This procedure follows the general methods of Ref.~\cite{Asimov:1985:GTT:2812.2906,Buja:2004,Cook2008} as recently described by two of us in the context of parton distribution fits \cite{Cook:2018mvr}.} To produce the animation we first  center all parameter directions, the centered points are then scaled  with the standard deviation in each direction within $S_{1\sigma}$, such that all directions have comparable scales.\footnote{Differences in scale that describe how well each direction is constrained in the fit are thus eliminated, and they will be discussed separately below.} We then search for appropriate projections that illustrate the separation between the BF point and the SM point.  Finally, to show these projections we center the view by subtracting the mean position of the points in $S_{1\sigma}$.

In this animation, the corresponding projection matrix for each view is shown as ``axes''. Notice that the axes are centered with respect to the point cloud and therefore the origin is somewhat shifted with respect to the BF point. 
Watching the animation builds intuition about the features of $S_{1\sigma}$, for example it illustrates how features observed in lower dimensional studies embed in the full space.

\begin{figure}[!h]
\center{\includegraphics[scale=.5]{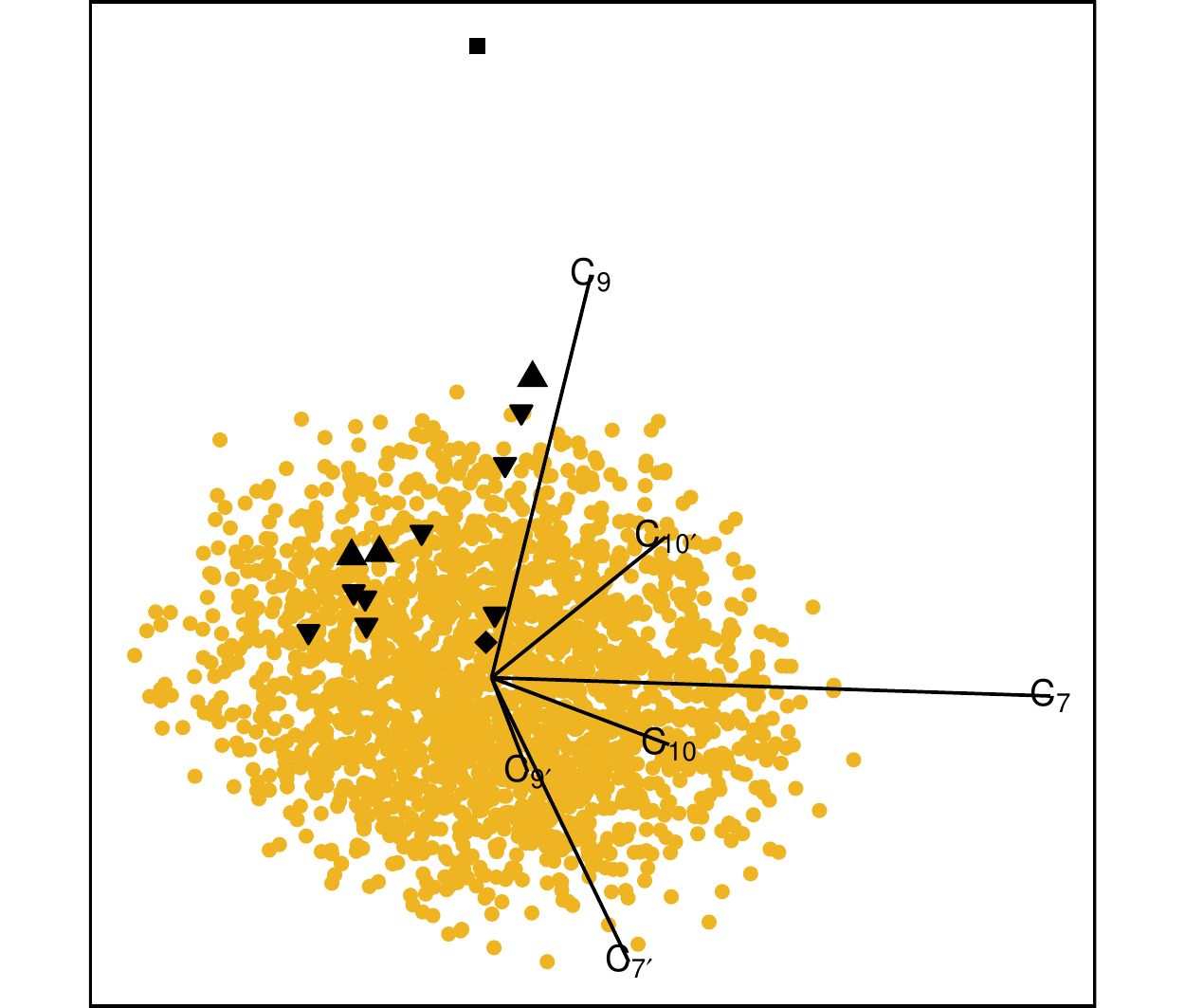}}
\caption{Visualization of parameter space in six dimensions via a general two dimensional projection. The set  $S_{1\sigma}$ of points within $1\sigma$ of the BF is shown in yellow, the black symbols mark the SM point (box); the BF point (diamond); one dimensional best fits (upwards pointing triangle); and two dimensional best fits (downwards pointing triangle). The  $S_{1\sigma}$ cloud is seen to be separated from the SM point mostly along the $C_9^{}$ direction.}
\label{fig:68center}
\end{figure}

Most notably in this example we observe how the SM point moves away from   $S_{1\sigma}$. The animation provides a striking picture of the separation between the BF region and the SM that is mostly along the $C^{}_9$ direction, a result that is well known in the literature.   We further illustrate this with a static picture in Figure~\ref{fig:68center}, where the projection has been selected for showing the large distance between the SM and BF point. For comparison we illustrate the positions of the SM, BF and selected one and two dimensional best fits from Ref.~\cite{Capdevila:2017bsm} (listed in the first column of Table~\ref{t;bfpulls}) as described in the caption.
Another interesting feature that can be seen in the animation is that  all these lower dimensional best fits are found roughly in the same half of $S_{1\sigma}$.

\subsection{Quadratic approximation}

The Hessian matrix is the standard tool to construct a quadratic approximation to the $\chi^2$ function in the vicinity of the global minimum.  The eigenvectors of this matrix correspond to the directions of the principal axes of the six-dimensional confidence level ellipsoids around the BF that occur in this approximation.  There are twelve sets of points $C_i$ defined by  the intersections between the $1\sigma$ confidence level hyper-ellipsoid and its principal axes that serve to gauge the behavior of the fit as one moves away from the BF within $S_{1\sigma}$. As these points are constructed through a singular value decomposition (SVD) of the Hessian matrix, we will refer to them as the SVD points in what follows.

The Hessian at the minimum, in the basis $(C_7, C_9, C_{10}, C_{7^\prime}, C_{9^\prime}, C_{10^\prime})$, is given by
\begin{eqnarray}
H=\left(
\begin{array}{cccccc}
 6606.7 & 211.7 & -1.3 & -84.1 & 26.6 & -67.8 \\
 211.7 & 62.9 & -5.2 & 38.0 & 9.9 & -20.2 \\
 -1.3 & -5.2 & 71.3 & 47.0 & -1.9 & -19.9 \\
 -84.1 & 38.0 & 47.0 & 5651.0 & 61.1 & -94.3 \\
 26.6 & 9.9 & -1.9 & 61.1 & 25.8 & -38.4 \\
 -67.8 & -20.2 & -19.9 & -94.3 & -38.4 & 89.5 \\
\end{array}
\right).
\end{eqnarray}
In diagonal form, $H_D={\rm diag}(6621, 5647, 115.6, 72.6, 44.7, 6.1)$, exhibits a hierarchy in its eigenvalues, corresponding to some directions being much more constrained than others. In terms of $H$ one has
\begin{eqnarray}
\chi^2\approx \chi^2_{\rm min} + \frac{1}{2}(C-C_{BF})^{\rm T} \cdot H\cdot (C-C_{BF})
\end{eqnarray}
where the deviations of the six Wilson coefficients $C_i$ from their BF values are written as the vector $(C-C^{BF})$. 
The SVD points are reproduced in Table~\ref{t:fits} in the appendix along with their corresponding $\Delta\chi^2$ (with respect to the BF).
The first column ``EV'' labels the SVD directions from 1 to 6 by decreasing eigenvalue, the $\pm$ refer to the two possible ways of moving along a particular direction away from the BF point.
Within the quadratic approximation to the $\chi^2$ function, all these twelve points have $\Delta\chi^2=7.1$ and lie precisely $1\sigma$ away from the BF. The exact values of the $\chi^2$ for these points are also shown in the table and are an indication of how well the approximation works in this case, they range approximately between $0.85~\sigma$ and $1.4~\sigma$.

\begin{figure}[!h]
\includegraphics[scale=.4]{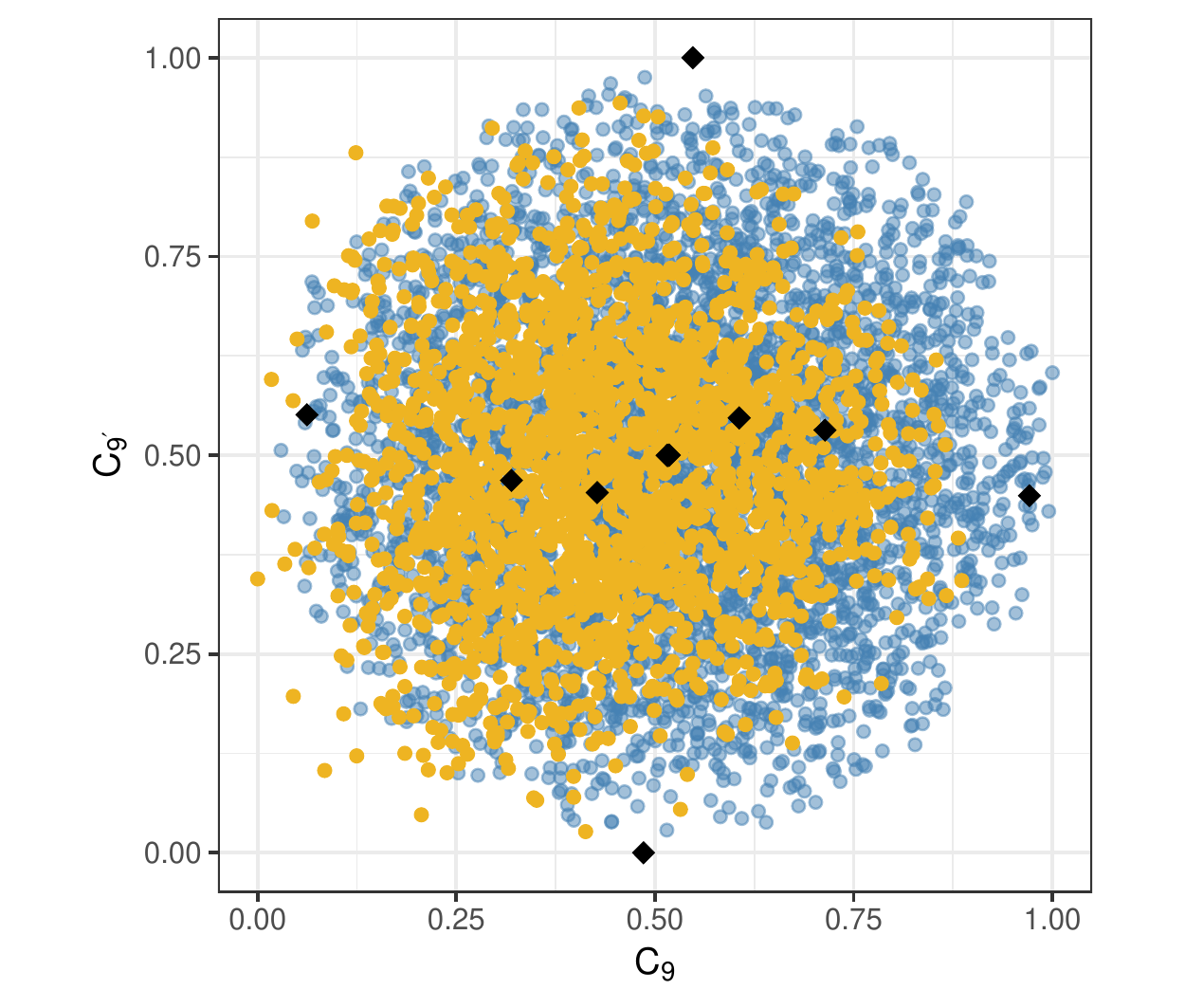}\includegraphics[scale=.4]{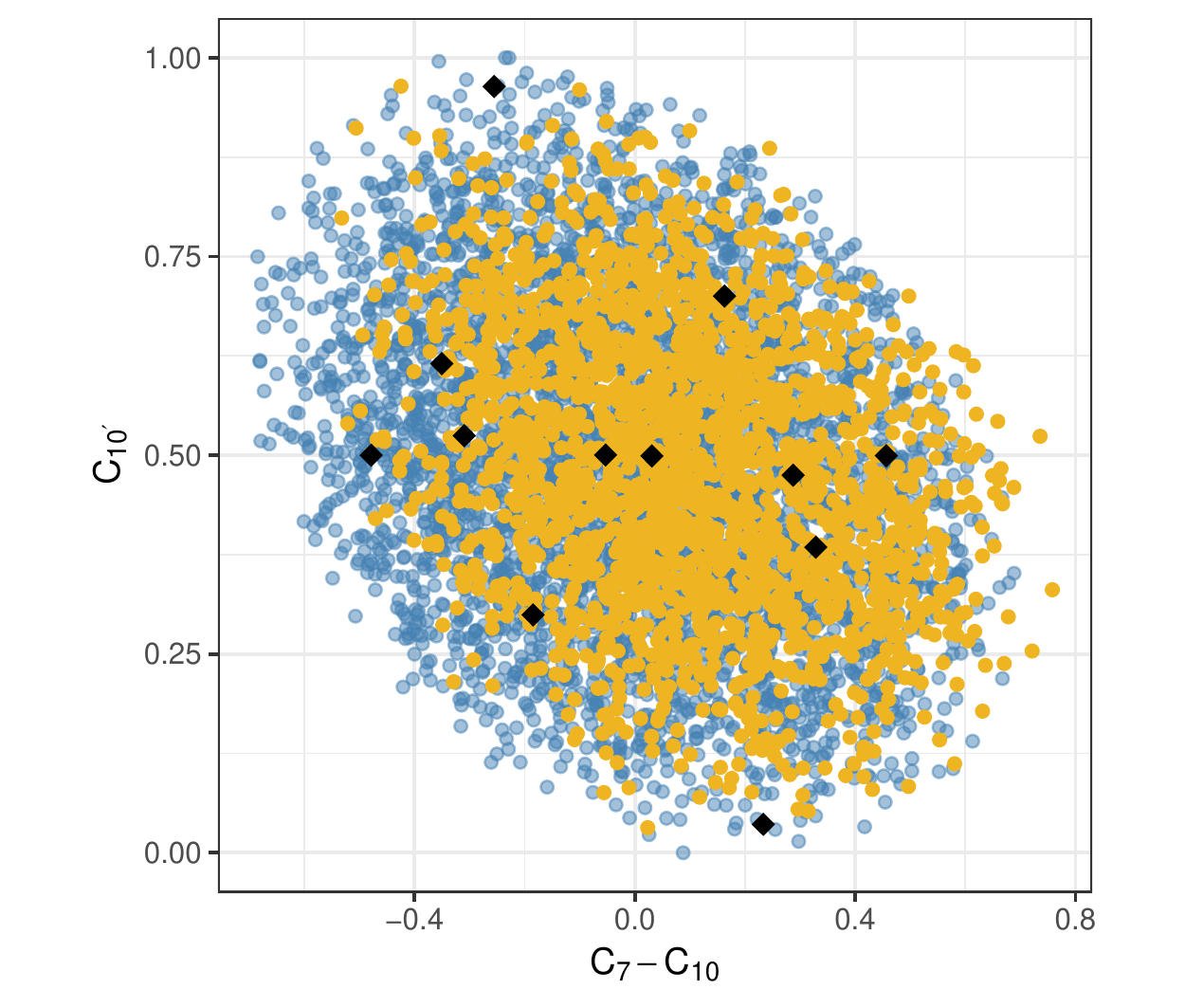}\includegraphics[scale=.4]{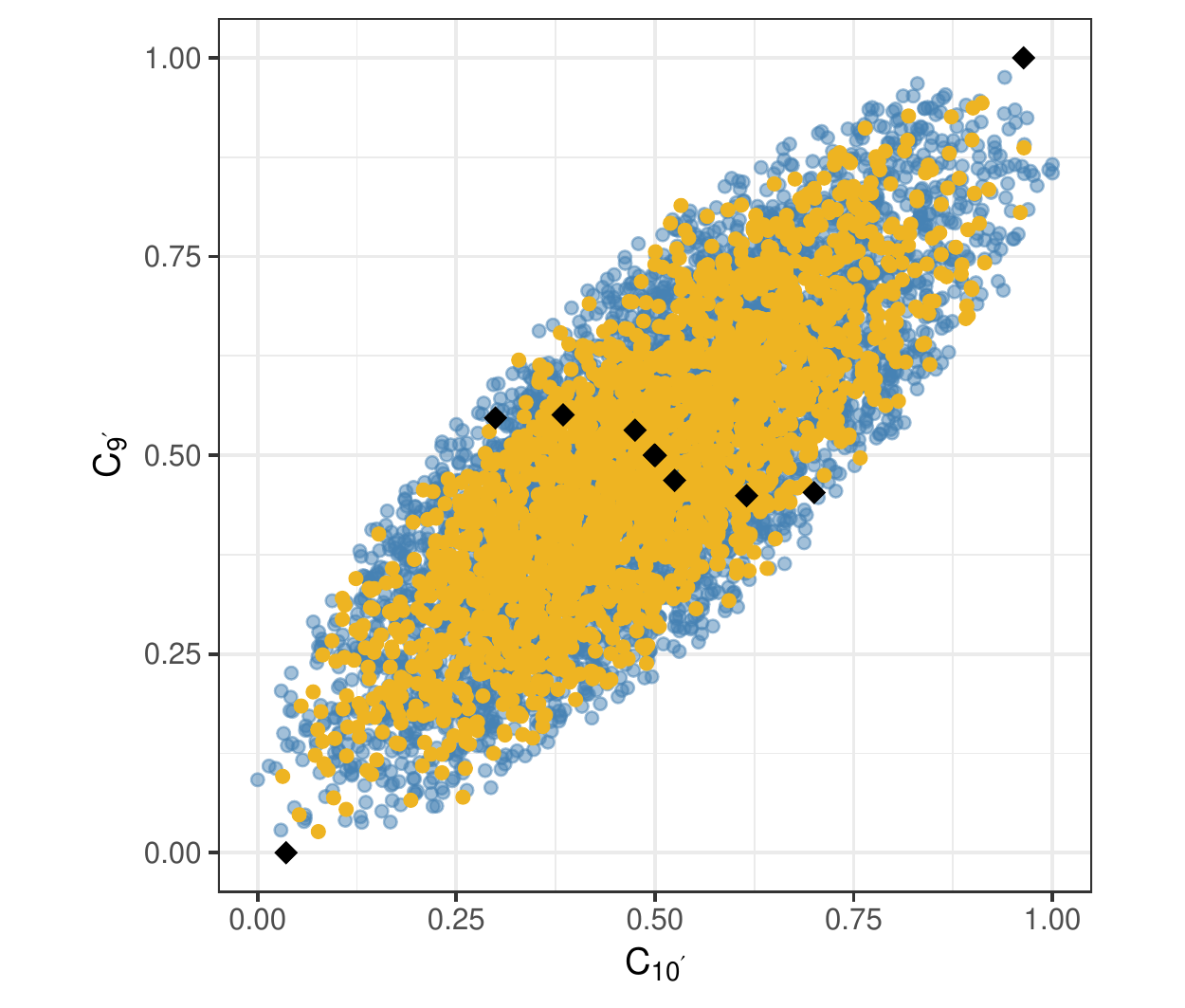}
\caption{Three projections of six-dimensional points in rescaled $C^{}_i$ space (i.e.\ each coefficient takes values between 0 and 1) $\in S_{1\sigma}$ for the exact numerical calculation  (yellow) and  the quadratic approximation (blue). The black diamonds are the twelve SVD points listed in Table~\ref{t:fits}, which sit on the boundary of this region.}
\label{fig:exactvssq}
\end{figure}

A  comparison between $S_{1\sigma}$ (yellow) and its quadratic approximation (blue) is shown as an animation \href{https://uschilaa.github.io/animations/shell/animation.html}{(here)}. To produce this visualization we have rescaled each parameter direction to range between 0 and 1 to facilitate comparisons. Three static projections are shown in Figure~\ref{fig:exactvssq}, where the black diamonds show the 12 SVD points, and the yellow (blue) regions are the corresponding projections of $S_{1\sigma}$ (quadratic approximation to $S_{1\sigma}$). While the approximation is seen to work reasonably well in most directions, the first two views illustrate some inadequacies.

This visualization also reveals new features of the fit, for example, the third view shows the projection onto $C^{}_{10^\prime}$ vs $C^{}_{9^\prime}$ illustrating a strong correlation found between these two parameters. 

The Hessian approximation also allows us to quantify the distance between the global best fit point, Eq.~\ref{bfp}, and certain one or two-parameter scenarios singled out previously in the literature which can be associated with simple NP models. The comparison of these scenarios against the SM was already presented in Ref.~\cite{Capdevila:2017bsm} in terms of the measure Pull$_{\rm SM}$, with a large value of this quantity indicating a large deviation from the SM.

This so-called Pull is a statistical measure, presented in units of Gaussian standard deviations ($\sigma$), quantifying the level of agreement between two different parametric hypotheses $H_0$ and $H_1$ ($H_0\subset H_1$) in describing a given data set. Let $\chi^2_{\text{min},H_0}$ and $\chi^2_{\text{min},H_1}$ be the minimum values of the $\chi^2$ statistic under $H_0$ and $H_1$, respectively (since $H_0$ is contained in the family of parametrizations described by $H_1$, then $\chi^2_{\text{min},H_0}\geq\chi^2_{\text{min},H_1}$). For large fits, Wilks Theorem~\cite{Wilks:1938dza} guarantees that $\Delta\chi^2_{H_0H_1}=\chi^2_{\text{min},H_0}-\chi^2_{\text{min},H_1}$ will follow a $\chi^2$ distribution with $n_\text{dof}=n_{\text{pars},{H_1}}-n_{\text{pars},{H_0}}$ degrees of freedom, being $n_{\text{pars},{H_1}}$ and $n_{\text{pars},{H_0}}$ the number of free parameters characterizing the two hypotheses. Then, the Pull comparing $H_0$ and $H_1$ reads,
\begin{equation}
\text{Pull}_{H_0H_1}=\sqrt{2}\,\text{Erf}^{-1}\left[F(\Delta\chi^2_{H_0H_1};n_\text{dof})\right],
\label{eq:pullh0h1}
\end{equation}
where $F$ is the $\chi^2$ cumulative distribution function (CDF).

In Table~\ref{t;bfpulls} we present one and two dimensional fits using the quadratic approximation to the $\chi^2$ function. For the sake of comparison, we analyse all the scenarios (without electronic NP contributions) already studied in Ref.~\cite{Capdevila:2017bsm}.
For each scenario, we assess its statistical significance with respect to both the SM and the BF, introducing $\text{Pull}_{\rm 6D}$ in analogy with Pull$_{\rm SM}$ to quantify the preference for the six-dimensional BF over the different 1D and 2D hypotheses considered.

\begin{table}[htp]
\begin{center}
\begin{tabular}[H]{|c|c|c|c|c|c|c|}%
\hline%
Scenario&Best fit \cite{Capdevila:2017bsm}&$\text{Pull}_\text{SM}$ \cite{Capdevila:2017bsm}&Best fit (Quad)&$\text{Pull}_\text{SM}^\text{quad}$&$\Delta\chi^2$&$\text{Pull}_\text{6D}^\text{quad}$\\%
\hline%
$C_{9}^{}$&{-}1.11&5.8&{-}1.00&5.6&8.4&1.5\\%
$C_{9}^{}=-C_{10}^{}$&{-}0.62&5.3&{-}0.65&5.5&9.3&1.7\\%
$C_{9}^{}=-C_{9'}$&{-}1.01&5.4&{-}0.93&5.4&9.9&1.8\\%
$(C_{9}^{},C_{10}^{})$&({-}1.01, 0.29)&5.7&({-}0.97, 0.36)&5.6&3.7&0.8\\%
$(C_{9}^{},C_{7'})$&({-}1.13, 0.01)&5.5&({-}1.01, 0.03)&5.4&6.5&1.4\\%
$(C_{9}^{},C_{9'})$&({-}1.15, 0.41)&5.6&({-}1.07, 0.46)&5.5&5.8&1.2\\%
$(C_{9}^{},C_{10'})$&({-}1.22, {-}0.22)&5.7&({-}1.09, {-}0.31)&5.6&4.5&0.9\\%
\begin{tabular}{@{}c@{}}$(C_{9}^{}=-C_{9'}$,\\ $C_{10}^{}=C_{10'})$\end{tabular}&({-}1.16, 0.38)&5.7&({-}1.00, 0.33)&5.7&3.4&0.7\\%
\begin{tabular}{@{}c@{}}$(C_{9}^{}=-C_{9'},$\\ $C_{10}^{}=-C_{10'})$\end{tabular}&({-}1.15, 0.01)&5.0&({-}0.90, 0.08)&5.1&9.3&1.9\\%
\begin{tabular}{@{}c@{}}$(C_{9}^{}=-C_{10}^{}$,\\ $C_{9'}=C_{10'})$\end{tabular}&({-}0.67, {-}0.10)&5.0&({-}0.65, 0.09)&5.2&9.2&1.9\\%
\begin{tabular}{@{}c@{}}$(C_{9}^{}=-C_{10}^{}$,\\$C_{9'}=-C_{10'})$\end{tabular}&({-}0.70, 0.28)&5.0&({-}0.65, 0.08)&5.2&8.8&1.8\\%
\hline%
\end{tabular}%
\end{center}
\caption{Comparison of certain low global $\chi^2$ one or two parameter fits to the global six-dimensional best fit point. We also include the comparison of these scenarios to the SM as presented in Ref.~\cite{Capdevila:2017bsm}. The listed Pull indicates that all these scenarios are better than the SM by at least $5~\sigma$ and that most of them are between $1~\sigma$ and $2~\sigma$ worse than the 6D BF.}
\label{t;bfpulls}
\end{table}%

\subsection{Hessian eigenvectors and the Wilson coefficients}

The SVD directions are mostly aligned with one or two of the Wilson coefficients as shown in Table~\ref{t:eignev}. Thus, to a good approximation there is a simple correspondence between a given SVD direction and the $C^{}_i$ as given in the table. The direction corresponding to the largest eigenvalue of the Hessian is that in which the $\chi^2$ function changes most rapidly near its minimum and is thus most strongly constrained by the data. In this case the first two directions have similarly large eigenvalues (more than 50 times larger than the rest) and, as can be seen from Table~\ref{t:eignev}, they correspond to the parameters $C_7$ and $C_{7'}$ to a very good approximation. The interpretation  that these two parameters are essentially fixed and there is very little room to play with them is compatible with statements made in the literature \cite{Descotes-Genon:2015uva}.
Interestingly a combination of $C_{10'}$ and other coefficients is the third most constrained direction, in particular we observe a large correlation with $C_{9'}$, see Figure~\ref{fig:exactvssq} (right panel), with a correlation coefficient of 0.82. Curiously,  this correlation approximates the pattern that would be expected from  right-handed currents.

The  constraints on $C_{9',10'}$ arise primarily from $P_1$ and $P_4^\prime$. As can already be seen in Table 1 of~\cite{Descotes-Genon:2015uva}, both $P_1$, in all its $q^2$ bins, and $P_4^{\prime}$ in the high end of the low-$q^2$ region, are very sensitive to these two coefficients, with the sensitivity to $C_{10'}$ generally more pronounced. As already observed in that reference,  these observables are within $1\sigma$ of the SM and thus restrict the range $C_{9'}$ and $C_{10'}$ can take. The coefficient $C_{10'}$ is further constrained by measurement of $Br(B_s\to\mu^+\mu^-)$, leaving little room for departures from 0.

\begin{table}[h!]
\centering
\begin{tabular}{|c|c|c|c|c|c|c|} \hline
eigenvector  &  $C_7$& $C_9$& $C_{10}$& $C_{7^\prime}$& $C_{9^\prime}$& $C_{10^\prime}$\\
\hline
1 & 0.996 & 0.0317 & -0.000796 & -0.0841 & 0.00334 & -0.00924 \\
 2 & -0.0836 & -0.01 & -0.00843 & -0.996 & -0.0114 & 0.0181 \\
 3 & 0.0192 & -0.267 & -0.306 & 0.023 & -0.361 & 0.839 \\
 4 & 0.0169 & -0.466 & 0.859 & -0.000316 & -0.192 & 0.0824 \\
 5 & 0.023 & -0.843 & -0.374 & 0.0015 & 0.243 & -0.3 \\
 6 & 0.000335 & 0.0212 & 0.166 & -0.0036 & 0.88 & 0.445 \\ \hline
\end{tabular}
\caption{Normalized eigenvectors of the Hessian matrix in the space of Wilson coefficients. The ordering corresponds to decreasing eigenvalue.}
\label{t:eignev}
\end{table}

\section{Metrics for Quantitative comparisons}\label{s:quanti}

Here we introduce several quantitative metrics to compare different sets of predictions and observations in the next section.

\subsection{Comparing predictions and observations}

In order to quantify the comparison between the different predictions and the experimental results we will use the following metrics
\begin{enumerate}
\item The Pull, which measures the difference between the prediction $T(p)$ for a given set of parameters $p$ and the observation ${\cal O}$ in terms of the uncertainty constructed by adding experimental and theoretical errors  in quadrature and ignoring correlations:\footnote{This Pull is defined for each observable, while the Pull quoted in Table~\ref{t;bfpulls} is defined in the parameter space of the six Wilson coefficients.}
\begin{equation}
{\rm Pull}(p)_i = \frac{T(p)_i - {\cal O}_i}{\sqrt{\Delta_{exp,i}^2 + \Delta_{T(p)_i}^2}}
\label{eq:residual}
\end{equation}
We will use the parameter sets, $p$, corresponding to the SM and the BF point in what follows.

\item The pull difference will be used to compare the BF  against the SM, quantified as
\begin{eqnarray}
\Delta({\rm Pull})_i &=& |{\rm Pull}({\rm SM})_i| - |{\rm Pull}({\rm BF})_i|, \nonumber \\
\Delta_\sigma({\rm Pull})_i &=& \bigg|\sum_{j}\sigma^{-1/2}_{ij} (T(SM) - {\cal O})_j\bigg| -
\bigg|\sum_{j}\sigma^{-1/2}_{ij} (T(BF) - {\cal O})_j\bigg|.
\end{eqnarray}
where $\sigma^{-1/2}$ is the square root of the inverse of the full covariance matrix (including both experimental and theoretical errors evaluated at the SM point).

Notice that these measures allow an explicit and systematic comparison between Pulls in the SM and BF scenarios. However, the distribution of $\Delta(Pull)$ will not follow a $\chi^2$ distribution (but is measuring differences in units of the total uncertainty).\footnote{Alternatively, one may wish to define a measure comparing how much each measurement contributes to the total $\chi^2$ in each scenario, i.e. comparing quadratic Pulls rather than linear ones. We find that such a definition gives qualitatively similar results here (an explicit comparison is given in the Appendix). Since our aim is a direct comparison between SM and BF predictions and measured values (rather than statistical interpretation of the results), we will not consider such a definition here.}

The absolute value ensures that a positive number indicates that the BF prediction is in better agreement with the observation, and a negative value signals better agreement of the SM prediction with the observation. $\Delta({\rm Pull})$ captures the improvement in matching the observed value for the BF as compared to the SM.
Notice that $\Delta({\rm Pull})$ and $\Delta_\sigma({\rm Pull})_i$ have different connotations.
While $\Delta({\rm Pull})$ measures the absolute preference for the BF over the SM for a given observable, $\Delta_\sigma({\rm Pull})_i$ corresponds to the difference in conditional Pull, taking into account the correlation with other observables.
\end{enumerate}

\subsection{Examining variations in the fit within $S_{1\sigma}$}

We turn our attention to variations in the predictions, ignoring agreement with experiment. The goal is to associate specific observables with particular SVD directions (and therefore specific NP parameters).  These metrics are thus constructed to single out large contributions to $\Delta\chi^2$ as the parameters move away from their best fit value. Several definitions are possible and we will use the following ones:

\begin{eqnarray}
\delta_i&=&\frac{(T_i-T_{BF,i})}{\sqrt{\Delta_{exp,i}^2+\Delta_{BF,i}^2}} \nonumber \\
\delta^\prime_i&=&\frac{(T_i-{\cal O}_i)}{\sqrt{\Delta_{exp,i}^2+\Delta_{i}^2}}-\frac{(T_{BF,i}-{\cal O}_i)}{\sqrt{\Delta_{exp,i}^2+\Delta_{BF,i}^2}}  \nonumber \\
\delta_{\sigma,i}&=&\sum_{l} \sigma_{il}^{-1/2}(T_{pt}-T_{BF})_l \nonumber \\
\tilde\delta_{\sigma,i}&=&\sum_{l} \frac{1}{\sqrt{\sigma^{-1}_{ii}}}\sigma_{il}^{-1}(T_{pt}-T_{BF})_l
\label{deltasdef}
\end{eqnarray}
where the index $i$ labels the observables and the different $\delta$s are all calculated for the SVD directions. The different definitions have the simple interpretations:
\begin{enumerate}
\item $\delta$ is a straightforward comparison between the points on the $1\sigma$ ellipsoid and the best fit. The difference is quantified in terms of one standard deviation that ignores correlations.
\item $\delta^\prime$ is a variation which takes into account the different theoretical errors at different points in parameter space. Note that $\delta^\prime\to\delta$ if the theoretical errors are the same. The comparison between $\delta$ and $\delta^\prime$ thus contains information about the variation in the theoretical errors.
\item $\delta_{\sigma}$ is a generalization of $\delta$ to include correlations between observables.  Neglecting the dependence of the theoretical errors on the parameters, we  define $\delta_{\sigma,i}$ such that
\begin{eqnarray}
\sum_i\delta_{\sigma,i}^2& =&\sum_i \left(\sigma_{il}^{-1/2}(T_{pt}-T_{BF})_l\right)\left(\sigma_{ik}^{-1/2}(T_{pt}-T_{BF})_k\right)\nonumber\\
&=&\sum_{ij}(T_{pt}-T_{BF})_i\sigma^{-1}_{ij}(T_{pt}-T_{BF})_j
\label{eq38}
\end{eqnarray}

We interpret this as the conditional definition of $\delta$.
\item Finally, $\tilde\delta_{\sigma}$  is a definition based on to the conditional variance\footnote{In the context of a multivariate Gaussian each individual variable will also be normal distributed, and the conditional variance for variable $i$ (i.e.\ the variance given the values observed for the other variables) is $1/\sigma^{-1}_{ii}$.}
which turns out to be approximately equal to  $\delta_{\sigma}$ for the points we discuss below.

A Table that illustrates the largest $\delta$'s for the different definitions is provided in Appendix~\ref{s:larged}.

\end{enumerate}

To provide some intuition for the interpretation of measures defined in terms of the covariance matrix, we present a short discussion of the two dimensional scenario in the Appendix~\ref{s:dSigma}.

It is important at this stage to reiterate that the fit procedure of Ref.~\cite{Capdevila:2017bsm}, which forms the basis of our analysis, uses a {\it single covariance matrix} with theory errors evaluated at the standard model point.

\subsection{Correlation between observables}\label{corrsec}

\begin{figure}[!h]
\includegraphics[width=\textwidth]{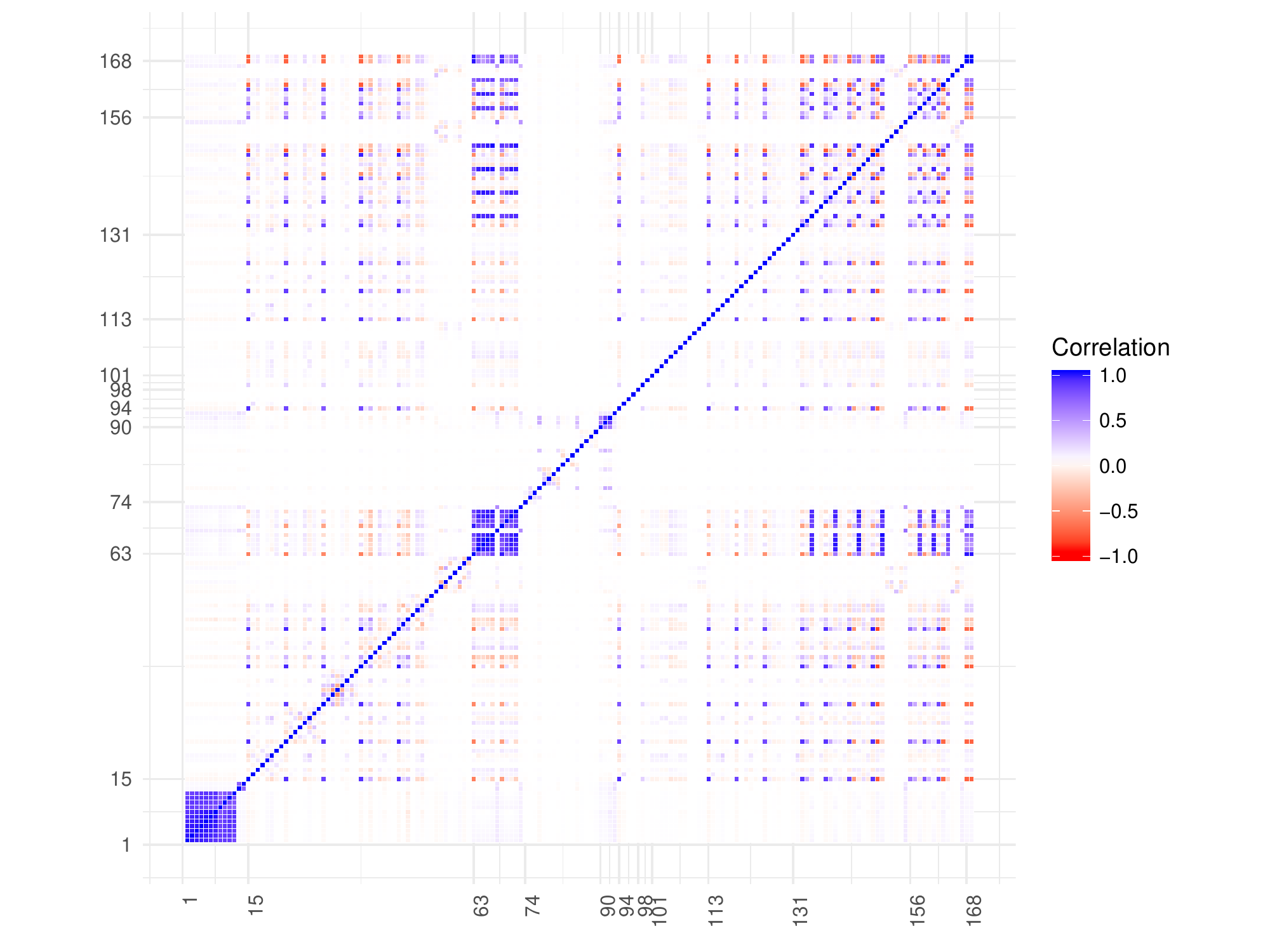}
\caption{Correlation map derived from the full covariance matrix. }
\label{fig:correlationMap}
\end{figure}

The covariance matrix is an important ingredient to the global fit which, at present,  mainly includes correlated theory uncertainties. These theory correlations arise, for example,  from definitions in terms of common coefficients,
common form factor dependencies among observables or from common  hadronic parameters.
There are also important  experimental correlations which we include when available, but note that not much information on error correlations has been released by the experimental collaborations. The resulting correlations between observables are illustrated in Figure~\ref{fig:correlationMap}. This figure shows, for instance, the correlation between various $B\to K^{\star}\mu^+\mu^-$ branching ratios (IDs 63-73), $F_L$ and $A_{FB}$ observables (a table listing all observables and their corresponding ID is presented in Appendix~\ref{s:idmatch}). This is expected since these three observables are related by $\sfrac{d\Gamma}{dq^2}$, the $q^2$ distribution of the process~\cite{Descotes-Genon:2013vna}.

We see also that large $q^2$ bins are correlated between themselves, but not so much with lower $q^2$ bins. This appears in the correlation map, for example, observables with IDs between 63 and 73 are highly correlated except for IDs 68 and 73 which are only correlated with each other. For LHCb measurements of angular observables (IDs 15-62) experimental correlations for each bin are strongest for $q^2 \in [2.5-4]$ (IDs 31-38, for example, stand out in the experimental correlation map).
See  Appendix \ref{s:2maps} for separate theoretical and experimental correlation maps.

\section{Quantitative comparisons}\label{s:quantiresults}

\subsection{Pull}

\begin{figure}[h!]
\includegraphics[width=.45\textwidth]{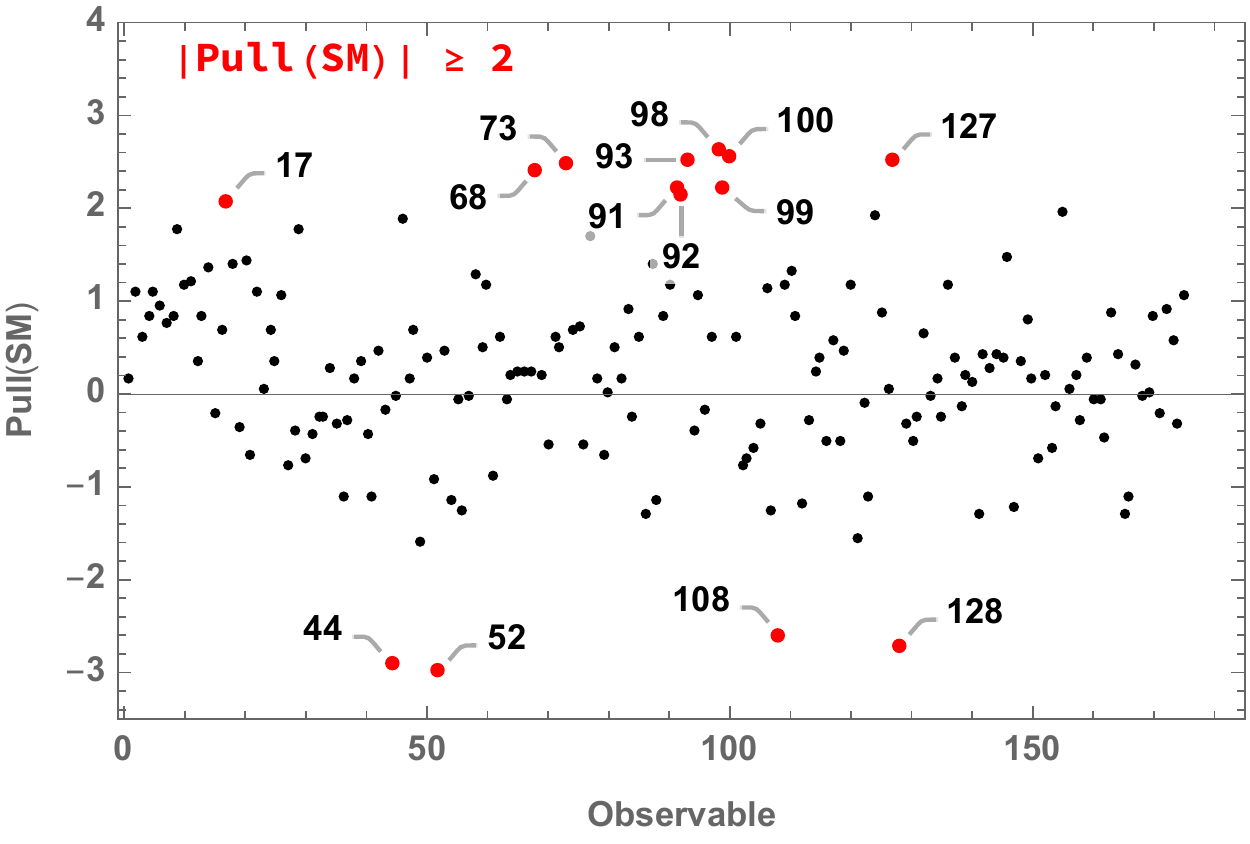}\hfill\includegraphics[width=.45\textwidth]{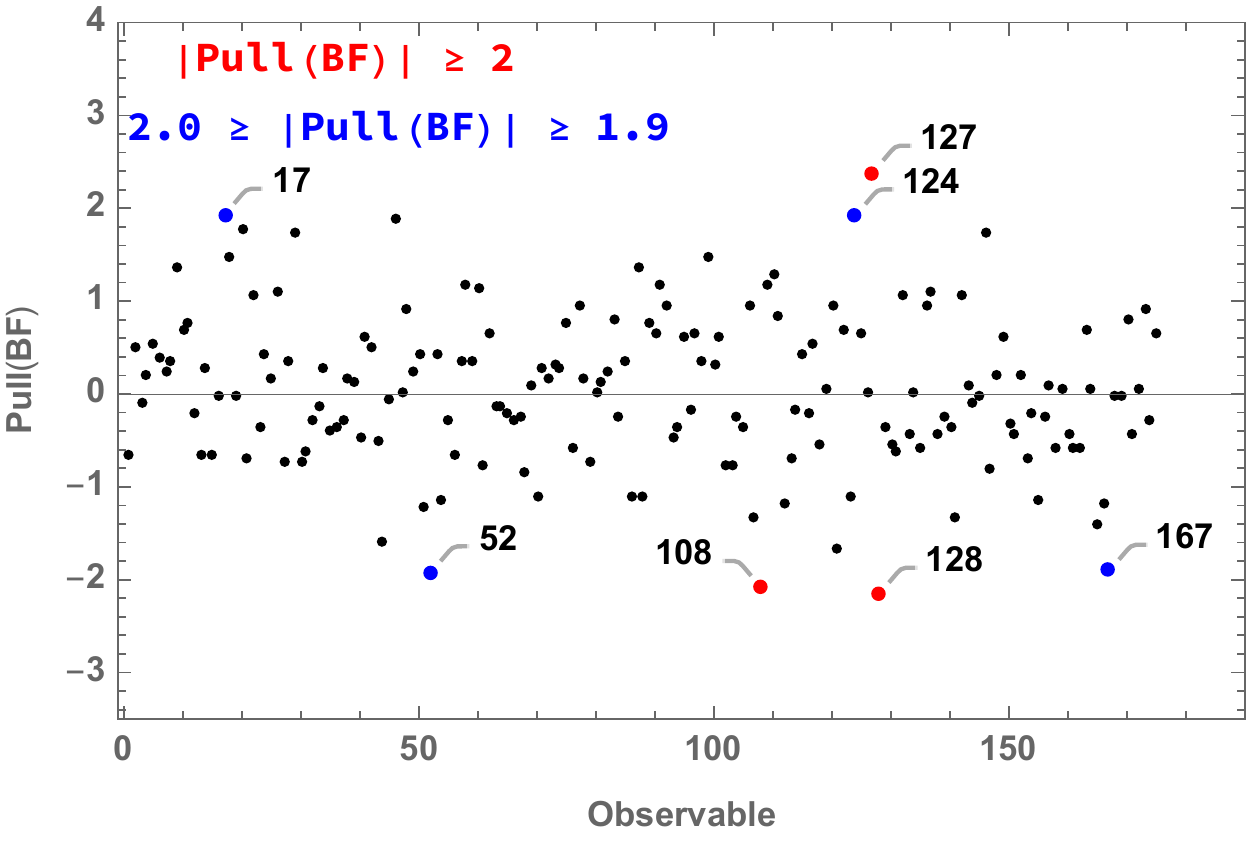}
\includegraphics[width=.45\textwidth]{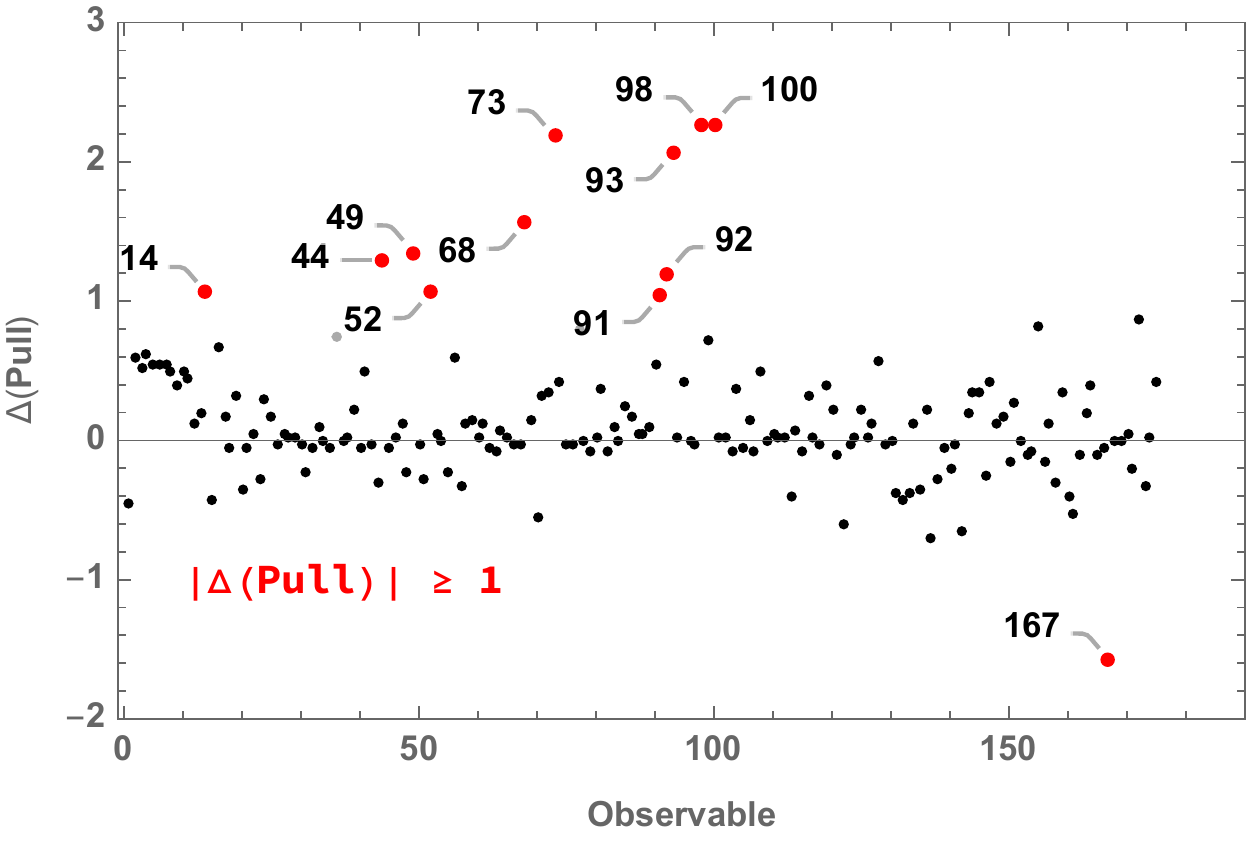}\hfill\includegraphics[width=.45\textwidth]{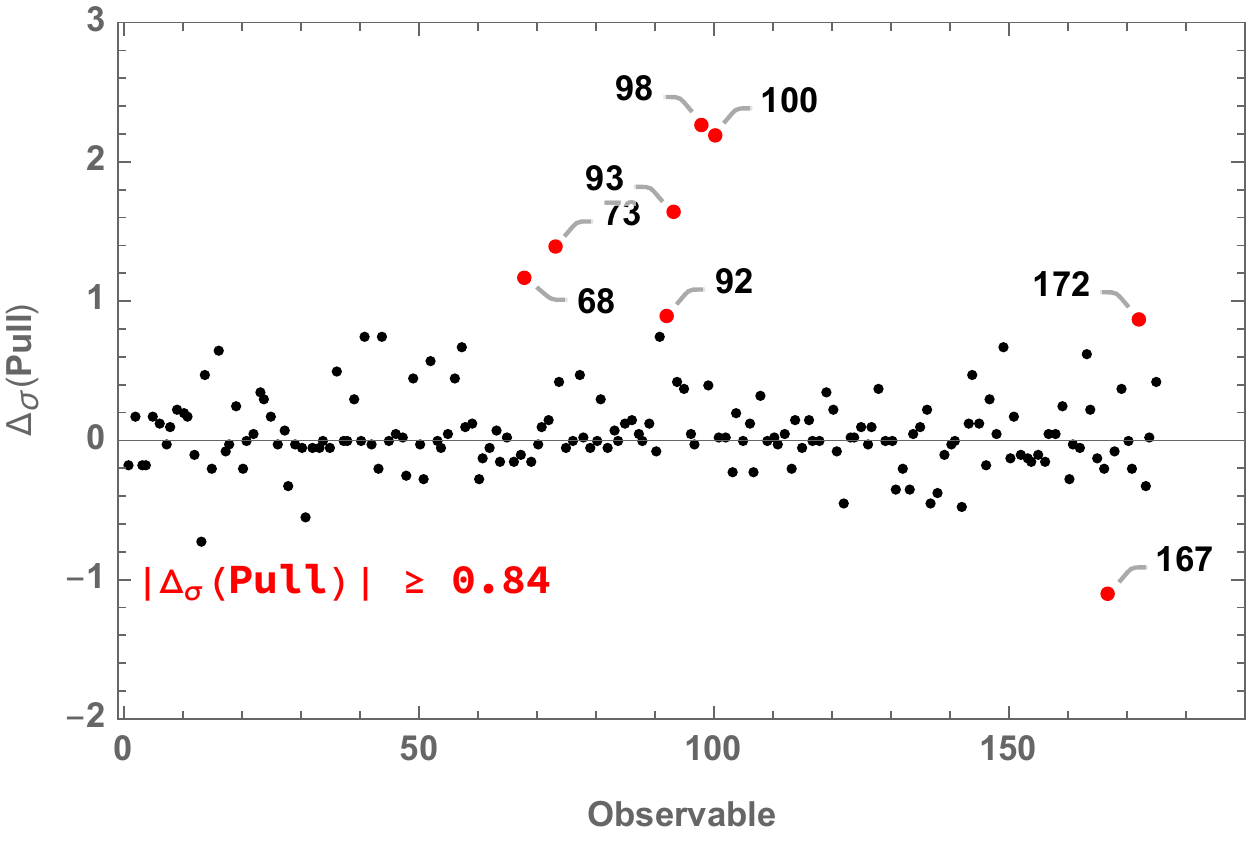}
\caption{The top panels show the Pull of each observable with respect to the SM (left) and the best fit (right). The bottom panels show $\Delta({\rm Pull})$ for all observables ignoring correlations (left) and with correlations (right). Values larger than 2 for the top panels (1 and 0.84 for the bottom) have been labeled in red and selected for discussion in the text.}
\label{fig:dpull}
\end{figure}

Here we compare all measurements directly to the SM and the BF respectively in the top two panels of Figure~\ref{fig:dpull}.
We have highlighted in red those observables with Pull greater than 2, i.e. those where the theory prediction differs from the measured value by more than $2\sigma$. 
The Pull(SM) metric simply updates known results from Ref.~\cite{Descotes-Genon:2015uva}: four of the largest over-predictions of the SM occur for BR and $R_{K^{(\star)}}$ LHCb measurements, whereas the largest under-predictions occur for $P_{5}^\prime$ measurements by three different experiments  (IDs 44, 52, 108, 128).

The top-right panel of Figure~\ref{fig:dpull} presents the Pull(BF) metric for all observables,
showing overall better agreement between predictions and measured values.
We note however that several tensions remain. 
Two of the $P_{5}^\prime$ measurements (IDs 108 and 128) are also above the BF prediction whereas an ATLAS $P_{4}^\prime$ (ID 127) falls below both the SM and BF predictions by more than $2\sigma$.

In the bottom two panels of Figure~\ref{fig:dpull} we show $\Delta({\rm Pull})$ for all observables.
Recall that this metric captures the improved agreement between BF prediction and measurement, as compared to the SM prediction, measured in units of the total uncertainty, and negative values signal preference for the SM. 
The results are shown both ignoring correlations (lower left panel) and including them (lower right panel).
The majority of the points are clustered at small values of  $\Delta({\rm Pull})$ (or $\Delta_\sigma({\rm Pull})$), indicating insignificant resolving power between the SM and the BF. The distribution shows, however, that even among these points with small $\Delta({\rm Pull})$ there is an average preference for positive values. This is of course just the statement that the global fit prefers the BF to the SM, but the distribution shows how much of this overall preference is built from many small differences that go in the same direction.

Points with values of $|\Delta({\rm Pull})|\geq1$ (left) and  $|\Delta_\sigma({\rm Pull})|\geq0.84$ (left) are highlighted in red in Figure~\ref{fig:dpull}.
The  particular cutoffs for this are rather arbitrary. For individual pulls, we have a statistical interpretation since $\Delta({\rm Pull})$ is normalised to the total uncorrelated errors.
When including correlations, we follow the argument sketched in App.~\ref{s:larged}.

$\Delta({\rm Pull})$ simply combines the two upper panels of Figure~\ref{fig:dpull} to highlight the observables where the BF is a significant improvement over the SM. 
This exercise shows, for example, that  branching ratio measurements and $R_K$ which have a large ${\rm Pull(SM)}$ also have the largest values of $\Delta({\rm Pull})$. On the other hand, the low-$q^2$ bin of the $R_{K^*}$ measurement, ID 99 has a large  ${\rm Pull(SM)}$ but not $\Delta({\rm Pull})$, indicating that the BF does not offer a significant improvement over the SM in this case. Amongst the angular observables,  $P_5^\prime$ measurements (ID 44, 52) stand out in both ${\rm Pull(SM)}$ and $\Delta({\rm Pull})$, but the latter also highlights a comparable improvement over the SM in $P_2$, ID 49.

To include the effect of correlations, we next compare $\Delta({\rm Pull})$ to $\Delta_\sigma({\rm Pull})$. The points that are singled out as large by both definitions are:
\begin{itemize}
\setlength{\itemsep}{.1\baselineskip}
\item 68: $10^7 \times Br(B^0 \to K^{0*}\mu\mu)$ [15-19] LHCb
\item 73: $10^7 \times Br(B^0 \to K^{+*}\mu\mu)$ [15-19] LHCb
\item 92:  $10^7\times Br(B_s \to \Phi\mu\mu) $  [5-8]  LHCb
\item 93: $10^7 \times Br(B_s \to \Phi\mu\mu)$ [15-18.8] LHCb
\item 98: $R_K(B^+ \to K^+)$ [1-6] LHCb
\item 100: $R_{K^*}(B^0 \to K^{0*})$ [1.1-6] LHCb
\item 167: $10^7\times Br(B \to K^* \mu\mu)$ [16-19]  CMS-7
\end{itemize}
Of  the observables in this list, (167) is the only one that has a  large preference for the SM over the BF.

There are several observables highlighted in the bottom-left panel as showing $1\leq |\Delta({\rm Pull})|\leq2$, that no longer stand out when correlations are included (bottom-right panel). This is the case for measurements of $P_5^\prime$ in the last two bins of the low-$q^2$ region (IDs 44 and 52). The largest value, $\Delta({\rm Pull})=1.3$, for a $P_5^\prime$ observable is found for the LHCb measurement in the bin [4,6] (ID 44). When correlations are included this drops to  $\Delta_\sigma({\rm Pull})=0.75$ and indeed no $P_5^\prime$ observables are singled out in the bottom right panel.\footnote{This conclusion is robust with respect to details of how the covariance matrix is included in the computation of the measure, concretely we have verified that the overall picture remains the same when using a definition as given for $\tilde{\delta}_{\sigma}$ in Eq.~\ref{eq38}.} The reason is that the correlations implied by the covariance matrix for this observable are more consistent with the SM than with the BF.  This information is captured and penalised in the conditional definition $\Delta_\sigma({\rm Pull})$.
Similar effects are insignificant for e.g.\ $R_K$ which is found to have negligible correlation with the other observables.
This of course does not imply that these observables are no longer relevant to the fit, as values of $\Delta_\sigma({\rm Pull})$ are still significantly larger than zero.
In other words, when calculating the differences in $\chi^2$ between the SM and BF point after dropping a subset of observables,
we find that the smaller set of observables singled out by $\Delta_\sigma({\rm Pull})$ already accounts for the largest differences,
while a smaller additional reduction in $\Delta \chi^2$ is observed when removing the full set singled out by $\Delta({\rm Pull})$.

Another observation that results from this comparison is that there is a systematic offset in $\Delta({\rm Pull})$ for the BR measurements in observables 1-14, which is no longer present when considering $\Delta_\sigma({\rm Pull})$, indicating that the effect is well described by the covariance matrix.

\subsection{The BF-SM direction at $1\sigma$ from the BF}

\begin{figure}[ht]
\center{
\includegraphics[width=0.45\textwidth] {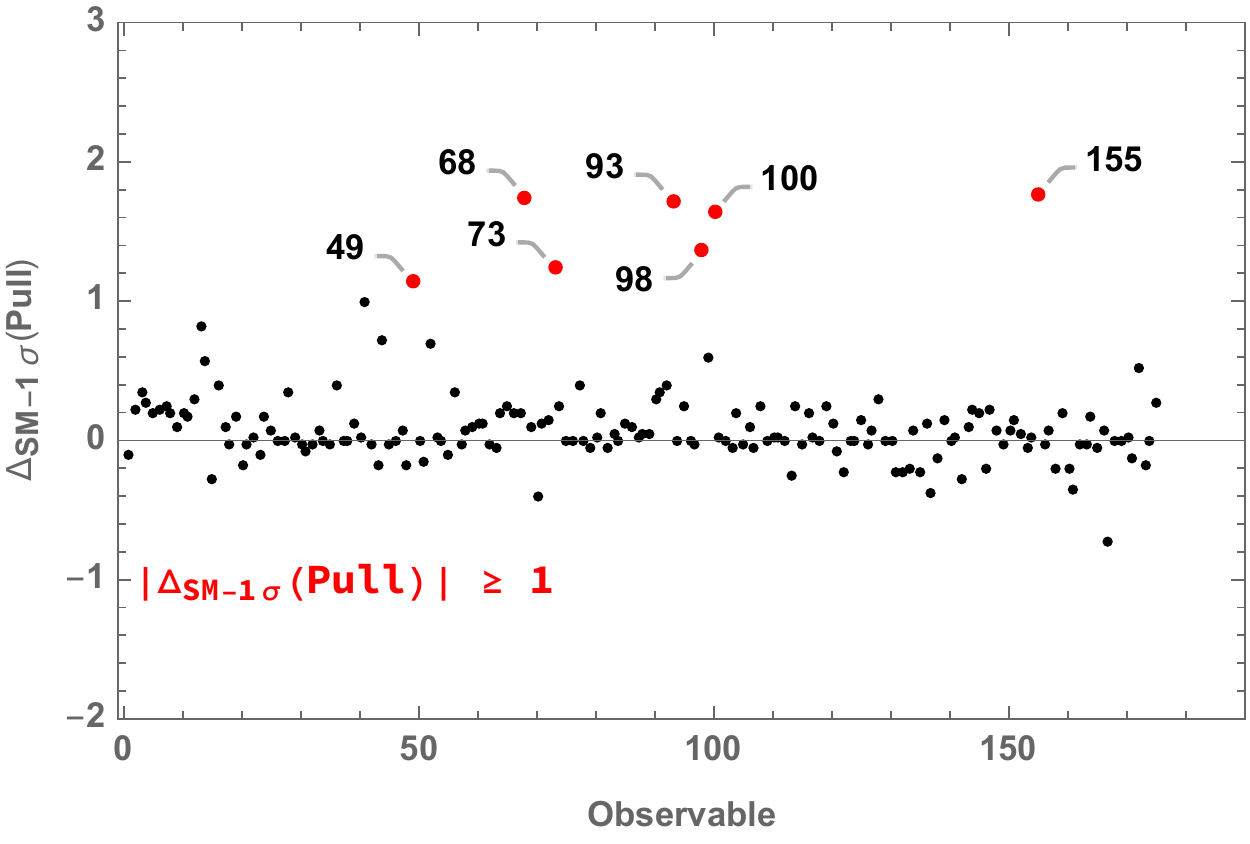}
\includegraphics[width=0.45\textwidth] {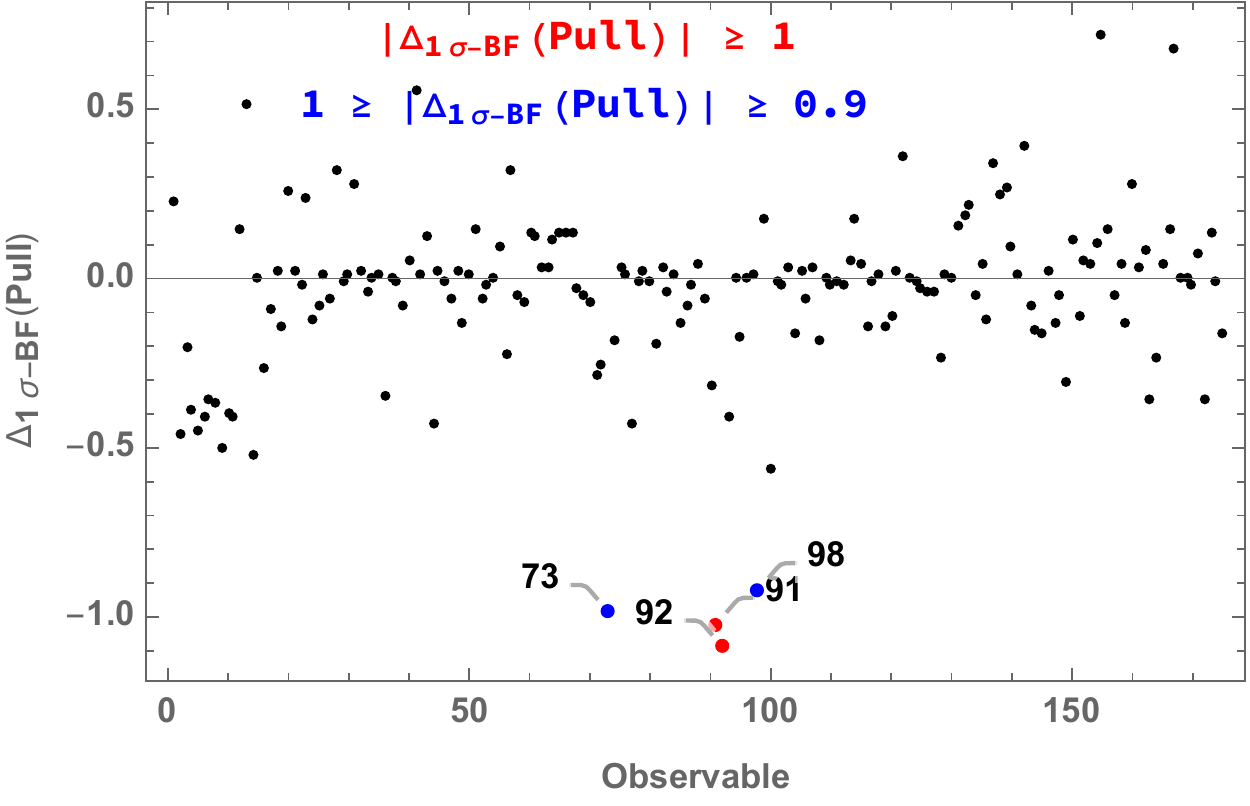}
}
\caption{Pull difference between the point along the  BF-SM direction at $1\sigma$ from the BF, and the SM (BF) in the left (right)  panel.}
\label{fig:newpt}
\end{figure}

The quadratic approximation of Section~\ref{s:bestfit} permits a quick estimate of the point within $S_{1\sigma}$ that lies in the BF-SM direction, it has parameters
\begin{eqnarray}
C_7 = 2.5 \times 10^{-3},~ C_{7^\prime}=0.01, ~
C_9 = -0.61, ~ C_{9^\prime\mu}=0.22, ~
C_{10\mu} = 0.20, ~ C_{10^\prime\mu}=-0.02.
\label{cpsm}
\end{eqnarray}
After evaluating the theory predictions at this point, we find it is exactly at $\Delta\chi^2=8.0$ from the BF. A comparison of the predictions at this point with both the SM and BF points displays the pattern observed as one moves from the SM towards the BF as can be seen in Figure~\ref{fig:newpt}.

Comparing $|\Delta_{{\rm SM}-1\sigma}({\rm Pull})|$ to $|\Delta({\rm Pull})|$ in Figure~\ref{fig:dpull}, 
we generally note reduced improvement over the SM as expected, except for observables 155 and 167 (where the reduced tension removes it as an outlier from Figure~\ref{fig:newpt}). Recall these are measurements
of $Br(B\to K^* \mu^+\mu^-)$ in the high $q^2$ region from CMS and ATLAS respectively.
This behaviour illustrates some internal tension in the fit, the measured value of 155 is best accommodated by smaller NP contributions than found in the BF point, while still showing significant deviation from the SM prediction. 
Considering now the Pull difference with the BF point (right panel), we see that this direction is most constrained by two LHCb measurements of $Br(B_s\to \Phi \mu^+\mu^-)$ (at intermediate $q^2$, observables 91 and 92), the LHCb measurement of $Br(B^0\to K^{+\ast}\mu\mu)$ at high $q^2$ (observable 73) and of $R_K$ (observable 98).

\subsection{Fit uncertainty}

Here we present a direct comparison between theory and experiment for all the observables. In addition to the usual theory and experimental errors, we include an estimate for the uncertainty in the fit. For this purpose we will follow the framework used to discuss uncertainties in global fits of parton distribution functions~\cite{Pumplin:2000vx,Pumplin:2001ct}.  The results are shown in Figure~\ref{fig:e-ranges} which shows the following for all the 175 observables listed in Appendix~\ref{s:idmatch}:
\begin{itemize}
\item The experimentally measured value of the observable with its error (black).
\item The SM prediction with the estimated theory error (green).
\item The BF prediction with the estimated theory error (brown).
\item Our estimate for the error in the fit calculated as the difference between the maximum deviations in theory predictions along one of the eigenvectors of $H$, within the $1\sigma$ region (purple).
\end{itemize}

\begin{figure}[!h]
\includegraphics[width=\textwidth]{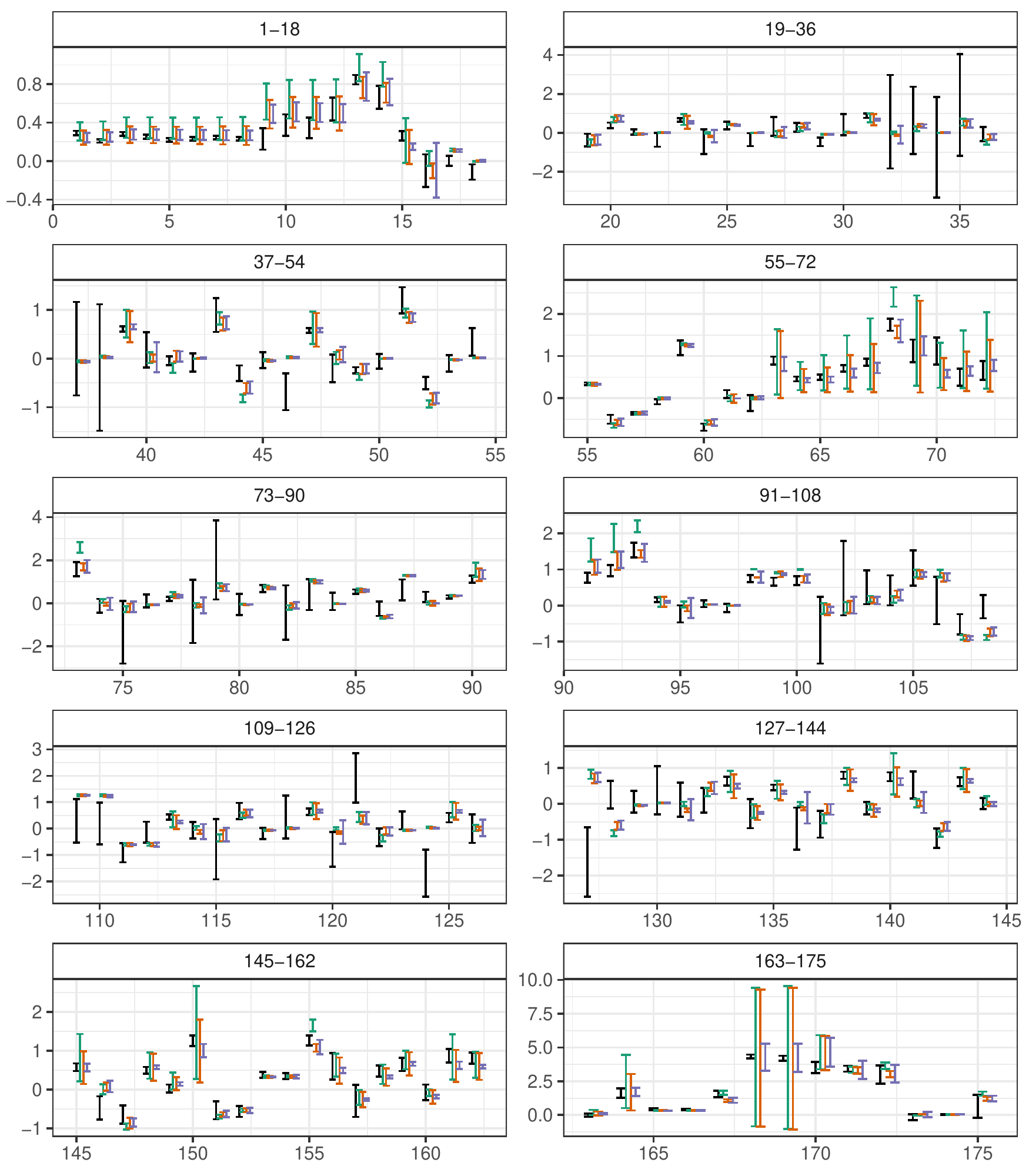}
\caption{Measured value (black), SM prediction (green), best fit (brown) and error in the fit (purple) for the 175 observables. }
\label{fig:e-ranges}
\end{figure}

Figure~\ref{fig:e-ranges} allows for a quick assessment of the overall picture. It is particularly interesting to study the prediction range within the $1\sigma$ BF region (the error in the fit). For example considering observable 99 we see that the discrepancy between the measured value and theory predictions cannot be resolved anywhere within this region, while other measurements may be better explained by points within $1 \sigma$ of the BF, for example the previously mentioned observable 155.

The observables with values  $|\delta_i|>1$ (or $|\delta_{\sigma i}|>0.84$) are listed separately in Table~\ref{t:largedel}, showing which parameter directions are associated with large fit uncertainties. 
Several branching ratios, as well as $R_{K^{(*)}}$ show large deviations in directions $\pm4,6$ which we saw in Table~\ref{t:eignev} refer mainly to $C_{10}^{}$ and $C_{9'}^{}$ respectively. Note that this is the case for observable 155 which thus points to smaller values of $C_{10}^{}$ than found in the BF point, and negative values of $C_{9'}^{}$. 
There are also a few large values along direction $\pm3$ which is dominated by $C_{10^\prime}^{}$. Three $P_1$ measurements at low $q^2$ show large deviations along directions $\pm 2$ which corresponds mostly to  $C_{7^\prime}^{}$. When correlations are included, two $P_2$ measurements single out direction $5$ which aligns mostly with $C_{9}^{}$.
Finally, the case of $\delta^\pm_1$ shows that the global fit permits a much larger variation of $C_7^{}$ than the experimental constraint from $B\to X_s\gamma$. It is important to recall at this stage that a large $\delta$ means that the one-sigma region around the BF contains variations in the predictions for that particular observable along the specified direction that are much larger than its corresponding experimental or theoretical errors. The corresponding observables are thus strongly constraining the fit in the given direction.

\begin{figure}[!h]
\center{\includegraphics[scale=.5]{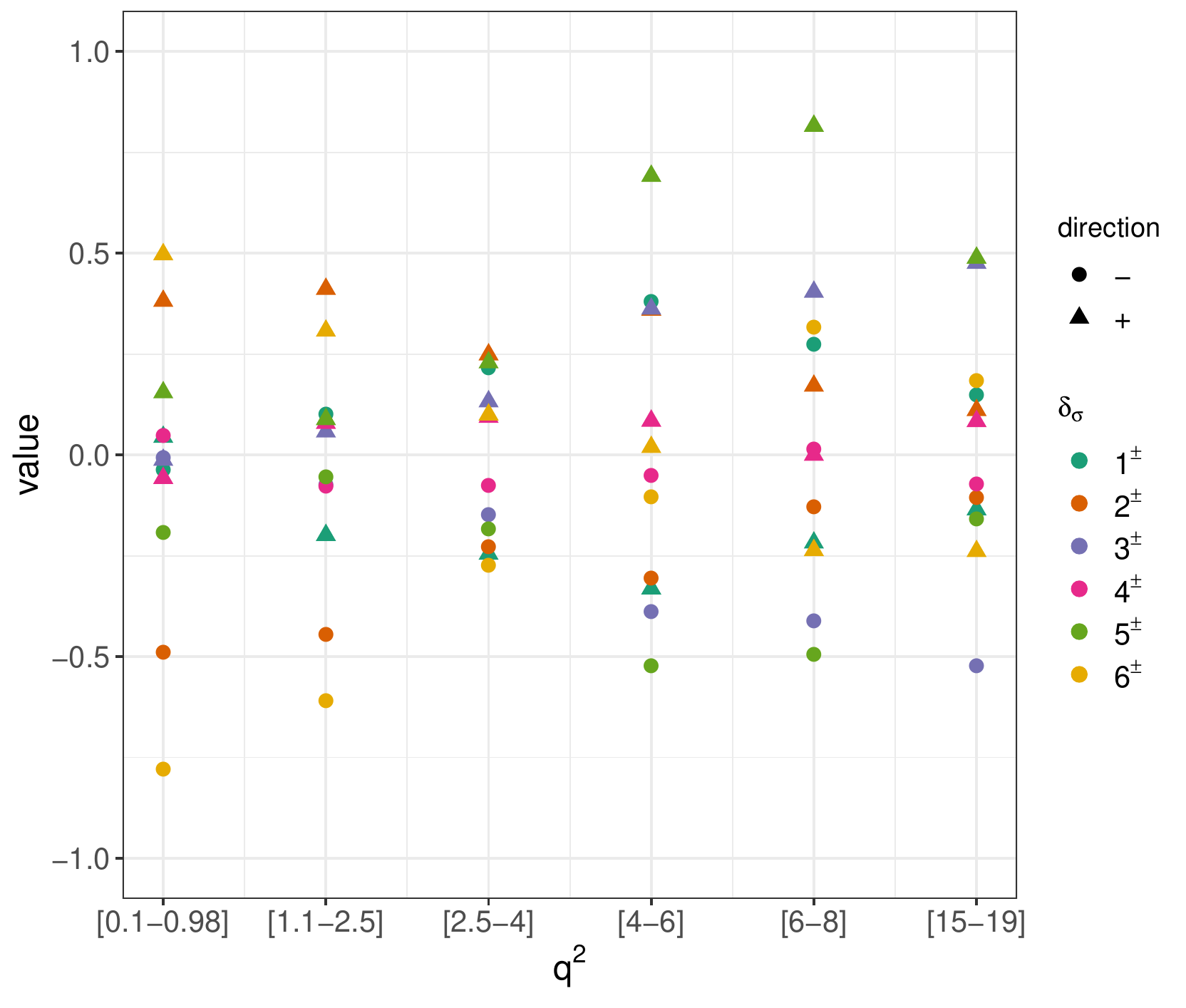}}
\caption{Distribution of $\delta_\sigma$ for the $P_5^\prime$ observable at LHCb in different $q^2$ bins.}
\label{f:deltap5}
\end{figure}

\begin{table}[htp]
\begin{center}
\[\begin{array}{|c|c|c|c|}
\hline
 \text{ID} & \text{Observable} & \text{$|\delta |$ $>$ 1} & |\delta _{\sigma }|\text{ $>$ 0.84} \\ \hline
 13 & \text{10${}^{7}$$\times $Br($B^+$ $\to $ $K^+\mu\mu$) [15,22] LHCb} &  \delta ^{4+}\text{  }\delta ^{4-}  & \text{} \\
 16 & \text{$P_1$(B $\to $ $K^*\mu\mu$) [0.1,0.98] LHCb} &  \delta ^{2+}\text{  }\delta ^{2-}  &  \delta ^{2+}\text{  }\delta ^{2-}  \\
 49 & \text{$P_2$(B $\to $ $K^*\mu\mu$) [6,8] LHCb} & \text{} &  \delta ^{5+}\text{  }\delta ^{5-}  \\
 {\color{red} 52} &  \text{$P_5^\prime$(B $\to $ $K^*\mu\mu$) [6,8] LHCb} & \text{} & {\color{red}  \delta ^{5+}\text{  }} \\
 57 & \text{$P_2$(B $\to $ $K^*\mu\mu$) [15,19] LHCb} & \text{} {\color{red} \delta ^{3+}\delta ^{5+}} &  \delta ^{5+}  \\
 68 & \text{10${}^{7}$$\times $Br($B^0$ $\to $ $K^{0*}\mu\mu$) [15,19] LHCb} &  \delta ^{3+}\text{  }\delta ^{4+}\text{  }\delta ^{4-}\text{  }\delta ^{6-}
 & \text{} \\
 74 & \text{$P_1$($B_s$ $\to $ $\Phi\mu\mu$) [0.1,2] LHCb} & \text{} &  \delta ^{2+}\text{  }\delta ^{2-}  \\
 93 & \text{10${}^{7}$$\times $Br($B_s$ $\to $ $\Phi\mu\mu$) [15,18.8] LHCb} &  {\color{red} \delta ^{3+}} \delta ^{4-}\text{  }\delta ^{6-}  & {\color{blue} \delta ^{6-}}  \\
 95 & \text{$P_1$(B $\to $ $K^\star ee$) [0.0020,1.120] LHCb} &  \delta ^{2+}\text{  }\delta ^{2-}  &  \delta ^{2+}\text{  }\delta ^{2-}  \\
 98 & \text{$R_K$($B^+$ $\to $ K+) [1,6] LHCb} &  {\color{red} \delta ^{3+}} \delta ^{4+}\text{  }\delta ^{4-}\text{  }\delta ^{6+}  &  \delta ^{3+}\text{  }{\color{blue}\delta ^{3-}}\text{  }\delta
^{4+}\text{  }\delta ^{4-}\text{  }\delta ^{6+}  \\
 100 & \text{$R_K^\star$ ($B^0 \to  K^{\star 0} $) [1.1,6] LHCb} & \text{} &  {\color{red} \delta ^{3+}} \delta ^{4-}\text{  }\delta ^{6-}  \\
 {\color{red} 114} & \text{$P_1$(B $\to $ $K^\star \mu\mu$) [0.04,2] ATLAS} &   &  {\color{red}\delta ^{2+}}  \\
 155 & \text{10${}^{7}$$\times $Br(B $\to $ $K^* \mu\mu$) [16,19] CMS8} & {\color{red}\delta ^{3+}}  {\color{blue}\delta ^{4+}}\text{  }\delta ^{4-}\text{  }\delta ^{6-}  & \text{} \\
 171 & \text{10${}^{4} $$\times$ Br(B $\to $ $X_s\gamma$)} &  \delta ^{1+}\text{  }\delta ^{1-}  &  \delta ^{1+}\text{  }\delta ^{1-}  \\
 172 & \text{10${}^{9}$$\times $Br($B_s$ $\to $ $\mu\mu$)} & {\color{red}\delta ^{3+}} \text{} &  \delta ^{3+}\text{  }\delta ^{6+}\text{  }{\color{blue}\delta ^{6-} } \\
 173 & \text{S(B $\to$ $K^* \gamma$)} & \text{} &  \delta ^{2+}\text{  }\delta ^{2-} \\  \hline
\end{array}\]
\end{center}
\caption{Observables from Figure~\ref{fig:e-ranges}  that present values $|\delta_i|>1$ or $|\delta_{\sigma i}|>0.84$. The figures in blue appear only with the SVD points calculated in the quadratic approximation for $\chi^2$ whereas the figures in red appear only when the exact $\chi^2$ is used instead.}
\label{t:largedel}
\end{table}%
A quick scan of Table~\ref{t:largedel} reveals that there are no large $\delta$s in the $P_5^\prime$ observables. Given the interest in this observable, we examine it  separately in Figure~\ref{f:deltap5} where we compare the different $q^2$ bins as they show different behaviour. There is a somewhat large variation in direction 6
 in the first bin, a larger variation with direction 5 for bins $[4,6]$ and $[6,8]$ and finally direction 3  is behind most of the variation for the last bin.

\subsection{Ranking observables}

Absolute values of $\delta$, or rather $\delta^2$, tell us how much each observable contributes to constraining a given direction in parameter space.
Therefore this ranking  gives an indication of how the eigendirections get constrained in the global fit, and the hierarchy in values of $\delta^2$ allows us to judge the relative importance of each observable. Without correlations we list the largest five values of  $\delta^2$ for each direction in Tables~\ref{randel1},~\ref{randel2}. Direction one (mostly $C_7^{}$) for example, is mostly constrained by $Br(B\to X_s\gamma)$ (ID 171), as can be seen both from Table~\ref{t:largedel} or Table~\ref{randel1}.
A somewhat different picture is obtained when taking into account correlations, see Tables~\ref{randel3},~\ref{randel4}, and direct comparison of the two gives indications of which observables are most sensitive to such effects.

Direction two is constrained predominantly by low $q^2$ bin observations of $P_1$,
and direction four is dominated by the single observable  98 (LHCb measurement of $R_K$), especially when taking into account correlation effects.
A very different picture is observed in direction three, which does not exhibit a large hierarchy in $\delta^2$. This
indicates that it is really the combination of multiple observables that constrains this direction.
Observable 98 is found to be relevant in constraining direction six, for which we find that the list of most sensitive observables is similar to that found for direction three.
Direction five  (mostly $C_9^{}$) is not especially constrained by a single observable, as indicated by the absence of a particularly large $\delta^{5\pm}$. The largest $\delta^2$ in this case occurs for 57 or $P_2$ (rather than
$P_5^\prime$ as one might have expected). Moreover, observable 57 is more constraining in direction 5+ than in 5-.

The most striking difference when including correlations occurs for observable 68 (a large $q^2$ bin measurement of $Br(B^0\to K^{\star 0}\mu^+\mu^-)$ by LHCb). While
it appears in the first few positions in the rankings without covariance in several directions (3, 4, 5, 6), it drops below our cutoff
when including covariance. As the correlation maps show, 68 is part of a group of highly correlated observables.
It is in particular strongly correlated with observables 73 and 155, which also drop out of the rankings when including correlations.

\begin{table}[ht]
\centering
\begin{tabular}{|c|c|c|c|c|c|c|c|c|c|c|c|}
\hline
\multicolumn{2}{|c|}{1+} & \multicolumn{2}{|c|}{1-} &
\multicolumn{2}{|c|}{2+} & \multicolumn{2}{|c|}{2-} &
\multicolumn{2}{|c|}{3+} & \multicolumn{2}{|c|}{3-}\\
\hline
ID & $\delta^2$ & ID & $\delta^2$ &
ID & $\delta^2$ & ID & $\delta^2$ &
ID & $\delta^2$ & ID & $\delta^2$ \\
\hline
171 & 4.07 & 171 & 5.03 & 16 & 2.48 & 16 & 2.28 & 68 & 1.02 & 57 & 0.87\\
170 & 0.58 & 170 & 0.74 & 95 & 1.25 & 95 & 1.19 & 57 & 0.89 & 98 & 0.85\\
41 & 0.56 & 41 & 0.52 & 114 & 0.83 & 114 & 0.74 & 155 & 0.85 & 68 & 0.80\\
90 & 0.34 & 90 & 0.46 & 173 & 0.74 & 173 & 0.74 & 172 & 0.85 & 172 & 0.69\\
49 & 0.31 & 49 & 0.39 & 74 & 0.73 & 74 & 0.72 & 98 ({\color{red} 93}) & 0.75 & 155 & 0.68\\
\hline
\end{tabular}
\caption{Ranking of observables by $\delta$ along directions 1,2,3. In red the changes when the SVD points are defined by the exact $1\sigma$ surface.}
\label{randel1}
\end{table}

\begin{table}[ht]
\centering
\begin{tabular}{|c|c|c|c|c|c|c|c|c|c|c|c|}
\hline
\multicolumn{2}{|c|}{4+} & \multicolumn{2}{|c|}{4-} &
\multicolumn{2}{|c|}{5+} & \multicolumn{2}{|c|}{5-} &
\multicolumn{2}{|c|}{6+} & \multicolumn{2}{|c|}{6-}\\
\hline
ID & $\delta^2$ & ID & $\delta^2$ &
ID & $\delta^2$ & ID & $\delta^2$ &
ID & $\delta^2$ & ID & $\delta^2$ \\
\hline
98 & 2.33 & 98 & 2.87 & 57 & 0.93 & 49 & 0.64 & 98 & 1.42 & 68 & 2.07\\
68 & 1.41 & 68 & 1.67 & 49 & 0.72 & 68 & 0.58 & 172 & 0.97 & 155 & 1.64\\
13 & 1.33 & 13 & 1.62 & 52 & 0.56 & 155 & 0.49 & 13 & 0.91 & 93 & 1.44\\
155 & 1.08 & 155 & 1.28 & 44 & 0.56 & 41 & 0.43 & 40 & 0.61 & 20  ({\color{red} 40}) & 0.80\\
93 & 0.93 & 93 & 1.10 & 171 & 0.35 & 93 & 0.42 & 19 & 0.54 & 73  ({\color{red} 20}) & 0.80\\
\hline
\end{tabular}
\caption{Ranking of observables by $\delta$ along directions 4,5,6. In red the changes when the SVD points are defined by the exact $1\sigma$ surface.}
\label{randel2}
\end{table}

\begin{table}[ht]
\centering
\begin{tabular}{|c|c|c|c|c|c|c|c|c|c|c|c|}
\hline
\multicolumn{2}{|c|}{1+} & \multicolumn{2}{|c|}{1-} &
\multicolumn{2}{|c|}{2+} & \multicolumn{2}{|c|}{2-} &
\multicolumn{2}{|c|}{3+} & \multicolumn{2}{|c|}{3-}\\
\hline
ID & $\delta^2$ & ID & $\delta^2$ &
ID & $\delta^2$ & ID & $\delta^2$ &
ID & $\delta^2$ & ID & $\delta^2$ \\
\hline
171 & 4.07 & 171 & 5.03 & 16 & 1.95 & 16 & 1.77 & 172 & 0.85 & 98 & 0.85\\
170 & 0.49 & 170 & 0.64 & 95 & 0.96 & 95 & 0.91 & 98 & 0.75 & 172 & 0.69\\
41 & 0.30 & 49 & 0.35 & 173 & 0.74 & 173 & 0.74 & 100 & 0.53 & 93 & 0.42\\
49 & 0.24 & 41 & 0.27 & 74 & 0.74 & 74 & 0.73 & 93 & 0.51 & 100 & 0.42\\
169 & 0.13 & 169 & 0.17 & 114 & 0.66 & 114 & 0.59 & 57 & 0.40 & 13 & 0.36\\
\hline
\end{tabular}
\caption{Ranking of observables by $\delta_{\sigma}$ in the first three directions.}
\label{randel3}
\end{table}

\begin{table}[ht]
\centering
\begin{tabular}{|c|c|c|c|c|c|c|c|c|c|c|c|}
\hline
\multicolumn{2}{|c|}{4+} & \multicolumn{2}{|c|}{4-} &
\multicolumn{2}{|c|}{5+} & \multicolumn{2}{|c|}{5-} &
\multicolumn{2}{|c|}{6+} & \multicolumn{2}{|c|}{6-}\\
\hline
ID & $\delta^2$ & ID & $\delta^2$ &
ID & $\delta^2$ & ID & $\delta^2$ &
ID & $\delta^2$ & ID & $\delta^2$ \\
\hline
98 & 2.34 & 98 & 2.88 & 49  ({\color{red} 57}) & 1.20 & 49 & 1.14 & 98 & 1.42 & 100 & 1.07\\
100 & 0.66 & 100 & 0.79 & 57  ({\color{red} 49}) & 1.15 & 41 & 0.47 & 172 & 0.97 & 172 & 0.78\\
172 & 0.51 & 13 & 0.61 & 52 & 0.66 & 171 & 0.37 & 19 & 0.60 & 93 & 0.71\\
13 & 0.50 & 172 & 0.61 & 44 & 0.48 & 44 ({\color{red} 57})  & 0.27 & 13 & 0.49 & 40 & 0.61\\
93 & 0.41 & 93 & 0.49 & 56 & 0.42 & 57 ({\color{red} 44})  & 0.26 & 40 & 0.45 & 20 & 0.61\\
\hline
\end{tabular}
\caption{Ranking of observables by $\delta_{\sigma}$ in the last three directions. In red the changes when the SVD points are defined by the exact $1\sigma$ surface.}
\label{randel4}
\end{table}

\subsection{Variance in the fit}

The rankings of the previous section provide information about observables in the parameter space defined by the eigendirections.
We have already seen that several observables are important in constraining multiple directions.
An alternative way of looking at this information is to study which parameter combinations result in the largest variance in
theory predictions. One approach is therefore to perform a principal component analysis (PCA) on the set of delta vectors.
PCA is an orthogonal linear transformation onto a coordinate system such that the first basis direction is aligned with the maximum variance in the data, the second basis is the direction of maximum variation orthogonal to the first coordinate, and the remaining bases are sequentially computed analogously. It can be used for dimension reduction as the first few principal components (PCs) capture most of the information.

For this we consider each observable as one data point with 12 parameters, the values of $\delta$ in the 12 shifted points.
The first two principal components, for example, provide the directions with largest variations, and plotting
the data points in these projections shows which observables dominate.
Different information is captured by looking at each observable in isolation (using  $\delta$) or in the context of correlations within the global fit (using $\delta_{\sigma}$), and we therefore reproduce this analysis for both cases.

For the PCA analysis the data should first be centered, i.e. the mean in each direction has to be subtracted. In our case, the mean values are close to zero so the effect of centering is not very large. We find very symmetric behavior: the main difference between plus/minus directions is just the sign of $\delta$.
This means that we can fully describe the 12 dimensional distribution in the space of the first six PCs. These six remaining PCs  are found to contain considerable variance in the distribution: whereas the first PC explains 31\% of the variance, the sixth one explains 8\% when correlations are ignored.
When correlations are kept the first PC explains 20\% of the variance and the sixth one explains 13\%.
This suggests that all six dimensions (i.e.\ WCs) still allow for considerable variance in the predictions of the observables (recall that this is measured relative to the errors).
The full rotation matrix transforming from delta space onto the first six PCs is given explicitly  in Table~\ref{tab:rot1}.
Notice the differences between the two rotation matrices, for example $\delta_3$ (mostly $C^{}_{10^\prime}$) is an important contribution to PC2 based on $\delta$, but not relevant in the first two PCs when considering $\delta_\sigma$ (we can already observe the large reduction of variance in that direction by comparing the rankings in $\delta_3^2$ of Table~\ref{randel1} and Table~\ref{randel3}).

\begin{figure}[ht]
\center{
\includegraphics[width=0.45\textwidth]{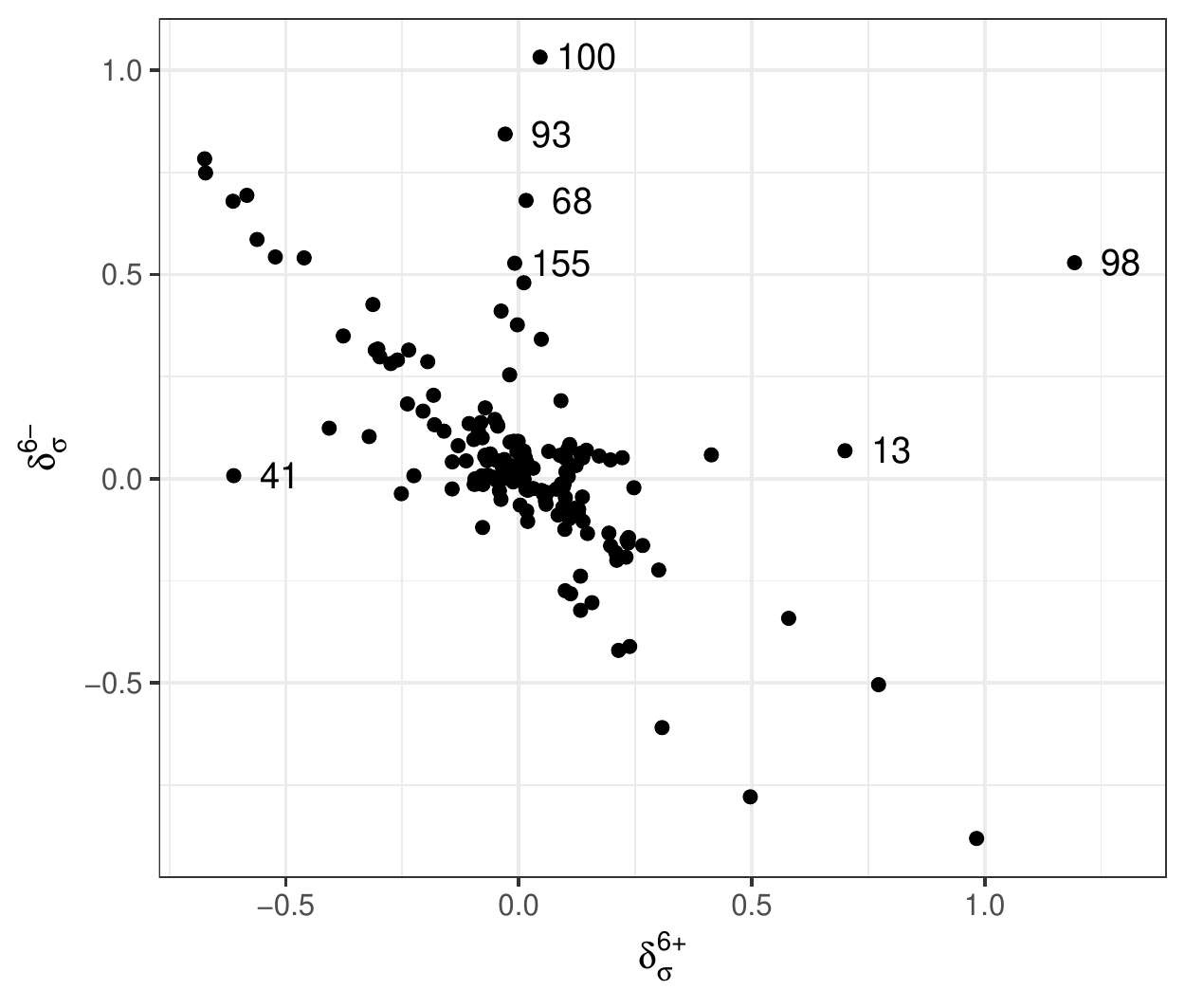}
\includegraphics[width=0.45\textwidth]{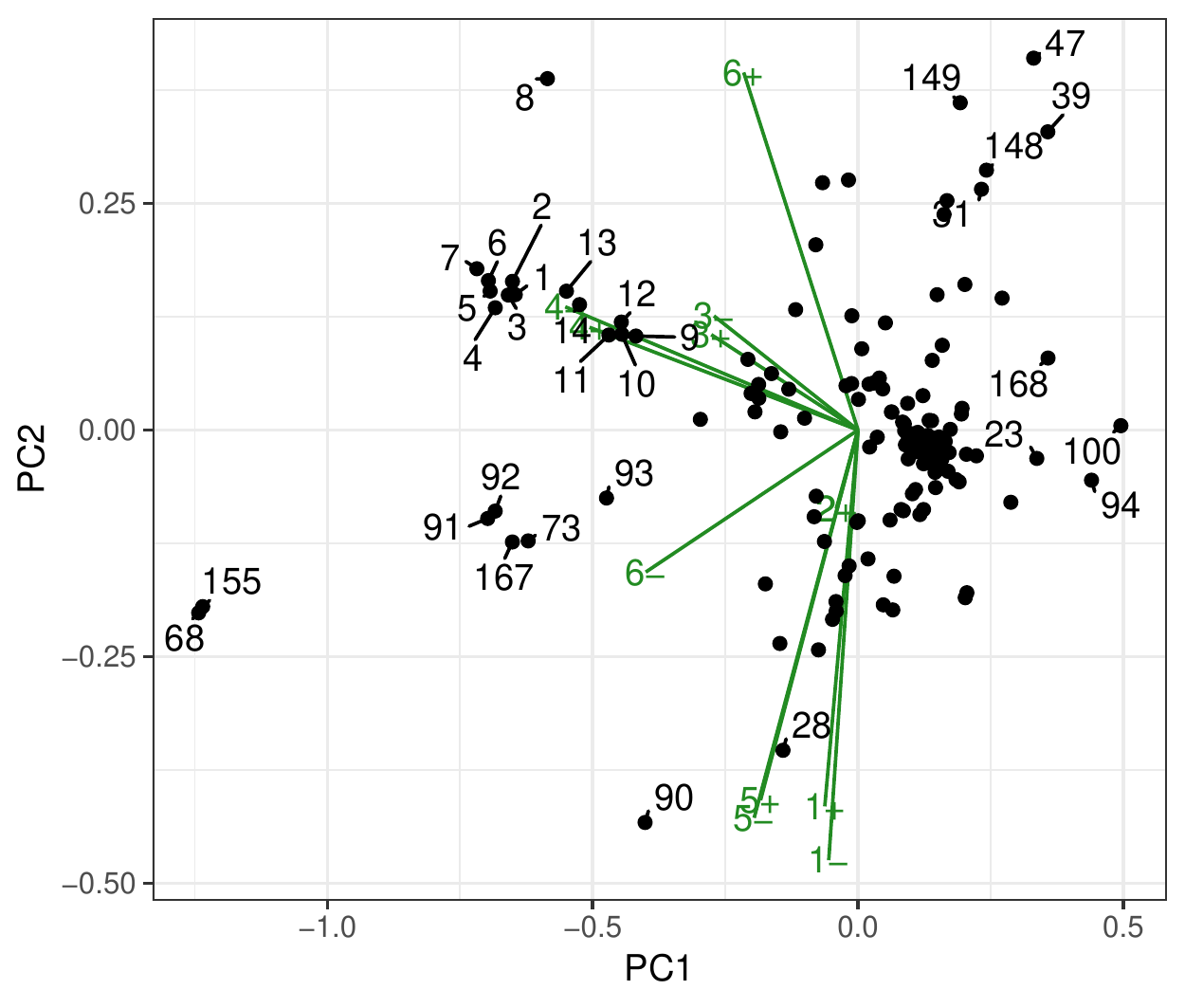}}
\caption{Comparison of distribution in $\delta_{\sigma}^{6+}$ and $\delta_{\sigma}^{6-}$, where observables with the largest deviation from anticorrelated behaviour are labeled (left panel).
Illustration of $\Delta(\delta_i^{\pm})$ in the space of the first two principal components capturing most of the total variance in the distribution (right panel).}
\label{fig:pca_deltadiff}
\end{figure}

We find that $\delta_6$ is the only direction which exhibits strongly asymmetric behavior:  for certain observables there are differences between the change in prediction in plus/minus directions, see Figure~\ref{fig:pca_deltadiff} (left). This figure compares the values in the two directions of $\delta^6_{\sigma}$  (a similar but more crowded picture is found plotting $\delta^6$), and shows as an extreme example observable 100, $R_{K^\star}$, for which the theory prediction varies significantly along one direction but not the opposite. $\delta^{6-}$ is the only one of the twelve points with a large negative $C_{9^\prime}$, and to a lesser extent $C_{10^\prime}$. 
\begin{table}
\begin{tabular}{|c|}
\hline
\\ \hline
$1+$ \\ $1-$ \\
$2+$ \\ $2-$ \\
$3+$ \\ $3-$ \\
$4+$ \\ $4-$ \\
$5+$ \\ $5-$ \\
$6+$ \\ $6-$ \\
\hline
\end{tabular}
\begin{tabular}{|rrrrrr|}
\hline
PC1 & PC2 & PC3 & PC4 & PC5 & PC6\\
\hline
-0.17 & -0.02 & -0.26 & 0.56 & -0.19 & 0.13\\
0.16 & 0.02 & 0.28 & -0.60 & 0.21 & -0.14\\
0.13 & 0.12 & 0.06 & -0.20 & -0.66 & -0.03\\
-0.13 & -0.12 & -0.04 & 0.12 & 0.66 & 0.00\\
-0.09 & 0.54 & 0.38 & 0.17 & 0.08 & 0.09\\
0.05 & -0.56 & -0.37 & -0.19 & -0.07 & -0.08\\
0.50 & 0.17 & -0.20 & 0.12 & 0.09 & -0.05\\
-0.54 & -0.18 & 0.24 & -0.14 & -0.10 & 0.08\\
0.17 & -0.13 & 0.19 & -0.05 & 0.01 & 0.73\\
-0.22 & 0.20 & -0.07 & 0.01 & -0.02 & -0.55\\
-0.13 & -0.42 & 0.52 & 0.26 & -0.11 & -0.17\\
-0.51 & 0.26 & -0.39 & -0.32 & -0.01 & 0.27\\
\hline
\end{tabular}
\begin{tabular}{|rrrrrr|}
\hline
PC1 & PC2 & PC3 & PC4 & PC5 & PC6\\
\hline
-0.35 & 0.09 & -0.21 & 0.36 & -0.38 & -0.07\\
0.37 & -0.09 & 0.24 & -0.40 & 0.40 & 0.08\\
-0.02 & -0.27 & 0.07 & 0.24 & 0.11 & 0.60\\
0.05 & 0.25 & -0.03 & -0.29 & -0.06 & -0.58\\
0.05 & -0.07 & -0.42 & -0.43 & -0.28 & 0.23\\
-0.05 & 0.09 & 0.45 & 0.41 & 0.25 & -0.22\\
0.24 & -0.38 & -0.24 & 0.19 & 0.04 & -0.23\\
-0.24 & 0.41 & 0.27 & -0.22 & -0.06 & 0.26\\
0.48 & 0.08 & 0.27 & 0.05 & -0.53 & 0.11\\
-0.42 & -0.07 & -0.23 & -0.16 & 0.39 & 0.00\\
0.08 & 0.67 & -0.18 & 0.11 & 0.16 & 0.23\\
-0.45 & -0.27 & 0.46 & -0.30 & -0.28 & -0.02\\
\hline
\end{tabular}
\caption{Rotation matrix between spaces of $\delta$s and principal components for definition 1 ($\delta$ left) and definition 3 ($\delta_\sigma$ right)}
\label{tab:rot1}
\end{table}

\begin{figure}[ht]
\center{\includegraphics[width=0.45\textwidth]{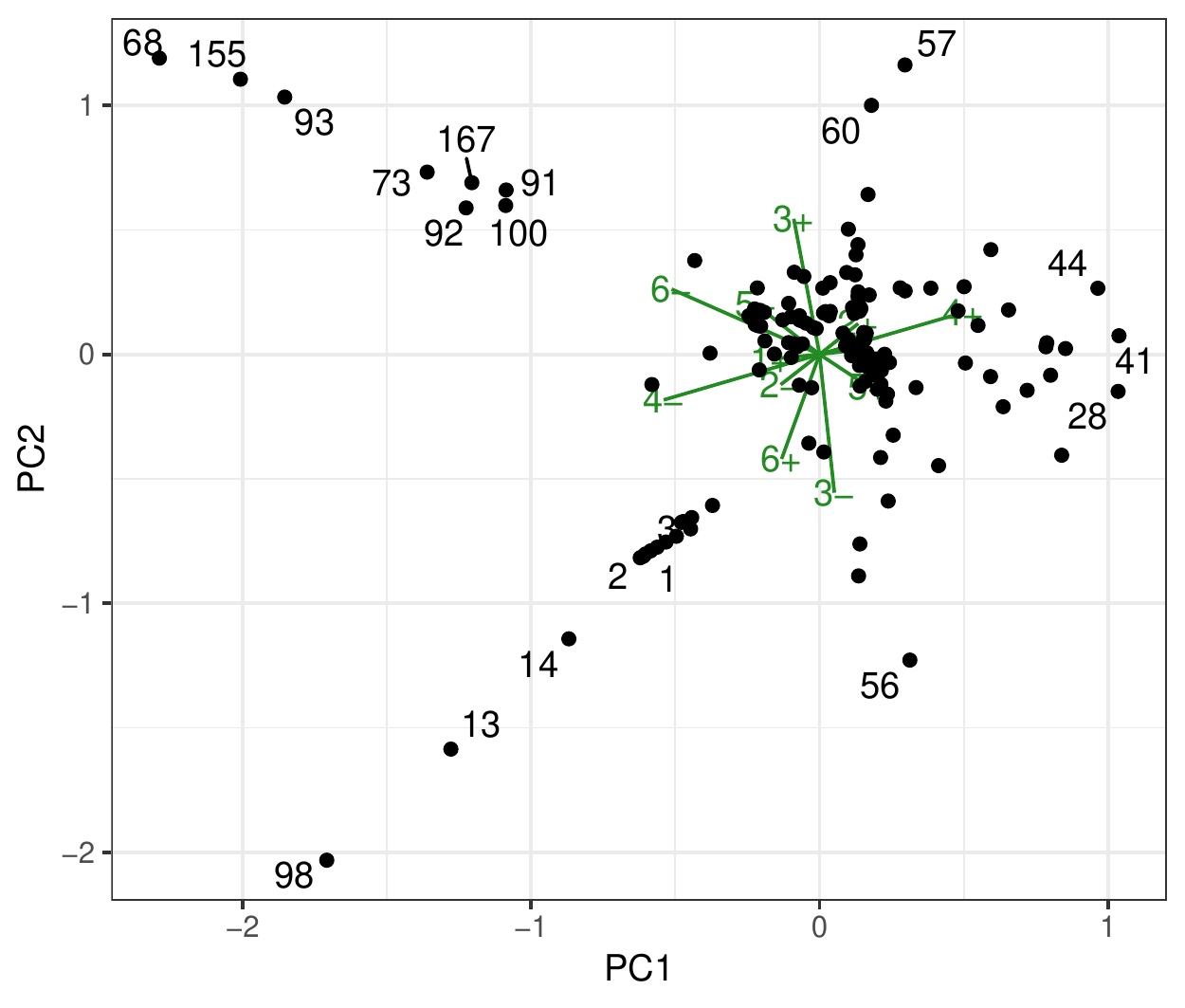}
\includegraphics[width=0.45\textwidth]{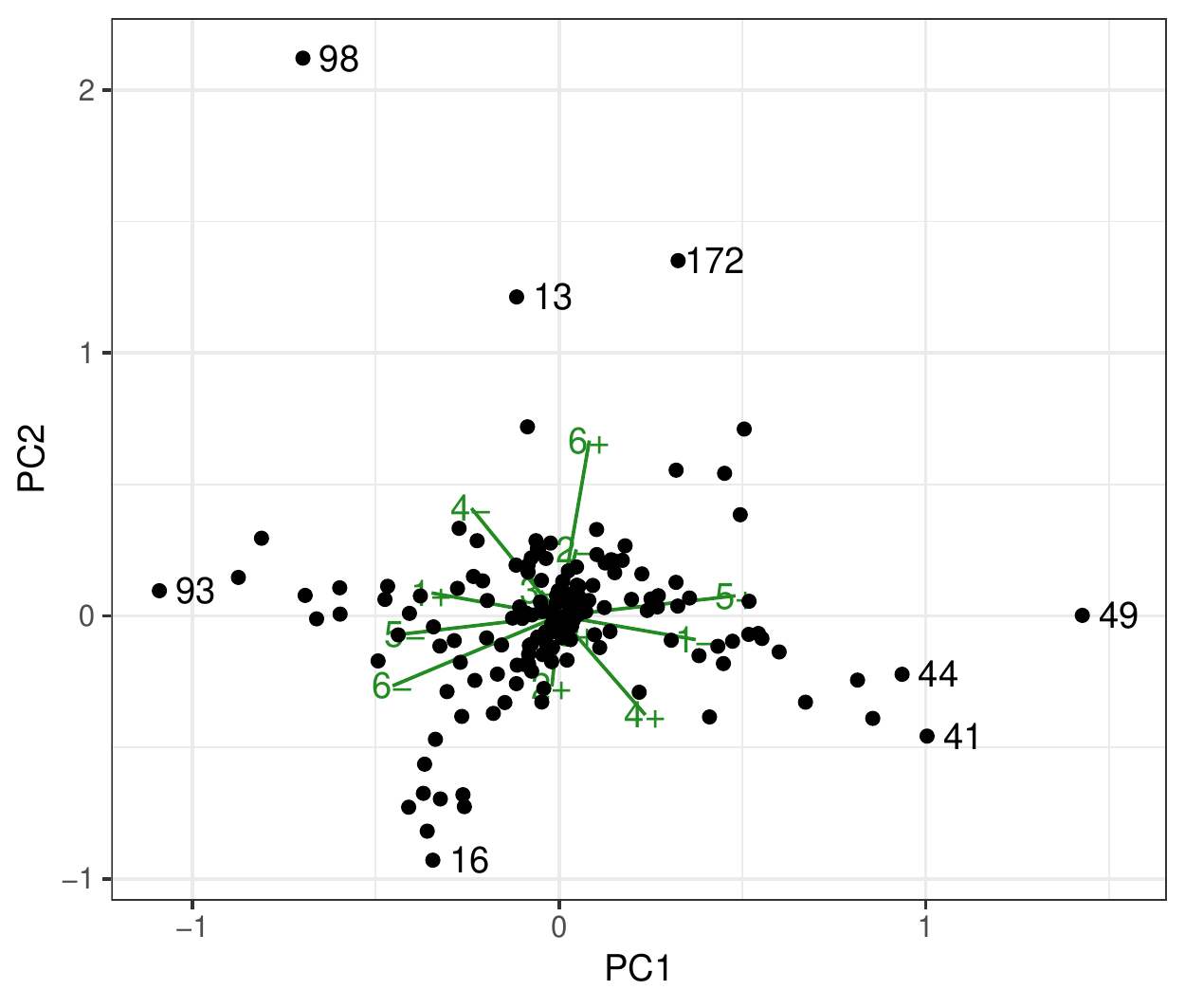}}
\caption{Observables in the space of principal components obtained by performing PCA on the distributions in $\delta$ (left) and $\delta_{\sigma}$ (right). The green arrows illustrate the projection onto the first principal components and are labeled by direction in $\delta$ space. They indicate how the original parameters relate to the projection.}
\label{fig:PC1_PC2}
\end{figure}

We now focus on the first two PCs to study which directions and observables are responsible for the largest variation.
To get an overview of the distribution of $\delta$s we show the projection of observables onto the first two principal components in Figure~\ref{fig:PC1_PC2}
in the form of so-called biplots.  These show the projected data points, as well as a visualisation of the projection in the form of labeled arrows pointing outwards from the center. This format makes it easy to relate directions on the projection to the original parameters.

When considering each observable in isolation (left view), clear trends can be observed. For example, observables aligned with direction $6-, 5-$ and anti-aligned with direction $5+$ are mainly branching ratio observations in bins of large $q^2$ (e.g. IDs 68, 93, ...).
There are differences in branching ratio observables depending on the final state: notably most observables with negative PC1 but positive PC2 correspond to decays into $K^*$, while decays into $K^+$ and $K^0$ appear to take negative values in PC2 (e.g. IDs 98, 13, 14).
Angular observables on the other hand show a very different behavior. For example observables 28, 41 and 44 are found to have the largest values of PC1. Large $q^2$ bins are different, e.g. IDs 56, 57, 60, take small positive values in PC1 but large absolute values in PC2.

The picture changes drastically when considering correlations (right view), where the relevance of large $q^2$ bins of branching ratio observables is no longer dominant. Note also the different effect that including the correlations has on the angular observables: for some of them the differences appear to be enhanced (e.g. IDs 16 and 49); yet others no longer stand out in this picture such as $P_5'$ (IDs 44 and 60).

It is further instructive to assess the impact of covariance in a more direct fashion, to this end we introduce a difference in terms of the absolute values,
\begin{equation}
\Delta(\delta_i^{\pm}) = |\delta_i^{\pm}| - |\delta_{\sigma i}^{\pm}|.
\end{equation}
With the aid of this metric, most of the differences can be explained in two dimensions as the first two principal components capture about 80\% of overall variation. We present the projection onto these first two PCs in Figure~\ref{fig:pca_deltadiff} (right). 
This difference shows the same behaviour of BR observables already observed by comparing the results in Fig.~\ref{fig:PC1_PC2}.
In addition we note large effects for a different group of observables, taking the largest positive values in PC1 and a range of values in PC2.
At larger values of PC2 these are dominantly measurements of $F_L$ (IDs 31, 39, 47, 148) and $A_{FB}$ (ID 149), while a number of other observables stand out at low values of PC2.

\subsection{Limitations of the Hessian approach}
\label{sec:limit}
As discussed above and in Appendix~\ref{sec:appA}, the quadratic approximation is found to be a good description to the full fit, but does not capture all details.\footnote{Note that this is expected by construction, the $\chi^2$ function of~\cite{Capdevila:2017bsm} is defined in the linearized or gaussian regime, see~\cite{Descotes-Genon:2015uva}. Asymmetries in the $\chi^2$ function are therefore induced by higher order terms in the expression of the observables as functions of the Wilson coefficients, which are expected to be small for $|\mathcal{C}_i^\text{NP}|\lesssim 1$.}
Most notably, asymmetries present in the exact $\chi^2$ function are not captured in the approximation,
and this leads to deviations of up to about $30\%$ in $\Delta(\chi^2)$ for some directions. 
As a consequence several SVD points are not exactly $1\sigma$ away from the BF point when measured with the exact $\chi^2$ function.

An alternative approach would be to use the Hessian for the identification of the SVD directions, but to define the SVD points by the intersection of said directions with the exact  $1\sigma$ surface. We have explicitly verified that we reach the same conclusions with both approaches, even though detailed quantitative results will of course be slightly different. A summary of these comparisons is given below. 

The observables with $\delta$ values above the cutoff are listed in Tables~\ref{t:largedel} and~\ref{deltasmix}, where blue (red) indicates they are only above the cutoff when evaluating the SVD points in the approximation (exact calculation). Several differences are found, most notably  along direction $3^+$, as may be expected from the results in Table~\ref{t:fits}. In those cases values of $\delta$ are close to the cutoff in both calculations, with differences typically smaller that $10\%$.

The change also affects the rankings shown in Tables~\ref{randel1}-\ref{randel4}, where differences in ordering when using the exact calculation to find the SVD points are given in red. While we find the exact values of $\delta$ in directions with the largest asymmetries to change notably (as expected), the effect in terms of hierarchies and ordering is very limited, and reordering only happens in a few instances where the absolute values of $\delta$ are very similar in both approaches.

Finally we have also re-evaluated the PCA. We find that the projections onto the first two principal components are similar in terms of how each of the $\delta$ directions contributes to the projection.  
Differences become more relevant beyond the first two principal components, since these directions carry less of the overall variance and thus are more sensitive to details. For the same reason these higher PC's are not important in the discussion.

These results confirm that the quadratic approximation is appropriate for the description of the fit to the level performed in this paper.

\section{Additional observables proposed to test lepton flavor universality}\label{s:newob}

Several additional observables have been proposed in the literature to test lepton flavor universality by directly comparing the distributions in modes with muons to those in modes with electrons~\cite{Capdevila:2016ivx}. In this section we assess their likely future impact on the global fit, pinpointing which of these best constrain each eigendirection. A list of the 48 new observables with their corresponding ID is provided in Appendix~\ref{s:idmatch}.

\begin{figure}[!h]
\includegraphics[width=\textwidth]{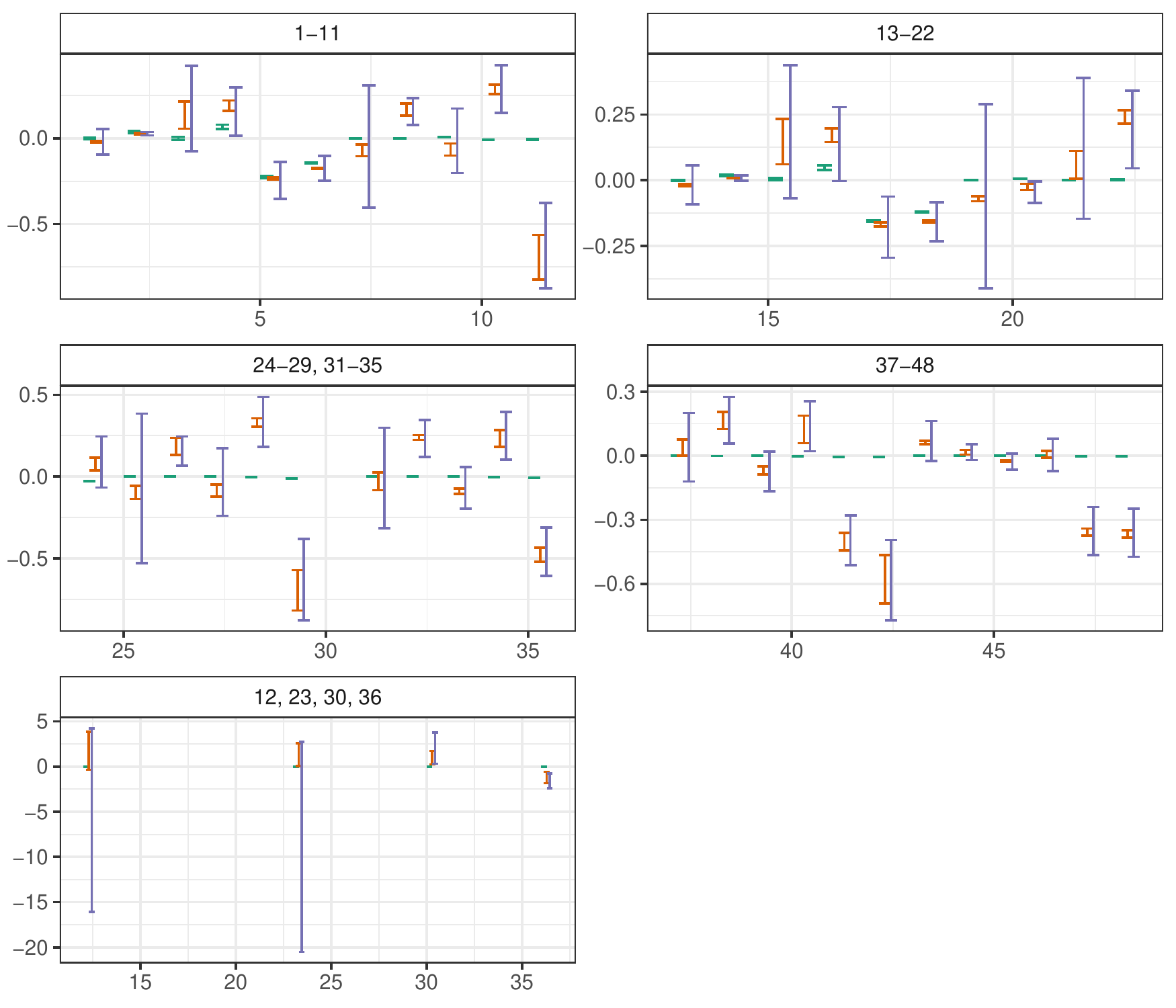}
\caption{Comparison of SM prediction (green), BF prediction (brown) and fit uncertainty (purple) for the proposed observables listed in Table~\ref{tab:qObs}.}
\label{fig:errors-qs}
\end{figure}

We begin with a direct comparison of the theoretical predictions for the SM along with their uncertainty (green), a theoretical prediction for the six dimensional BF (brown) and the uncertainty in the fit calculated as before (purple) in Figure~\ref{fig:errors-qs}.
Estimates of the experimental sensitivities expected for measurements of $Q$ observables have been presented by the Belle II collaboration, see~\cite{Kou:2018nap}, Table 67.

Theoretical errors are larger for predictions for the BF point compared to SM predictions. This is because long distance non-perturbative effects, one of the main sources of hadronic uncertainties, cancel in the SM.
In the presence of NP that distinguishes between muons and electrons, these uncertainties get reintroduced proportionally to the size of NP.
A small set of observables have a pole in their $q^2$-spectrum, which causes predictions within bins containing the pole to become unstable and their errors to diverge. These particular observables, which show large uncertainties are shown separately in Figure~\ref{fig:errors-qs}.

These two figures already indicate that most of these proposed observables will increase the constraining power of the global fit, as their uncertainties are much smaller than the current uncertainty in the fit. Of course, this will require measurements with experimental errors comparable in size (or smaller) than the corresponding theory errors.

\begin{figure}[!h]
\includegraphics[width=\textwidth]{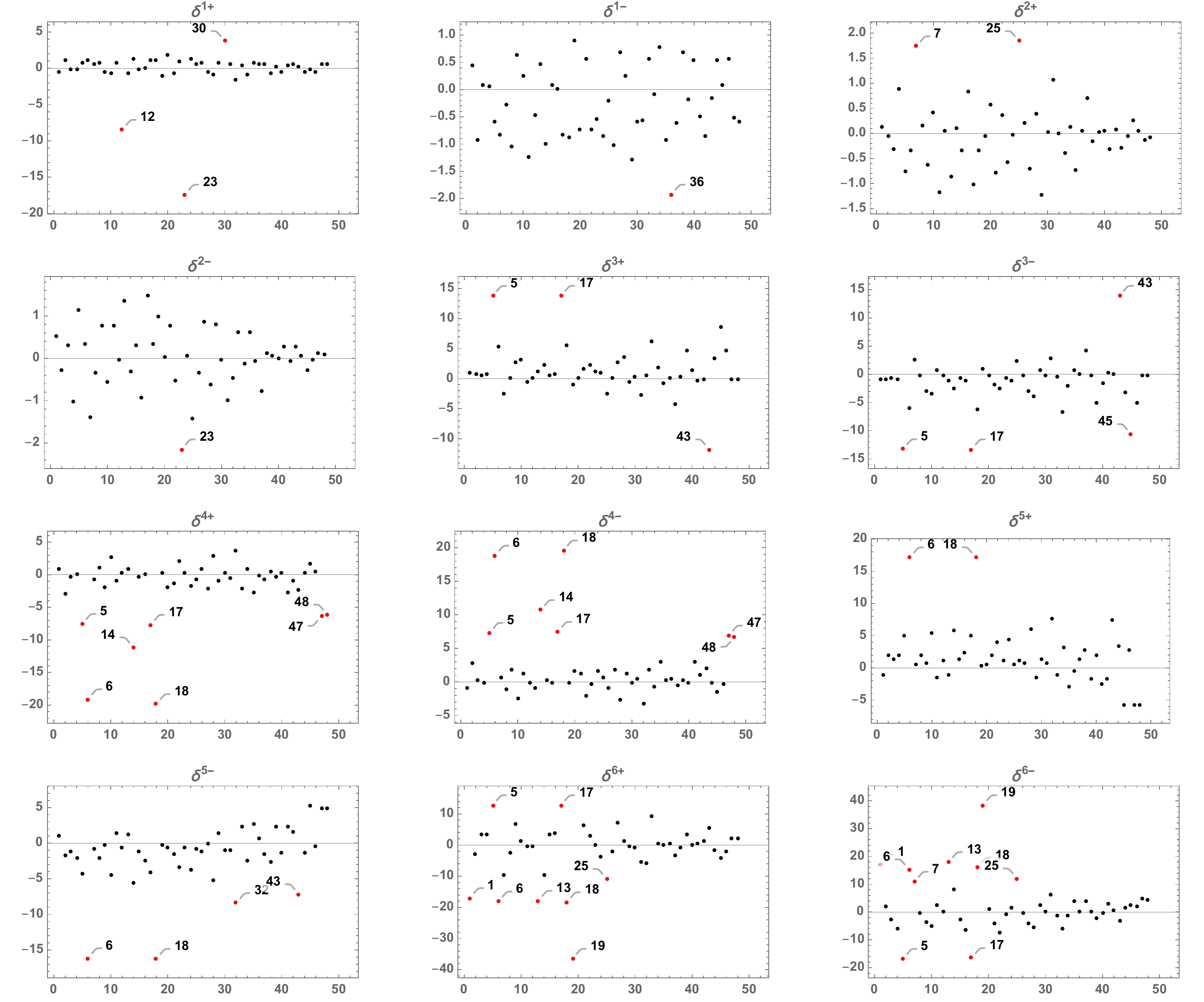}
\caption{Value of $\delta$ for the proposed observables listed in Table~\ref{tab:qObs} for the 12 SVD directions. The largest ones for each case are highlighted in red and labeled.}
\label{fig:deltas-qs}
\end{figure}

We can be a bit more specific by associating with each of these observables the corresponding directions which they can constrain. This is shown in Figure~\ref{fig:deltas-qs}, quantified with the metric $\delta_i$. The most sensitive observables to each of the 12 directions are marked in red and labelled in the figure.

There are a handful of observables that stand out regarding their potential to further constrain the parameter space: these are $Q_1$, $B_5$ and $B_{6s}$.
As a function of Wilson coefficients, $Q_1$ only depends on $C_{7',9',10'}$, and is one of the most stringent tests for the search of right-handed currents.
Since the current fit is poorly constrained in the $C_{9'}$ direction, this results in large values of $\delta^6$ for $Q_1$ (IDs 1, 7, 13, 19, 25)  in Figure~\ref{fig:deltas-qs}.
The case of $B_{5,6s}$ is rather different. These two observables, particularly in the very low-$q^2$ region, provide direct access to $C_{10}$ and this shows up as large $\delta^4$ (IDs 5, 6, 17, 18, 47, 48) in Figure~\ref{fig:deltas-qs}. Surprisingly, some large-$q^2$ bins of these observables also show large $\delta^4$, at the same level of some of their low-$q^2$ counterparts, suggesting high sensitivity to $C_{10}$ also in this region.

This illustrates the promising opportunities for setting more precise constraints on the Wilson coefficients (as also stated in \cite{Alguero:2018nvb}). As a caveat we note that our observations could change when correlations are included.

\section{Summary and Conclusions}\label{s:con}

In this paper we establish a framework to quantify the importance of individual observables in the results of a global multidimensional fit. 
We apply our framework to the global fit to $b\to s \ell^+\ell^-$ mediated observables with the six free parameters $C^\text{NP}_7$, $C^\text{NP}_{7^\prime}$,
$C^\text{NP}_{9\mu}$, $C^\text{NP}_{9^\prime\mu}$, $C^\text{NP}_{10\mu} $ and $C^\text{NP}_{10^\prime\mu}$ of Ref.~\cite{Capdevila:2017bsm}
(and for notational simplicity have dropped the indices NP and $\mu$).

We began with a direct visualisation of the one sigma region around the BF point and its position relative to the SM in six dimensions.
We then provided a quadratic approximation to the global fit in parameter space based on the Hessian matrix of second derivatives at the BF point.
This construction was used to find twelve points characterizing the $1\sigma$ contour.
These 12 points were used to assess the fit uncertainty for each observable in the fit.
In addition they represent $1\sigma$ shifts along well defined directions in parameter space and are thus representative of parameter directions.
In Section~\ref{s:quanti} we defined quantitative metrics to evaluate the relative importance of each of the 175 observables to the global fit.
Section~\ref{s:quantiresults} presented a systematic study of these measures, comparing both the SM to the BF point and the BF point to the set of 12 representative points, thereby illustrating the interplay between observable and parameter space.
Throughout this discussion we have emphasised the role of correlated errors, which we found to be important in the evaluation of single observables.
Finally Section~\ref{s:newob} applied the same framework to assess the likely impact of future measurements on the global fit.

The coefficient $C^\text{NP}_{9\mu}$ is of particular interest, as it has been singled out by lower dimensional fits, always finding a large deviation from the SM in this coefficient.
Indeed among the six parameters considered in the global fit, only a negative $C^\text{NP}_{9\mu}$ presents the correct patterns for explaining some of the most striking anomalies, increasing the predictions for the $P_{5}^\prime$ observable while reducing the predicted values of $R_K$ and $R_{K^*}$, and it is therefore expected to play a major role in the global fit.
Our visualisation of the six dimensional BF region confirms that the cloud of points within $1\sigma$ of the BF is clearly separated from the SM mostly along the $C_{9\mu}^\text{NP}$ direction.
At one sigma from the BF, in the direction of the SM, $C_{9\mu}^\text{NP}$ is still 60\% as large and still the largest of the NP parameters.
The importance of $R_K$ and $R_{K^*}$ can be appreciated in Figure~\ref{fig:newpt}, which shows that after moving $1\sigma$ from the BF towards the SM,
these observables (98, 100) still stand out: they have a large $\Delta$(Pull) preference for the NP point and (especially 98) already shows a large Pull against this move from the BF.

Correlated uncertainties (which at present include mostly the ones in the theory), play an important role in the discussion.
They reduce the preference for the BF over the SM for angular observables when considering conditional measures such as $\Delta_\sigma({\rm Pull})$ in place of the
isolated metric Pull(SM), used exclusively in previous works.
While Pull(SM) shows the largest discrepancies with the SM for BR, $R_{K^{(\star)}}$ and $P_{5}^\prime$ observables,
we find that the preference of the BF over the SM for some of the discrepancies is less apparent with the inclusion of correlations.
This is because patterns in deviations between predictions and measurements may be more/less consistent with expectations from correlated uncertainties depending on the parameter point, and this information is needed to understand the impact of each observable on the total $\chi^2$.
While correlation effects are negligible for the considered $R_{K^{(\star)}}$ observables,
important correlations are found for the uncertainties of angular observables.
In particular for the $P_{5}^\prime$ they reduce the Pull difference between BF and SM point when measured by $\Delta_\sigma({\rm Pull})$. 
We further note that some of the $P_{5}^\prime$ observables also disagree with the BF at more than the $2\sigma$ level. As pointed out in~\cite{Capdevila:2017bsm}, the anomaly in $P_5^\prime$
is best described by a larger negative $C_{9\mu}^\text{NP}$ ($\simeq -1.8$) than the one obtained in a global fit.

More generally our work illustrates some of the internal tensions in the fit, for example through a global discussion of the Pull for all observables. We also show for the first time a comprehensive picture of all observables entering the fit, comparing SM and BF predictions to the measured value, while also showing experimental, theoretical as well as fit uncertainties, see Figure~\ref{fig:e-ranges}.

A main focus of this study was then to consider each of the eigendirections of the Hessian matrix to asssociate the most sensitive observables to corresponding directions in parameter space.
For this discussion directions are labeled from most to least constrained, with each direction being dominated by one of the six considered Wilson coefficients, see Table~\ref{t:eignev}.
Our key findings for each direction can be summarized as follows (for ease of comparison we list ID numbers used in the Figures with each observable):
\begin{itemize}
\item Along the most constrained direction (direction one, corresponding to $C_{7}^\text{NP}$), the most sensitive observable is by far $B\to X_s\gamma$ (171),
which has a much smaller error than the variation allowed by the global fit as can be seen in  Figure~\ref{fig:e-ranges}. This is confirmed with large values of the $\delta$ measures defined in Equation~\ref{deltasdef}.

\item  Direction two (corresponding to $C_{7^\prime}^\text{NP}$) is also strongly constrained by a single observable, the low $q^2$ measurements of $P_1$ (16).
Again, this is a feature in Figure~\ref{fig:e-ranges} where the fit allows a much larger uncertainty for 16 than its experimental (or theoretical) error, and confirmed by large values of $\delta$.

\item A different picture is found for direction three (mostly $C^\text{NP}_{10^\prime\mu}$), for which several observables show comparable sensitivities,
demonstrating that the constraint accumulates from many small contributions in the same direction.
Note that in this case correlations become more important, as can be seen by comparing the rankings in $\delta$ and $\delta_{\sigma}$, and neglecting correlated contributions may result in overestimating the sensitivity of particular observables, most notably in this case for measurements of large $q^2$ bins of $Br(B \to K^\star \mu\mu)$ (IDs 68 and 155).

\item Direction four (mostly $C^\text{NP}_{10\mu}$) is particularly sensitive to the LFUV ratio $R_{K}$ (98) and can be further probed by the proposed observables $B_{5,6s}$.
Note that within the current six dimensional fit the tension with the SM and the fit projection along this direction is about $1\sigma$, making such future probes especially interesting.

\item Direction five (which mostly corresponds to $C_{9\mu}^\text{NP}$) exhibits a similar behavior of not being especially constrained by a single observable. Interestingly $P_2$ (41, 49, 57) observables are found to
be most sensitive to shifts in this direction, while neither $R_K$ (98) nor $R_{K^\star}$ (100) appear in our rankings.
Measurements of $P_{5}^\prime$ (44, 52) are found to be somewhat less sensitive than those of $P_2$ in our comparison, indicating that the normalisation to the uncertainties is important here.

\item The least constrained direction (direction six) is mostly aligned with $C^\text{NP}_{9^\prime\mu}$. In this case the parameter points of the $+(-)$ directions are quite different, with $C^\text{NP}_{9^\prime\mu} = 1.72 (-0.72)$.
As a consequence we observe very different sensitivities between the $\pm$ directions, which is illustrated e.g. in the left panel of Figure~\ref{fig:pca_deltadiff}.
Considering the fit uncertainties for future observables we find large potential of constraining this direction by measurements of $Q_1$.

\end{itemize}

We have thus cataloged sensitivities of (current and future) observables and related them to specific direction in the six dimensional parameter space considered here,
which can be used to inform future theoretical and experimental studies.
Note also that the Hessian approximation given in this paper allows for quick estimates of $\Delta\chi^2$ close to the BF point,
and the set of representative points listed in Table~\ref{t:fits} can be used to estimate current fit uncertainties on additional observables not considered here.

\section*{Acknowledgments} This work was supported in part by the Australian Government through the Australian Research Council. BC acknowledges financial support from the grant FPA 2017-86989-P and Centro de Excelencia Severo Ochoa SEV-2012-0234. We thank Dianne Cook for help with R and with visualization and Joaquim Matias, S\'ebastien Descotes-Genon and Ulrik Egede for useful conversations. G.V. thanks the CERN theory group for their hospitality and partial support while this work was completed.

\appendix

\section{Intersection of the one-sigma ellipsoid with its principal axes}
\label{sec:appA}
SVD produces the twelve points shown in Table~\ref{t:fits}. The eigenvalues are numbered in decreasing order. For comparison we list in the table the $\chi^2$ difference between these points and the best fit calculated numerically and using the quadratic approximation where $\Delta\chi^2=7.1$ exactly (this is the condition used to find the twelve points). The column  $\Delta\chi^2_{\rm exact}$ lists the number extracted from the code of Ref.~\cite{Capdevila:2017bsm}. The last column is used below to quantify a cut-off for `large' $\delta_\sigma$.

\begin{table}[htp]
\begin{center}
\begin{tabular}{|c|c|c|c|c|c|c|c|c|} \hline
EV &$C^{}_7$ & $C^{}_{9}$ & $C^{}_{10}$ &
$C^{}_{7^\prime}$ & $C^{}_{9^\prime}$ & $C^{}_{10^\prime}$ &
$\Delta\chi^2_{\rm exact}$ &  $\sum\delta_{\sigma}^2$  \\ \hline
 1$^+$ & 0.05 & -1.06 & 0.34 & 0.01 & 0.37 & -0.04 & 6.8 & 7.1 \\
 1$^-$ & -0.04 & -1.07 & 0.34 & 0.02 & 0.37 & -0.04 & 8.1 & 8.5 \\
 2$^+$ & 0.00 & -1.07 & 0.34 & -0.03 & 0.37 & -0.04 & 6.5 & 7.3\\
 2$^-$ & 0.01 & -1.06 & 0.34 & 0.07 & 0.38 & -0.04 & 7. & 7.1 \\
 3$^+$ & 0.01 & -1.16 & 0.23 & 0.03 & 0.25 & 0.25 & 4.9 & 7.2 \\
 3$^-$ & 0.00 & -0.97 & 0.45 & 0.01 & 0.50 & -0.33 & 8.9 & 7.2 \\
 4$^+$ & 0.01 & -1.27 & 0.72 & 0.02 & 0.29 & 0.00 & 7.7 & 6.6 \\
 4$^-$ & 0.00 & -0.86 & -0.04 & 0.02 & 0.46 & -0.08 & 6.3 & 7.7 \\
 5$^+$ & 0.02 & -1.54 & 0.13 & 0.02 & 0.51 & -0.21 & 4.9 & 9.6 \\
 5$^-$ & -0.01 & -0.59 & 0.55 & 0.02 & 0.24 & 0.13 & 9.1 & 6.7 \\
 6$^+$ & 0.00 & -1.03 & 0.59 & 0.01 & 1.70 & 0.64 & 7.4 & 10.0 \\
 6$^-$ & 0.00 & -1.10 & 0.09 & 0.02 & -0.97 & -0.72 & 8.6 & 11.5 \\
\hline
\end{tabular}
\end{center}
\caption{Parameter sets obtained with SVD  along with their correspondent exact $\Delta\chi^2$ for displacements (both ways $+/-$) away from the BF along the six eigenvector directions ($\Delta\chi^2=7.1$ for these points in the quadratic approximation). The eigenvectors are ordered by decreasing value of the corresponding eigenvalues of H.}
\label{t:fits}
\end{table}%

The small differences between the $\Delta\chi^2_{\rm exact}$ and 7.1 gives us an indication of uncertainty in our discussion, and we note how this is slightly different along the different SVD directions. For comparison purposes we show in Table~\ref{t:fitsex} the same information with points chosen at the intersections of the SVD directions with the exact $1\sigma$ surface.

\begin{table}[htp]
\begin{center}
\begin{tabular}{|c|c|c|c|c|c|c|c|} \hline
EV &$C^{}_7$ & $C^{}_{9}$ & $C^{}_{10}$ &
$C^{}_{7^\prime}$ & $C^{}_{9^\prime}$ & $C^{}_{10^\prime}$ &
$\Delta\chi^2_{\rm quad}$   \\ \hline
1$^+$ & 0.05 & -1.06 & 0.34 & 0.01 & 0.37 & -0.04 & 7.5 \\
 1$^-$ & -0.04 & -1.07 & 0.34 & 0.02 & 0.37 & -0.04 & 6.3 \\
 2$^+$ & 0.00 & -1.07 & 0.34 & -0.03 & 0.37 & -0.04 & 7.8 \\
 2$^-$ & 0.01 & -1.06 & 0.34 & 0.07 & 0.38 & -0.04 & 7.2 \\
 3$^+$ & 0.01 & -1.17 & 0.22 & 0.03 & 0.23 & 0.30 & 9.6 \\
 3$^-$ & 0.00 & -0.98 & 0.44 & 0.01 & 0.49 & -0.30 & 5.5 \\
 4$^+$ & 0.01 & -1.26 & 0.70 & 0.02 & 0.29 & -0.01 & 6.4 \\
 4$^-$ & 0.00 & -0.85 & -0.06 & 0.02 & 0.46 & -0.08 & 7.9 \\
 5$^+$ & 0.02 & -1.60 & 0.10 & 0.02 & 0.53 & -0.23 & 9.1 \\
 5$^-$ & -0.01 & -0.66 & 0.52 & 0.02 & 0.26 & 0.10 & 5.1 \\
 6$^+$ & 0.00 & -1.03 & 0.59 & 0.01 & 1.70 & 0.63 & 6.8 \\
 6$^-$ & 0.00 & -1.09 & 0.11 & 0.02 & -0.87 & -0.67 & 6.0 \\
\hline
\end{tabular}
\end{center}
\caption{Parameter sets obtained from the intersections of the SVD directions with the exact  $1\sigma$ surface (hence $\Delta\chi^2_{\rm exact}=7.1$).}
\label{t:fitsex}
\end{table}%

\begin{figure}[ht]
\center{\includegraphics[width=0.45\textwidth]{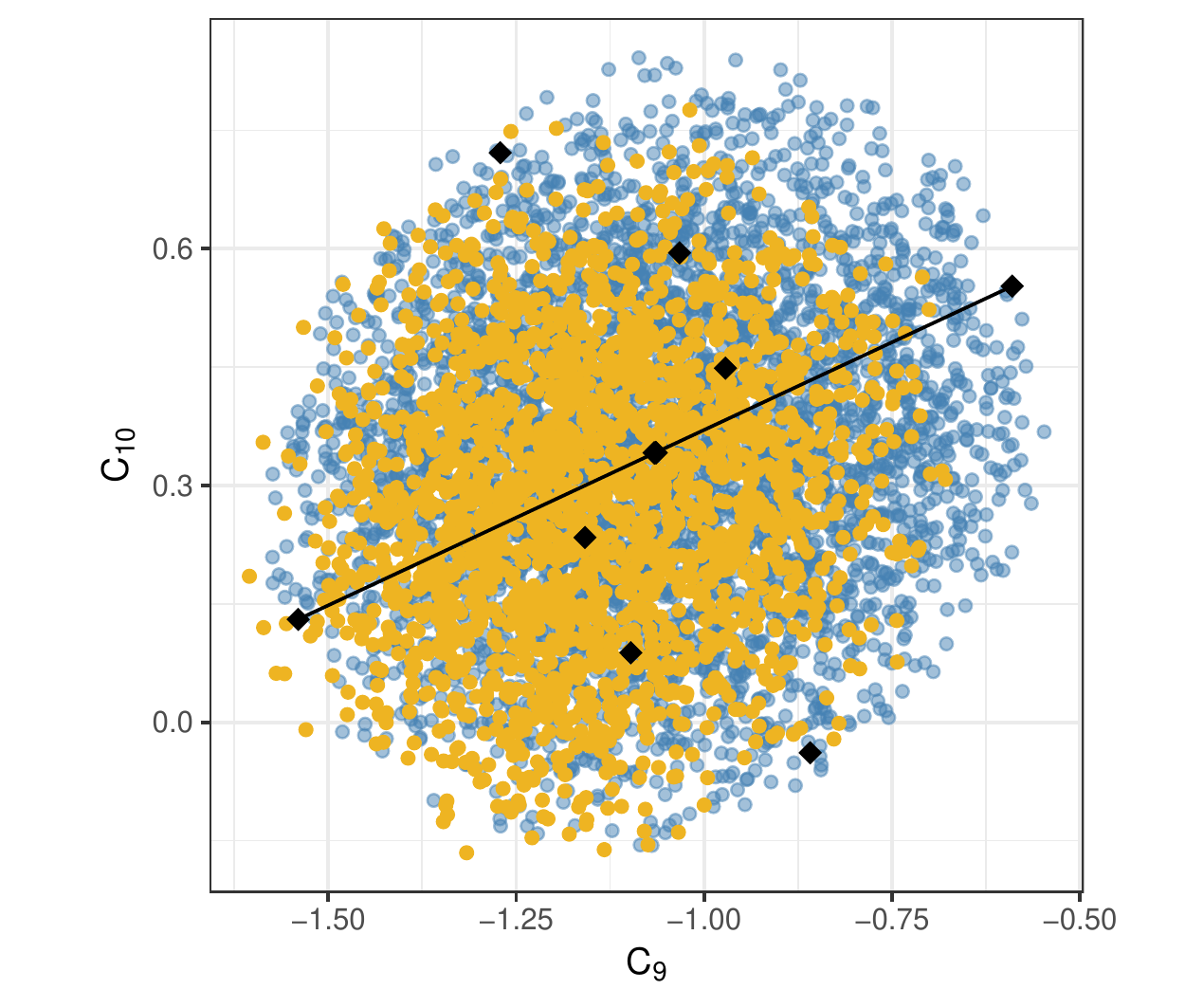}
\includegraphics[width=0.45\textwidth]{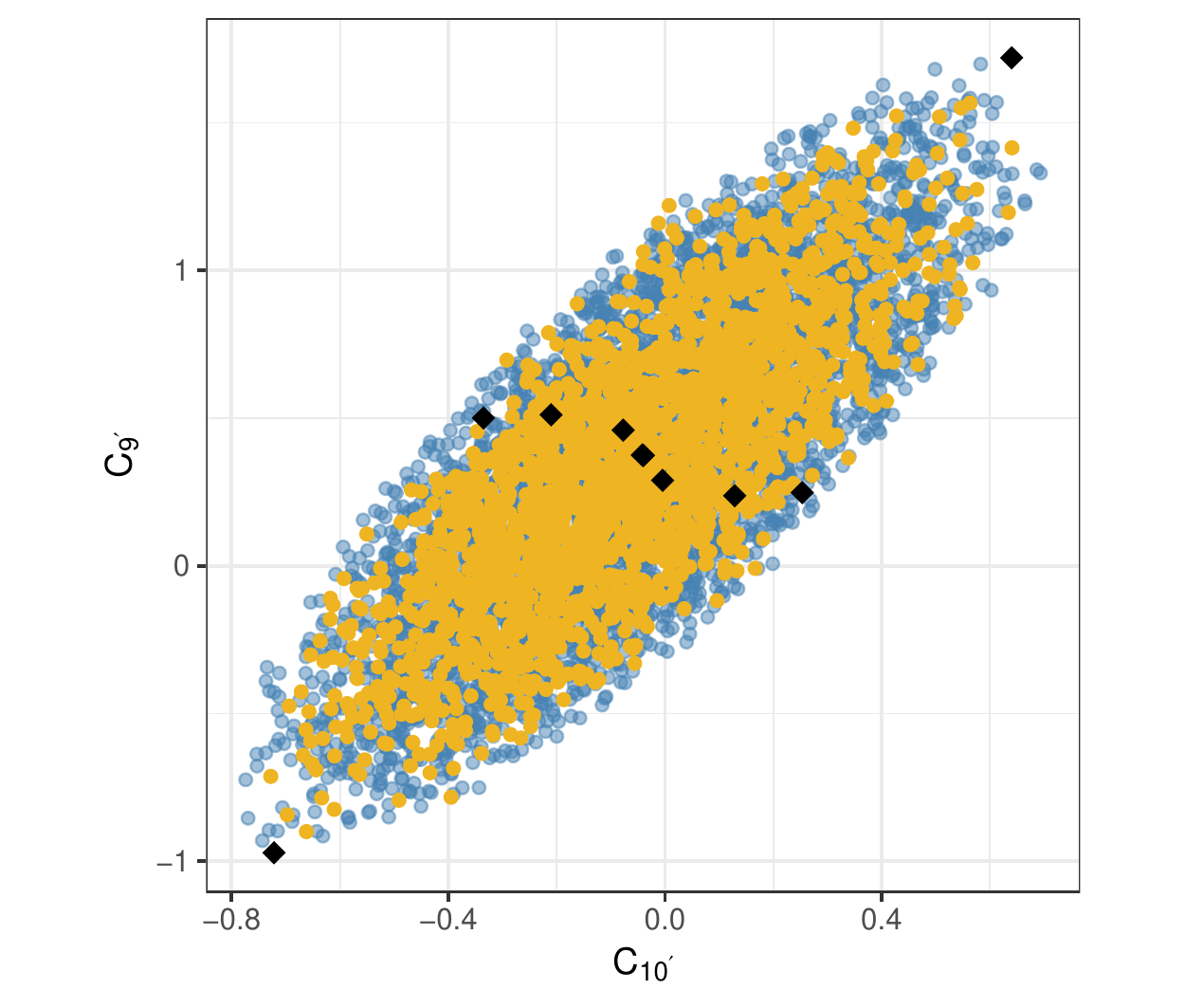}}
\caption{Sample points representing the quadratic approximation (blue) and the full fit (yellow), together with the 12 SVD points, projected onto $C_9 - C_{10}$ (left) and $C^{}_{9^\prime} - C^{}_{10^\prime}$ (right). The line in the left plot connects the SVD points in direction 5, illustrating how the asymmetric shape of the true $\chi^2$ distribution is not fully captured along this direction. The right plot highlights the importance of parameter correlations captured by the SVD points but not when varying one coefficient at a time.}
\label{fig:illSVD}
\end{figure}

Of course, the SVD  directions are only defined in terms of the Hessian matrix, and the shape of the full $1\sigma$ surface may differ significantly from the assumed hyper-ellipsoid. While in this case the true shape is adequately described by the approximation (see Figure~\ref{fig:exactvssq}), some  deviations occur, as illustrated in the left plot of Figure~\ref{fig:illSVD} which shows direction 5 in the $C_9 - C_{10}$ plane. In this case the asymmetry of the true $1\sigma$ contour results in large differences  (as quantified in Table~\ref{t:fits}) in $\Delta \chi^2$ between the quadratic approximation and the full fit. 
As summarised in Section~\ref{sec:limit}, this does not significantly alter our conclusions.

Figure~\ref{fig:illSVD} (right) shows the projection onto $C^{}_{9^\prime} - C^{}_{10^\prime}$ and illustrates the importance of the SVD directions.
The correlations between parameter directions in the fit encode how combinations of parameters are constrained by the measurements.
Important correlations are found between $C^{}_{9^\prime}$ and $C^{}_{10^\prime}$, and the shape is indeed captured by the SVD points shown in black.
On the other hand, selecting the intersection of WC directions with the $1\sigma$ surface will result in loss of information,
the selected points would not be representative of the surface envelope.\footnote{Note also that the selected basis in terms of Wilson coefficients is arbitrary, and different models will generate different patterns. While we do not consider it meaningful to map the directions to any high-scale model (they capture how parameter combinations are constrained by the data, while for modelling purposes the main interest is in the BF point) the directions are physically meaningful in how they enter predictions for the considered observables.}

\section{Samples of largest deltas}\label{s:larged}

The specific value that makes a given $\delta$ large, and thus interesting, is arbitrary. Here we compare the different definitions introduced in Eq.~\ref{deltasdef}.  The definition of $\delta$ and $\delta^\prime$ makes it reasonable to set the interest cutoff when they are equal to one, as this number corresponds to one standard deviation as measured by the total uncorrelated error. On the other hand, the definition $\delta_\sigma$  can be related to the calculation of $\Delta\chi^2$ as shown in the last column of Table~\ref{t:fits} with the quantity $\sum_i\delta_\sigma^2$. Since we are comparing points that lie within $\Delta\chi^2=7.1$ (in the quadratic or Hessian approximation), we define the cutoff for a `large' $\delta_\sigma$ in this case as $\sqrt{0.71}\sim 0.84$. This choice singles out those observables that by themselves contribute 10\% or more of the shift in $\chi^2$. Note that this approximation is not as good for directions $5^+,6^\pm$, where a better cut-off might be 1. The subjectivity of this cut-off is mitigated with the introduction of the section  with rankings in the text.
A list of $\delta$s above these cutoffs are presented in Table~\ref{deltasmix}.

\begin{table}[htp]
\begin{center}
\begin{tabular}{|c|c|c|c|c|}\hline
  {$\#$} &$ {|\delta|  > 1} $&$ |\delta ^\prime|{ >1} $&$ |\tilde\delta _{\sigma }|{ > 0.84} $&$ |\delta _{\sigma }|{ >
   0.84} $ \\  \hline $
 13 $ & $  \delta ^{4+}{  }\delta ^{4-}  $ & $  \delta ^{4+}{  }\delta ^{4-}  $ & $ {} $ & $ {} $  \\ $
 16 $ & $  \delta ^{2+}{  }\delta ^{2-}  $ & $  \delta ^{2+}{  }\delta ^{2-}  $ & $  \delta ^{2+}{  }\delta ^{2-}  $ & $  \delta ^{2+}{
   }\delta ^{2-}  $  \\ $
 20 $ & $ {} $ & $  \delta ^{6+}  $ & $ {} $ & $ {} $  \\ $
 49 $ & $ {} $ & $ {} $ & $  \delta ^{5+}{  }{\color{blue}\delta ^{5-}}  $ & $  \delta ^{5+}{  }\delta ^{5-}  $  \\ $
 {\color{red} 52} $ & $ {} $ & $ {} $ & $  {  }  $ & $  {\color{red} \delta ^{5+}} $  \\ $
 56 $ & $ {} $ & $ {} $ & $  \delta ^{5+}  $ & $ {} $  \\ $
 57 $ & $ {\color{red} \delta ^{3+}{  }\delta ^{5+}}{} $ & $ {\color{red} \delta ^{3+}{  }\delta ^{5+}} $ & $  \delta ^{5+}  $ & $  \delta ^{5+}  $  \\ $
 68 $ & $  \delta ^{3+}{  }\delta ^{4+}{  }\delta ^{4-}{  }\delta ^{6-}  $ & $  \delta ^{3+}{  }\delta ^{4+}{  }\delta ^{4-}{
    }\delta ^{6-}  $ & $ {} $ & $ {} $ \\ $
 74 $ & $ {} $ & $ {} $ & $  \delta ^{2+}{  }\delta ^{2-}  $ & $  \delta ^{2+}{  }\delta ^{2-}  $  \\ $
 93 $ & $   {\color{red} \delta ^{3+}}\delta ^{4-}{  }\delta ^{6-}  $ & $  \delta ^{4-}{  }\delta ^{6-}  $ & $ {\color{red} \delta ^{3+}}\ {} $ & $ {\color{blue}\delta ^{6-} }$  \\ $
 95 $ & $  \delta ^{2+}{  }\delta ^{2-}  $ & $  \delta ^{2+}{  }\delta ^{2-}  $ & $  \delta ^{2+}  $ & $  \delta ^{2+}{  }\delta ^{2-}  $  \\ $
 98 $ & $ {\color{red} \delta ^{3+}}\ \delta ^{4+}{  }\delta ^{4-}{  }\delta ^{6+}  $ & $  {\color{red} \delta ^{3+}}\ \delta ^{4+}{  }\delta ^{4-}{  }\delta ^{6+}  $ & $  \delta ^{3+}{
   }{\color{blue}\delta ^{3-}}{  }\delta ^{4+}{  }\delta ^{4-}{  }\delta ^{6+}  $ & $  \delta ^{3+}{  }{\color{blue}\delta ^{3-}}{  }\delta ^{4+}{
   }\delta ^{4-}{  }\delta ^{6+}  $  \\ $
 100 $ & $ {} $ & $ {} $ & $   {\color{red} \delta ^{3+}} \delta ^{4-}{  }\delta ^{6-}  $ & $  {\color{red} \delta ^{3+}} \delta ^{4-}{  }\delta ^{6-}  $  \\ $
 {\color{red}114} $ & $ {} $ & $ {} $ & $  $ & $  {\color{red} \delta ^{2+}} $  \\ $
 155 $ & $ {\color{red} \delta ^{3+}}  {\color{blue} \delta ^{4+}}{  }\delta ^{4-}{  }\delta ^{6-}  $ & $  {\color{red} \delta ^{3+}}  \delta ^{4+}{  }\delta ^{4-}{  }\delta ^{6-}  $ & $ {} $ & $ {} $  \\ $
 171 $ & $  \delta ^{1+}{  }\delta ^{1-}  $ & $  \delta ^{1+}{  }\delta ^{1-}  $ & $  \delta ^{1+}{  }\delta ^{1-}  $ & $  \delta ^{1+}{
   }\delta ^{1-}  $  \\ $
 172 $ &$ {\color{red} \delta ^{3+}}   {} $ & $  {\color{red} \delta ^{3+}}  {} $ & $  \delta ^{3+}{  }\delta ^{6+}{  }{\color{blue}\delta ^{6-} } $ & $  \delta ^{3+}{  }\delta ^{6+}{  }{\color{blue}\delta ^{6-} } $  \\ $
 173 $ & $ {} $ & $ {} $ & $  \delta ^{2+}{  }\delta ^{2-}  $ & $  \delta ^{2+}{  }\delta ^{2-}  $  \\
\hline \end{tabular}
\end{center}
\caption{Specific $\delta$s (using all four definitions) that are larger than 1 (uncorrelated errors), or 0.84 (correlated errors). The figures in blue appear only with the SVD points calculated in the quadratic approximation for $\chi^2$ whereas the figures in red appear only when the exact $\chi^2$ is used instead.}
\label{deltasmix}
\end{table}%

The columns for $\delta$ and $\delta^\prime$ are nearly identical because the estimate of  theoretical errors varies little between the BF and the twelve SVD points. The sole exception shown in the Table for $\delta^{6+\prime}>1$ occurs for observable 20, for which $\delta^{6+}=0.57$. The difference arises from the estimate for the theoretical error, at the BF it is just over 60\% larger than the corresponding estimate at the point $6+$. Our numerical calculations imply that with the current level of precision in the parametrization of  hadronic uncertainties (mainly form factors) the theoretical error estimates have a non-trivial dependence on the NP contributions to the Wilson coefficients. These errors are computed via a multivariate gaussian scan over all the nuisance parameters, i.e. form factors, decay constants, CKM matrix elements, etc. As explained in~\cite{Descotes-Genon:2015uva}, a rescaling of the errors on the nuisance parameters is needed to ensure gaussian behavior. The rescaling factor is most relevant for some observables such as the $Br(B\to K^*\mu\mu)$ computed for $q^2$ bins exceeding 8 $\text{GeV}^2$ (ID 164), while it has almost no impact on observables that are  less sensitive to form factors. We have checked explicitly that once the factor ensuring gaussian behavior is found,  the results are independent of the exact numerical choice for this rescaling.
Finally, the reasonable agreement between the columns $\delta_\sigma$ and $\tilde \delta_\sigma$ confirms that the latter is a fair approximation to the former.

\section{Quadratic differences between SM and BF Pull}

\begin{figure}[!h]
\center\includegraphics[width=0.5\textwidth]{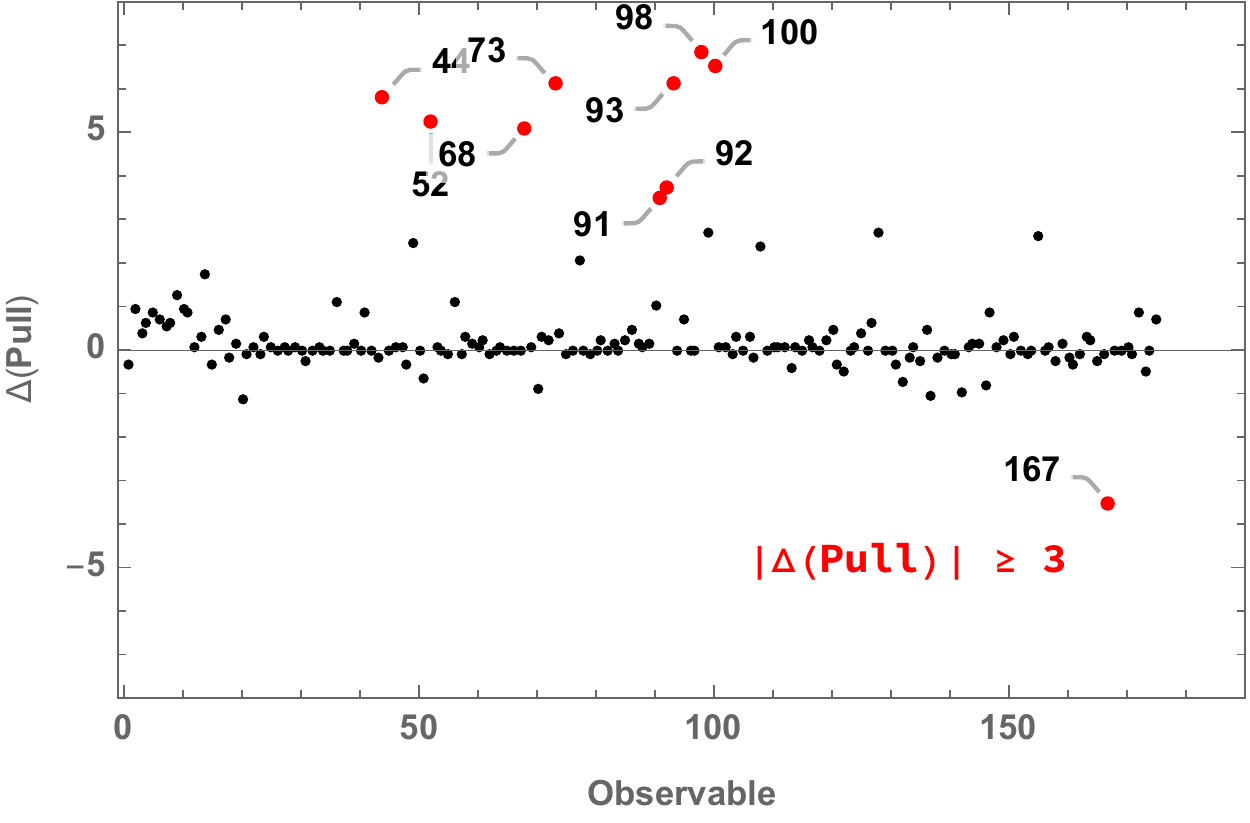}
\caption{$\Delta_2({\rm Pull})$ for all observables.}
\label{fig:pull2}
\end{figure}

In Section~\ref{s:quantiresults} we use the linear $\Delta (Pull)$ to compare the SM and BF point Pull for each observable. As noted, we may consider a quadratic definition
\begin{equation}
\Delta_2({\rm Pull})_i = {\rm Pull}({\rm SM})_i^2 - {\rm Pull}({\rm BF})_i^2
\end{equation}
which more accurately captures how each observable contributes differently to the $\chi^2$ function (in the absence of correlated uncertainties).
Figure~\ref{fig:pull2} shows $\Delta_2({\rm Pull})$ for each observable.
Comparing to the corresponding results in Figure~\ref{fig:dpull} we note that the qualitative picture remains the same, but larger differences are emphasised when considering $\Delta_2({\rm Pull})$. Notably small systematic effects seen in Figure~\ref{fig:dpull} are no longer visible in Figure~\ref{fig:pull2}. In addition, observable 14 no longer stands out when considering $\Delta_2({\rm Pull})$, which is because the Pull is close to one for the SM prediction, and close to zero for the BF prediction. In this case the details of the definition become important.

\section{Interpretation of measures defined in terms of the covariance matrix}
\label{s:dSigma}

The definition of e.g. $\Delta_{\sigma}$(Pull) and similar measures is not immediately intuitive. Here we present the explicit expression in a two dimensional scenario to aid with the interpretation.

In general the variance-covariance matrix can be decomposed as as
\begin{equation}
\sigma_{ij} = \Sigma_{ik} \rho_{kl} \Sigma_{lj}
\end{equation}
where $\Sigma_{ik} = diag(\sigma_1, \sigma_2, ..., \sigma_N)$ and $\rho$ the correlation matrix, i.e. with entries $\rho_{ij}$ equal to $1$ if $i=j$ and $\rho_{ij} = \rho_{ji} \in [-1, 1]$.
The inverse of the variance-covariance matrix is therefore given as
\begin{equation}
\sigma^{-1} = \Sigma^{-1} \rho^{-1} \Sigma^{-1}
\end{equation}
where $\Sigma^{-1} = diag(1/\sigma_1, 1/\sigma_2, ..., 1/\sigma_N)$ and in general the inverse correlation matrix is given via the adjugate matrix as
\begin{equation}
\rho^{-1} = \frac{1}{\det{\rho}} Adj(\rho).
\label{eq:invCorr}
\end{equation}
In two dimensions the correlation matrix and its inverse can be written as
\begin{equation}
\rho =
\begin{pmatrix}
1 & \varrho \\
\varrho & 1
\end{pmatrix},
\rho^{-1} = \frac{1}{1-\varrho^2}
\begin{pmatrix}
1 & -\varrho\\
-\varrho & 1
\end{pmatrix}
\end{equation}

Inverting the correlation matrix introduced a relative minus sign for the off-diagonal terms entering the $\chi^2$ function, 
which now takes the form
\begin{equation}
\chi^2 = \frac{1}{\sigma_1^2} \frac{1}{1-\varrho^2} D_1^2 +
\frac{1}{\sigma_2^2} \frac{1}{1-\varrho^2} D_2^2 -
2 \frac{\varrho}{\sigma_1 \sigma_2} \frac{1}{1-\varrho^2} D_1 D_2,
\label{eq:chi2_2d}
\end{equation}
and we have introduced the shorthand notation $D_i = T_i - O_i$ for the difference between theory prediction and observed value for observable $i$.
We can recover the uncorrelated definition by setting $\varrho=0$. Alternatively, in the presence of correlations, the overall factor $\frac{1}{1-\varrho^2}$ captures the additional information content available in that case. We see that the third term in Eq.~\ref{eq:chi2_2d} will depend on the signs of the $D_i$ relative to $\rho$. If both theory predictions  for the considered parameter point differ from the observed value in the same direction, i.e.\ the $D_i$ are either both positive or both negative, the sign of the third term will be determined by that of the correlation coefficient $\varrho$, leading to two possible scenarios:
\begin{itemize}
\item $\varrho > 0$:
This indicates that the patterns in $D_i$ are consistent with the correlation obtained by varying the nuisance parameters, 
as a result the overall value of the $\chi^2$ function is reduced.
\item $\varrho < 0$:
Negative correlation is not consistent with our assumptions about the $D_i$, this situation will therefore result in an overall
increase in the value of the $\chi^2$ function.
\end{itemize}
Analogous considerations hold if the $D_i$ have opposite sign, where $\varrho < 0$ will result in reduced values of $\chi^2$ and $\varrho > 0$ will increase the value.

We now want to capture this information on observable-by-observable level, e.g.\ in our definition of the Pull. To avoid working with the square root of the matrix let us consider the following definition for the Pull of observable $i$ (corresponding to using $\tilde\delta_\sigma$):
\begin{equation}
\mathrm{Pull}_i = \sum_j \frac{1}{\sqrt{\sigma^{-1}_{ii}}} \sigma^{-1}_{ij} D_j.
\end{equation}
For our 2-d example, for the first observable we obtain
\begin{equation}
\mathrm{Pull}_1 = \frac{1}{\sqrt{1-\varrho^2}} (\frac{D_1}{\sigma_1} - \frac{\varrho D_2}{\sigma_2})
\end{equation}

Indeed this results in the same behaviour observed for the $\chi^2$ function, i.e. for $\varrho=0$ we recover the uncorrelated definition, elsewise the total absolute value of the Pull will be increased/decreased if the pattern in the $D_i$ is inconsistent/consistent with that of the correlation matrix $\rho$, with the additional contribution being weighted according to the correlation obtained via the nuisance parameters.

Notice that while the factor used in the definition of the Pull is somewhat arbitrary, here we recover the $\chi^2$ function as
\begin{equation}
\chi^2 = \sum_i \frac{D_i}{\tilde\sigma_i} \mathrm{Pull}_i
\end{equation}
i.e. in analogy to our definition without correlation, where $\tilde\sigma_i = 1/\sqrt{\sigma_{ii}^{-1}}$ the conditional variance. We recall that numerically the results of $\delta_{\sigma}$ defined in terms of the square root of the covariance matrix, and those found for $\tilde{\delta}_{\sigma}$ are similar, though the exact expression is more complicated already for the two dimensional scenario.

Finally we note that when going beyond two dimensions additional effects need to be taken into account, such as simultaneous correlation between groups of observables. While this yields somewhat more complicated expressions (obtained via the inversion of higher dimensional matrices), the overall picture is similar to that shown by the two dimensional example.

\section{Experimental and theoretical correlation matrices}\label{s:2maps}

\begin{figure}[!h]
\includegraphics[width=\textwidth]{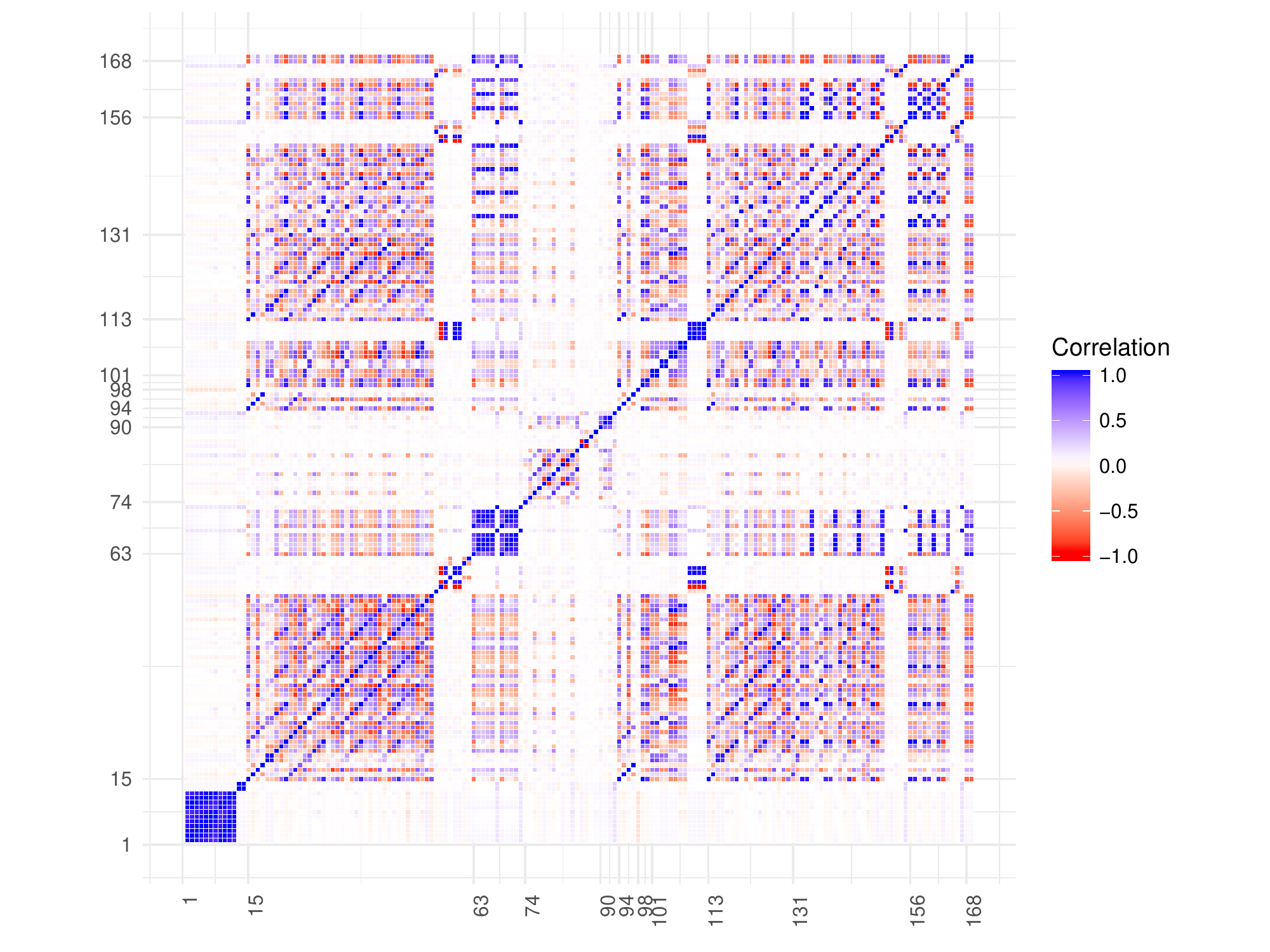}\\
\includegraphics[width=\textwidth]{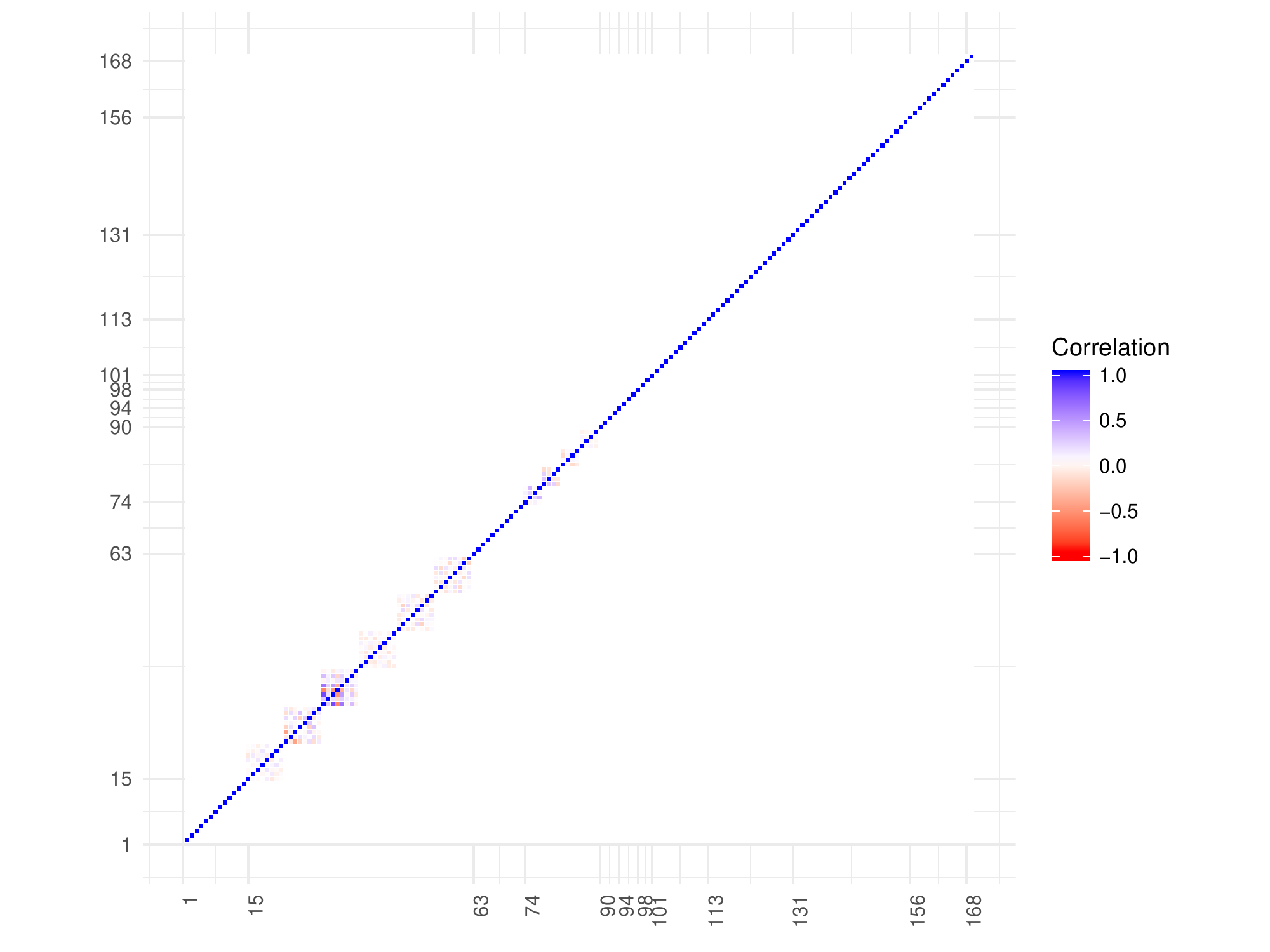}
\caption{Correlation maps derived from the covariance matrix of theory errors (top panel) and experimental errors (bottom panel). }
\label{fig:correlationMap-th-ex}
\end{figure}

We reproduce here the information in Section~\ref{corrsec} but showing separately the theoretical and experimental correlations between observables, see Figure~\ref{fig:correlationMap-th-ex}.

\section{Observables used in the fit}\label{s:idmatch}

In Tables~\ref{t:obs1},~\ref{t:obsc} we list the 175 observables included in the fit along with their corresponding ID used throughout our paper. The additional observables that have been proposed in the literature are listed in Table~\ref{tab:qObs}.

\begin{table}[htp]
 \resizebox{.45\textwidth}{!}{
\begin{tabular}{|c|c|c|c|c|c|}  \hline
ID & Observable & Exp.  \\ \hline
1 & $10^7\times Br(B^+  \to K^+ \mu\mu) [0.1-0.98] $ & LHCb \cite{Aaij:2014pli}  \\
2 & $10^7\times Br(B^+  \to K^+ \mu\mu) [1.1-2] $ & LHCb \cite{Aaij:2014pli} \\
3 & $10^7\times Br(B^+  \to K^+ \mu\mu) [2-3] $ & LHCb \cite{Aaij:2014pli} \\
4 & $10^7\times Br(B^+  \to K^+ \mu\mu) [3-4] $ & LHCb \cite{Aaij:2014pli} \\
5 & $10^7\times Br(B^+  \to K^+ \mu\mu) [4-5] $ & LHCb \cite{Aaij:2014pli} \\
6 & $10^7\times Br(B^+  \to K^+ \mu\mu) [5-6] $ & LHCb \cite{Aaij:2014pli} \\
7 & $10^7\times Br(B^+  \to K^+ \mu\mu) [6-7] $ & LHCb \cite{Aaij:2014pli} \\
8 & $10^7\times Br(B^+  \to K^+ \mu\mu) [7-8] $ & LHCb \cite{Aaij:2014pli} \\
9 & $10^7\times Br(B^0  \to K^0 \mu\mu) [0.1-2] $ & LHCb \cite{Aaij:2014pli} \\
10 & $10^7\times Br(B^0  \to K^0 \mu\mu) [2-4] $ & LHCb \cite{Aaij:2014pli} \\
11 & $10^7\times Br(B^0  \to K^0 \mu\mu) [4-6] $ & LHCb \cite{Aaij:2014pli} \\
12 & $10^7\times Br(B^0  \to K^0 \mu\mu) [6-8] $ & LHCb \cite{Aaij:2014pli} \\
13 & $10^7\times Br(B^+  \to K^+ \mu\mu) [15-22] $ & LHCb \cite{Aaij:2014pli} \\
14 & $10^7\times Br(B^0  \to K^0 \mu\mu) [15-22] $ & LHCb \cite{Aaij:2014pli} \\
15 & $F_L(B \to K^* \mu\mu) [0.1-0.98] $ & LHCb \cite{Aaij:2015oid} \\
16 & $P_1(B \to K^* \mu\mu) [0.1-0.98] $ & LHCb \cite{Aaij:2015oid} \\
17 & $P_2(B \to K^* \mu\mu) [0.1-0.98] $ & LHCb \cite{Aaij:2015oid} \\
18 & $P_3(B \to K^* \mu\mu) [0.1-0.98] $ & LHCb \cite{Aaij:2015oid} \\
19 & $P_4^\prime(B \to K^* \mu\mu) [0.1-0.98] $ & LHCb \cite{Aaij:2015oid} \\
20 & $P_5^\prime(B \to K^* \mu\mu) [0.1-0.98] $ & LHCb \cite{Aaij:2015oid} \\
21 & $P_6^\prime(B \to K^* \mu\mu) [0.1-0.98] $ & LHCb \cite{Aaij:2015oid} \\
22 & $P_8^\prime(B \to K^* \mu\mu) [0.1-0.98] $ & LHCb \cite{Aaij:2015oid} \\
23 & $F_L(B \to K^* \mu\mu) [1.1-2.5] $ & LHCb \cite{Aaij:2015oid} \\
24 & $P_1(B \to K^* \mu\mu) [1.1-2.5] $ & LHCb \cite{Aaij:2015oid} \\
25 & $P_2(B \to K^* \mu\mu) [1.1-2.5] $ & LHCb \cite{Aaij:2015oid} \\
26 & $P_3(B \to K^* \mu\mu) [1.1-2.5] $ & LHCb \cite{Aaij:2015oid} \\
27 & $P_4^\prime(B \to K^* \mu\mu) [1.1-2.5] $ & LHCb \cite{Aaij:2015oid} \\
28 & $P_5^\prime(B \to K^* \mu\mu) [1.1-2.5] $ & LHCb \cite{Aaij:2015oid} \\
29 & $P_6^\prime(B \to K^* \mu\mu) [1.1-2.5] $ & LHCb \cite{Aaij:2015oid} \\
30 & $P_8^\prime(B \to K^* \mu\mu) [1.1-2.5] $ & LHCb \cite{Aaij:2015oid} \\
31 & $F_L(B \to K^* \mu\mu) [2.5-4] $ & LHCb \cite{Aaij:2015oid} \\
32 & $P_1(B \to K^* \mu\mu) [2.5-4] $ & LHCb \cite{Aaij:2015oid} \\
33 & $P_2(B \to K^* \mu\mu) [2.5-4] $ & LHCb \cite{Aaij:2015oid} \\
34 & $P_3(B \to K^* \mu\mu) [2.5-4] $ & LHCb \cite{Aaij:2015oid} \\
35 & $P_4^\prime(B \to K^* \mu\mu) [2.5-4] $ & LHCb \cite{Aaij:2015oid} \\
36 & $P_5^\prime(B \to K^* \mu\mu) [2.5-4] $ & LHCb \cite{Aaij:2015oid} \\
37 & $P_6^\prime(B \to K^* \mu\mu) [2.5-4] $ & LHCb \cite{Aaij:2015oid} \\
38 & $P_8^\prime(B \to K^* \mu\mu) [2.5-4] $ & LHCb \cite{Aaij:2015oid} \\
39 & $F_L(B \to K^* \mu\mu) [4-6] $ & LHCb \cite{Aaij:2015oid} \\
40 & $P_1(B \to K^* \mu\mu) [4-6] $ & LHCb \cite{Aaij:2015oid} \\  \hline
\end{tabular} }
 \resizebox{.45\textwidth}{!}{
\begin{tabular}{|c|c|c|c|c|c|}  \hline
ID & Observable & Exp  \\ \hline
41 & $P_2(B \to K^* \mu\mu) [4-6] $ & LHCb \cite{Aaij:2015oid} \\
42 & $P_3(B \to K^* \mu\mu) [4-6] $ & LHCb \cite{Aaij:2015oid} \\
43 & $P_4^\prime(B \to K^* \mu\mu) [4-6] $ & LHCb \cite{Aaij:2015oid} \\
44 & $P_5^\prime(B \to K^* \mu\mu) [4-6] $ & LHCb \cite{Aaij:2015oid} \\
45 & $P_6^\prime(B \to K^* \mu\mu) [4-6] $ & LHCb \cite{Aaij:2015oid} \\
46 & $P_8^\prime(B \to K^* \mu\mu) [4-6] $ & LHCb \cite{Aaij:2015oid} \\
47 & $F_L(B \to K^* \mu\mu) [6-8] $ & LHCb \cite{Aaij:2015oid} \\
48 & $P_1(B \to K^* \mu\mu) [6-8] $ & LHCb \cite{Aaij:2015oid} \\
49 & $P_2(B \to K^* \mu\mu) [6-8] $ & LHCb \cite{Aaij:2015oid} \\
50 & $P_3(B \to K^* \mu\mu) [6-8] $ & LHCb \cite{Aaij:2015oid} \\
51 & $P_4^\prime(B \to K^* \mu\mu) [6-8] $ & LHCb \cite{Aaij:2015oid} \\
52 & $P_5^\prime(B \to K^* \mu\mu) [6-8] $ & LHCb \cite{Aaij:2015oid} \\
53 & $P_6^\prime(B \to K^* \mu\mu) [6-8] $ & LHCb \cite{Aaij:2015oid} \\
54 & $P_8^\prime(B \to K^* \mu\mu) [6-8] $ & LHCb \cite{Aaij:2015oid} \\
55 & $F_L(B \to K^* \mu\mu) [15-19] $ & LHCb \cite{Aaij:2015oid} \\
56 & $P_1(B \to K^* \mu\mu) [15-19] $ & LHCb \cite{Aaij:2015oid} \\
57 & $P_2(B \to K^* \mu\mu) [15-19] $ & LHCb \cite{Aaij:2015oid} \\
58 & $P_3(B \to K^* \mu\mu) [15-19] $ & LHCb \cite{Aaij:2015oid} \\
59 & $P_4^\prime(B \to K^* \mu\mu) [15-19] $ & LHCb \cite{Aaij:2015oid} \\
60 & $P_5^\prime(B \to K^* \mu\mu) [15-19] $ & LHCb \cite{Aaij:2015oid} \\
61 & $P_6^\prime(B \to K^* \mu\mu) [15-19] $ & LHCb \cite{Aaij:2015oid} \\
62 & $P_8^\prime(B \to K^* \mu\mu) [15-19] $ & LHCb \cite{Aaij:2015oid} \\
63 & $10^7\times Br(B^0  \to K^{0 *}\mu\mu) [0.1-0.98] $ & LHCb \cite{Aaij:2016flj} \\
64 & $10^7\times Br(B^0  \to K^{0 *}\mu\mu) [1.1-2.5] $ & LHCb \cite{Aaij:2016flj} \\
65 & $10^7\times Br(B^0  \to K^{0 *}\mu\mu) [2.5-4] $ & LHCb \cite{Aaij:2016flj} \\
66 & $10^7\times Br(B^0  \to K^{0 *}\mu\mu) [4-6] $ & LHCb \cite{Aaij:2016flj} \\
67 & $10^7\times Br(B^0  \to K^{0 *}\mu\mu) [6-8] $ & LHCb \cite{Aaij:2016flj} \\
68 & $10^7\times Br(B^0  \to K^{0 *}\mu\mu) [15-19] $ & LHCb \cite{Aaij:2016flj} \\
69 & $10^7\times Br(B^0  \to K^{+ *}\mu\mu) [0.1-2] $ & LHCb \cite{Aaij:2014pli} \\
70 & $10^7\times Br(B^0  \to K^{+ *}\mu\mu) [2-4] $ & LHCb \cite{Aaij:2014pli} \\
71 & $10^7\times Br(B^0  \to K^{+ *}\mu\mu) [4-6] $ & LHCb \cite{Aaij:2014pli} \\
72 & $10^7\times Br(B^0  \to K^{+ *}\mu\mu) [6-8] $ & LHCb \cite{Aaij:2014pli} \\
73 & $10^7\times Br(B^0  \to K^{+ *}\mu\mu) [15-19] $ & LHCb \cite{Aaij:2014pli} \\
74 & $P_1(B_s \to \Phi\mu\mu) [0.1-2] $ & LHCb \cite{Aaij:2015esa} \\
75 & $P_4^\prime(B_s \to \Phi\mu\mu) [0.1-2] $ & LHCb \cite{Aaij:2015esa} \\
76 & $P_6^\prime(B_s \to \Phi\mu\mu) [0.1-2] $ & LHCb \cite{Aaij:2015esa} \\
77 & $F_L(B_s \to \Phi\mu\mu) [0.1-2] $ & LHCb \cite{Aaij:2015esa} \\
78 & $P_1(B_s \to \Phi\mu\mu) [2-5] $ & LHCb \cite{Aaij:2015esa} \\
79 & $P_4^\prime(B_s \to \Phi\mu\mu) [2-5] $ & LHCb \cite{Aaij:2015esa} \\
80 & $P_6^\prime(B_s \to \Phi\mu\mu) [2-5] $ & LHCb \cite{Aaij:2015esa} \\  \hline
\end{tabular} }
\caption{List of observables used in the fit.}
\label{t:obs1}
\end{table}%
\begin{table}[htp]
 \resizebox{.45\textwidth}{!}{
\begin{tabular}{|c|c|c|c|c|c|}  \hline
ID & Observable & Exp  \\ \hline
81 & $F_L(B_s \to \Phi\mu\mu) [2-5] $ & LHCb \cite{Aaij:2015esa} \\
82 & $P_1(B_s \to \Phi\mu\mu) [5-8] $ & LHCb \cite{Aaij:2015esa} \\
83 & $P_4^\prime(B_s \to \Phi\mu\mu) [5-8] $ & LHCb \cite{Aaij:2015esa} \\
84 & $P_6^\prime(B_s \to \Phi\mu\mu) [5-8] $ & LHCb \cite{Aaij:2015esa} \\
85 & $F_L(B_s \to \Phi\mu\mu) [5-8] $ & LHCb \cite{Aaij:2015esa} \\
86 & $P_1(B_s \to \Phi\mu\mu) [15-18.8] $ & LHCb \cite{Aaij:2015esa} \\
87 & $P_4^\prime(B_s \to \Phi\mu\mu) [15-18.8] $ & LHCb \cite{Aaij:2015esa} \\
88 & $P_6^\prime(B_s \to \Phi\mu\mu) [15-18.8] $ & LHCb \cite{Aaij:2015esa} \\
89 & $F_L(B_s \to \Phi\mu\mu) [15-18.8] $ & LHCb \cite{Aaij:2015esa} \\
90 & $10^7\times Br(B_s \to \Phi\mu\mu) [0.1-2] $ & LHCb \cite{Aaij:2015esa} \\
91 & $10^7\times Br(B_s \to \Phi\mu\mu) [2-5] $ & LHCb \cite{Aaij:2015esa} \\
92 & $10^7\times Br(B_s \to \Phi\mu\mu) [5-8] $ & LHCb \cite{Aaij:2015esa} \\
93 & $10^7\times Br(B_s \to \Phi\mu\mu) [15-18.8] $ & LHCb \cite{Aaij:2015esa} \\
94 & $F_L(B \to K^* ee) [0.0020-1.120] $ & LHCb \cite{Aaij:2015dea} \\
95 & $P_1(B \to K^* ee) [0.0020-1.120] $ & LHCb \cite{Aaij:2015dea} \\
96 & $P_2(B \to K^* ee) [0.0020-1.120] $ & LHCb \cite{Aaij:2015dea} \\
97 & $P_3(B \to K^* ee) [0.0020-1.120] $ & LHCb \cite{Aaij:2015dea} \\
98 & $R_K(B^+  \to K^+ ) [1-6] $ & LHCb \cite{Aaij:2014ora} \\
99 & $R_{K^*} (B^0  \to K^{0 *}) [0.045-1.1] $ & LHCb \cite{Aaij:2017vbb} \\
100 & $R_{K^*} (B^0  \to K^{0 *}) [1.1-6] $ & LHCb \cite{Aaij:2017vbb} \\
101 & $P_4^\prime(B \to K^* ee) [0.1-4] $ & Belle \cite{Wehle:2016yoi} \\
102 & $P_4^\prime(B \to K^* \mu\mu) [0.1-4] $ & Belle \cite{Wehle:2016yoi} \\
103 & $P_5^\prime(B \to K^* ee) [0.1-4] $ & Belle \cite{Wehle:2016yoi} \\
104 & $P_5^\prime(B \to K^* \mu\mu) [0.1-4] $ & Belle \cite{Wehle:2016yoi} \\
105 & $P_4^\prime(B \to K^* ee) [4-8] $ & Belle \cite{Wehle:2016yoi} \\
106 & $P_4^\prime(B \to K^* \mu\mu) [4-8] $ & Belle \cite{Wehle:2016yoi} \\
107 & $P_5^\prime(B \to K^* ee) [4-8] $ & Belle \cite{Wehle:2016yoi} \\
108 & $P_5^\prime(B \to K^* \mu\mu) [4-8] $ & Belle \cite{Wehle:2016yoi} \\
109 & $P_4^\prime(B \to K^* ee) [14.18-19] $ & Belle \cite{Wehle:2016yoi} \\
110 & $P_4^\prime(B \to K^* \mu\mu) [14.18-19] $ & Belle \cite{Wehle:2016yoi} \\
111 & $P_5^\prime(B \to K^* ee) [14.18-19] $ & Belle \cite{Wehle:2016yoi} \\
112 & $P_5^\prime(B \to K^* \mu\mu) [14.18-19] $ & Belle \cite{Wehle:2016yoi} \\
113 & $F_L(B \to K^* \mu\mu) [0.04-2] $ & ATLAS \cite{ATLAS:2017dlm} \\
114 & $P_1(B \to K^* \mu\mu) [0.04-2] $ & ATLAS \cite{ATLAS:2017dlm} \\
115 & $P_4^\prime(B \to K^* \mu\mu) [0.04-2] $ & ATLAS \cite{ATLAS:2017dlm} \\
116 & $P_5^\prime(B \to K^* \mu\mu) [0.04-2] $ & ATLAS \cite{ATLAS:2017dlm} \\
117 & $P_6^\prime(B \to K^* \mu\mu) [0.04-2] $ & ATLAS \cite{ATLAS:2017dlm} \\
118 & $P_8^\prime(B \to K^* \mu\mu) [0.04-2] $ & ATLAS \cite{ATLAS:2017dlm} \\
119 & $F_L(B \to K^* \mu\mu) [2-4] $ & ATLAS \cite{ATLAS:2017dlm} \\
120 & $P_1(B \to K^* \mu\mu) [2-4] $ & ATLAS \cite{ATLAS:2017dlm} \\
121 & $P_4^\prime(B \to K^* \mu\mu) [2-4] $ & ATLAS \cite{ATLAS:2017dlm} \\
122 & $P_5^\prime(B \to K^* \mu\mu) [2-4] $ & ATLAS \cite{ATLAS:2017dlm} \\
123 & $P_6^\prime(B \to K^* \mu\mu) [2-4] $ & ATLAS \cite{ATLAS:2017dlm} \\
124 & $P_8^\prime(B \to K^* \mu\mu) [2-4] $ & ATLAS \cite{ATLAS:2017dlm} \\
125 & $F_L(B \to K^* \mu\mu) [4-6] $ & ATLAS \cite{ATLAS:2017dlm} \\
126 & $P_1(B \to K^* \mu\mu) [4-6] $ & ATLAS \cite{ATLAS:2017dlm} \\
127 & $P_4^\prime(B \to K^* \mu\mu) [4-6] $ & ATLAS \cite{ATLAS:2017dlm} \\
\hline
\end{tabular}}
 \resizebox{.45\textwidth}{!}{
\begin{tabular}{|c|c|c|c|c|c|}  \hline
ID & Observable & Exp  \\ \hline
128 & $P_5^\prime(B \to K^* \mu\mu) [4-6] $ & ATLAS \cite{ATLAS:2017dlm} \\
129 & $P_6^\prime(B \to K^* \mu\mu) [4-6] $ & ATLAS \cite{ATLAS:2017dlm} \\
130 & $P_8^\prime(B \to K^* \mu\mu) [4-6] $ & ATLAS \cite{ATLAS:2017dlm} \\
131 & $P_1(B \to K^* \mu\mu) [1-2] $ & CMS8 \cite{CMS:2017ivg} \\
132 & $P_5^\prime(B \to K^* \mu\mu) [1-2] $ & CMS8 \cite{CMS:2017ivg} \\
133 & $F_L(B \to K^* \mu\mu) [1-2] $ & CMS8 \cite{Khachatryan:2015isa} \\
134 & $A_{FB}(B \to K^* \mu\mu) [1-2] $ & CMS8 \cite{Khachatryan:2015isa} \\
135 & $10^7\times Br(B \to K^* \mu\mu) [1-2] $ & CMS8 \cite{Khachatryan:2015isa} \\
136 & $P_1(B \to K^* \mu\mu) [2-4.3] $ & CMS8 \cite{CMS:2017ivg} \\
137 & $P_5^\prime(B \to K^* \mu\mu) [2-4.3] $ & CMS8 \cite{CMS:2017ivg} \\
138 & $F_L(B \to K^* \mu\mu) [2-4.3] $ & CMS8 \cite{Khachatryan:2015isa} \\
139 & $A_{FB}(B \to K^* \mu\mu) [2-4.3] $ & CMS8 \cite{Khachatryan:2015isa} \\
140 & $10^7\times Br(B \to K^* \mu\mu) [2-4.3] $ & CMS8 \cite{Khachatryan:2015isa} \\
141 & $P_1(B \to K^* \mu\mu) [4.3-6] $ & CMS8 \cite{CMS:2017ivg} \\
142 & $P_5^\prime(B \to K^* \mu\mu) [4.3-6] $ & CMS8 \cite{CMS:2017ivg} \\
143 & $F_L(B \to K^* \mu\mu) [4.3-6] $ & CMS8 \cite{Khachatryan:2015isa} \\
144 & $A_{FB}(B \to K^* \mu\mu) [4.3-6] $ & CMS8 \cite{Khachatryan:2015isa} \\
145 & $10^7\times Br(B \to K^* \mu\mu) [4.3-6] $ & CMS8 \cite{Khachatryan:2015isa} \\
146 & $P_1(B \to K^* \mu\mu) [6-8.68] $ & CMS8 \cite{CMS:2017ivg} \\
147 & $P_5^\prime(B \to K^* \mu\mu) [6-8.68] $ & CMS8 \cite{CMS:2017ivg} \\
148 & $F_L(B \to K^* \mu\mu) [6-8.68] $ & CMS8 \cite{Khachatryan:2015isa} \\
149 & $A_{FB}(B \to K^* \mu\mu) [6-8.68] $ & CMS8 \cite{Khachatryan:2015isa} \\
150 & $10^7\times Br(B \to K^* \mu\mu) [6-8.68] $ & CMS8 \cite{Khachatryan:2015isa} \\
151 & $P_1(B \to K^* \mu\mu) [16-19] $ & CMS8 \cite{CMS:2017ivg} \\
152 & $P_5^\prime(B \to K^* \mu\mu) [16-19] $ & CMS8 \cite{CMS:2017ivg} \\
153 & $F_L(B \to K^* \mu\mu) [16-19] $ & CMS8 \cite{Khachatryan:2015isa} \\
154 & $A_{FB}(B \to K^* \mu\mu) [16-19] $ & CMS8 \cite{Khachatryan:2015isa} \\
155 & $10^7\times Br(B \to K^*  \mu\mu) [16-19] $ & CMS8 \cite{Khachatryan:2015isa} \\
156 & $F_L(B \to K^* \mu\mu) [1-2] $ & CMS7 \cite{Chatrchyan:2013cda} \\
157 & $A_{FB}(B \to K^* \mu\mu) [1-2] $ & CMS7 \cite{Chatrchyan:2013cda} \\
158 & $10^7\times Br(B \to K^* \mu\mu) [1-2] $ & CMS7 \cite{Chatrchyan:2013cda} \\
159 & $F_L(B \to K^* \mu\mu) [2-4.3] $ & CMS7 \cite{Chatrchyan:2013cda} \\
160 & $A_{FB}(B \to K^* \mu\mu) [2-4.3] $ & CMS7 \cite{Chatrchyan:2013cda} \\
161 & $10^7\times Br(B \to K^* \mu\mu) [2-4.3] $ & CMS7 \cite{Chatrchyan:2013cda} \\
162 & $F_L(B \to K^* \mu\mu) [4.3-8.68] $ & CMS7 \cite{Chatrchyan:2013cda} \\
163 & $A_{FB}(B \to K^* \mu\mu) [4.3-8.68] $ & CMS7 \cite{Chatrchyan:2013cda} \\
164 & $10^7\times Br(B \to K^* \mu\mu) [4.3-8.68] $ & CMS7 \cite{Chatrchyan:2013cda} \\
165 & $F_L(B \to K^* \mu\mu) [16-19] $ & CMS7 \cite{Chatrchyan:2013cda} \\
166 & $A_{FB}(B \to K^* \mu\mu) [16-19] $ & CMS7 \cite{Chatrchyan:2013cda} \\
167 & $10^7\times Br(B \to K^* \mu\mu) [16-19] $ & CMS7 \cite{Chatrchyan:2013cda} \\
168 & $10^5\times Br(B^0  \to K^{0 *}\gamma) $ & \cite{Amhis:2016xyh} \\
169 & $10^5\times Br(B^+  \to K^{+ *}\gamma) $ & \cite{Amhis:2016xyh} \\
170 & $10^5\times Br(B_s \to \Phi\gamma) $ & \cite{Amhis:2016xyh} \\
171 & $10^4\times Br(B \to X_s\gamma) $ & \cite{Amhis:2014hma} \\
172 & $10^9\times Br(B_s \to \mu\mu) $ & \cite{Aaij:2017vad} \\
173 & $S(B \to K^*  \gamma) $ & \cite{Asner:2010qj} \\
174 & $A_I(B \to K^*  \gamma) $ & \cite{Asner:2010qj} \\
175 & $10^6\times Br(B \to X_s \mu\mu)[1-6] $ & \cite{Lees:2013nxa} \\  & &  \\ \hline
\end{tabular} }
\caption{List of observables used in the fit continued.}
\label{t:obsc}
\end{table}%

\begin{table}
\center
\begin{tabular}{|c|c|}
\hline
ID & Observable\\ \hline
1 & $Q_1(B \to K^*) [0.045,1.1]$ \\
2 & $Q_2(B \to K^*) [0.045,1.1]$ \\
3 & $Q_4(B \to K^*) [0.045,1.1]$ \\
4 & $Q_5(B \to K^*) [0.045,1.1]$ \\
5 & $B_5(B \to K^*) [0.045,1.1]$ \\
6 & $B_{6s}(B \to K^*) [0.045,1.1]$ \\
7 & $Q_1(B \to K^*) [1.1, 6]$ \\
8 & $Q_2(B \to K^*) [1.1, 6]$ \\
9 & $Q_4(B \to K^*) [1.1, 6]$ \\
10 & $Q_5(B \to K^*) [1.1, 6]$ \\
11 & $B_5(B \to K^*) [1.1, 6]$ \\
12 & $B_{6s}(B \to K^*) [1.1, 6]$ \\
13 & $Q_1(B \to K^*) [0.1, 0.98]$ \\
14 & $Q_2(B \to K^*) [0.1, 0.98]$ \\
15 & $Q_4(B \to K^*) [0.1, 0.98]$ \\
16 & $Q_5(B \to K^*) [0.1, 0.98]$ \\
17 & $B_5(B \to K^*) [0.1, 0.98]$ \\
18 & $B_{6s}(B \to K^*) [0.1, 0.98]$ \\
19 & $Q_1(B \to K^*) [1.1, 2.5]$ \\
20 & $Q_2(B \to K^*) [1.1, 2.5]$ \\
21 & $Q_4(B \to K^*) [1.1, 2.5]$ \\
22 & $Q_5(B \to K^*) [1.1, 2.5]$ \\
23 & $B_5(B \to K^*) [1.1, 2.5]$ \\
24 & $B_{6s}(B \to K^*) [1.1, 2.5]$ \\ \hline
\end{tabular}
\begin{tabular}{|c|c|} \hline
ID & Observable \\ \hline
25 & $Q_1(B \to K^*) [2.5, 4]$ \\
26 & $Q_2(B \to K^*) [2.5, 4]$ \\
27 & $Q_4(B \to K^*) [2.5, 4]$ \\
28 & $Q_5(B \to K^*) [2.5, 4]$ \\
29 & $B_5(B \to K^*) [2.5, 4]$ \\
30 & $B_{6s}(B \to K^*) [2.5, 4]$ \\
31 & $Q_1(B \to K^*) [4, 6$ \\
32 & $Q_2(B \to K^*) [4, 6]$ \\
33 & $Q_4(B \to K^*) [4, 6]$ \\
34 & $Q_5(B \to K^*) [4, 6]$ \\
35 & $B_5(B \to K^*) [4, 6]$ \\
36 & $B_{6s}(B \to K^*) [4, 6]$ \\
37 & $Q_1(B \to K^*) [6, 8]$ \\
38 & $Q_2(B \to K^*) [6, 8]$ \\
39 & $Q_4(B \to K^*) [6, 8]$ \\
40 & $Q_5(B \to K^*) [6, 8]$ \\
41 & $B_5(B \to K^*) [6, 8]$ \\
42 & $B_{6s}(B \to K^*) [6, 8]$ \\
43 & $Q_1(B \to K^*) [15, 19]$ \\
44 & $Q_2(B \to K^*) [15, 19]$ \\
45 & $Q_4(B \to K^*) [15, 19]$ \\
46 & $Q_5(B \to K^*) [15, 19]$ \\
47 & $B_5(B \to K^*) [15, 19]$ \\
48 & $B_{6s}(B \to K^*) [15, 19]$ \\ \hline
\end{tabular}
\caption{List of proposed future observables.}
\label{tab:qObs}
\end{table}

\bibliography{biblio}

\end{document}